\documentstyle[psfig]{mn}

\newif\ifAMStwofonts

\newcommand{\be}{\begin{equation}}
\newcommand{\ee}{\end{equation}}
\newcommand{\bea}{\begin{eqnarray}}
\newcommand{\eea}{\end{eqnarray}}
\newcommand{\x}{${\bf x}$}
\newcommand{\mx}{{\bf x}}
\renewcommand{\v}{${\bf v}$}
\newcommand{\mv}{{\bf v}}
\newcommand{\q}{${\bf q}$}
\newcommand{\mq}{{\bf q}}
\renewcommand{\S}{${\bf S}$}
\newcommand{\mS}{{\bf S}}
\newcommand{\s}{${\bf s}$}
\newcommand{\ms}{{\bf s}}
\newcommand{\mpc}{$h^{-1}\; {\rm Mpc}$}

\ifoldfss
  \ifCUPmtlplainloaded \else
    \NewTextAlphabet{textbfit} {cmbxti10} {}
    \NewTextAlphabet{textbfss} {cmssbx10} {}
    \NewMathAlphabet{mathbfit} {cmbxti10} {} 
    \NewMathAlphabet{mathbfss} {cmssbx10} {} 
  \fi
  \ifAMStwofonts
    \ifCUPmtlplainloaded \else
      \NewSymbolFont{upmath} {eurm10}
      \NewSymbolFont{AMSa} {msam10}
      \NewMathSymbol{\upi}     {0}{upmath}{19}
      \NewMathSymbol{\umu}     {0}{upmath}{16}
      \NewMathSymbol{\upartial}{0}{upmath}{40}
      \NewMathSymbol{\leqslant}{3}{AMSa}{36}
      \NewMathSymbol{\geqslant}{3}{AMSa}{3E}

       \let\ge=\geqslant
    \fi
  \fi
\fi 

\ifnfssone
  \newmathalphabet{\mathit}
  \addtoversion{normal}{\mathit}{cmr}{m}{it}
  \addtoversion{bold}{\mathit}{cmr}{bx}{it}
  \newmathalphabet{\mathbfit} 
  \addtoversion{normal}{\mathbfit}{cmr}{bx}{it}
  \addtoversion{bold}{\mathbfit}{cmr}{bx}{it}
  \newmathalphabet{\mathbfss} 
  \addtoversion{normal}{\mathbfss}{cmss}{bx}{n}
  \addtoversion{bold}{\mathbfss}{cmss}{bx}{n}
  \ifAMStwofonts
    \ifCUPmtlplainloaded \else
      \UseAMStwoboldmath
      \makeatletter
      \new@mathgroup\upmath@group
      \define@mathgroup\mv@normal\upmath@group{eur}{m}{n}
      \define@mathgroup\mv@bold\upmath@group{eur}{b}{n}
      \edef\UPM{\hexnumber\upmath@group}
      \new@mathgroup\amsa@group
      \define@mathgroup\mv@normal\amsa@group{msa}{m}{n}
      \define@mathgroup\mv@bold\amsa@group{msa}{m}{n}
      \edef\AMSa{\hexnumber\amsa@group}
      \makeatother
      \mathchardef\upi="0\UPM19
      \mathchardef\umu="0\UPM16
      \mathchardef\upartial="0\UPM40
      \mathchardef\leqslant="3\AMSa36
      \mathchardef\geqslant="3\AMSa3E

       \let\ge=\geqslant
    \fi
  \fi
\fi 

\ifnfsstwo
  \DeclareMathAlphabet{\mathbfit}{OT1}{cmr}{bx}{it}
  \SetMathAlphabet\mathbfit{bold}{OT1}{cmr}{bx}{it}
  \DeclareMathAlphabet{\mathbfss}{OT1}{cmss}{bx}{n}
  \SetMathAlphabet\mathbfss{bold}{OT1}{cmss}{bx}{n}
  \ifAMStwofonts
    \ifCUPmtlplainloaded \else
      \DeclareSymbolFont{UPM}{U}{eur}{m}{n}
      \SetSymbolFont{UPM}{bold}{U}{eur}{b}{n}
      \DeclareSymbolFont{AMSa}{U}{msa}{m}{n}
      \DeclareMathSymbol{\upi}{0}{UPM}{"19}
      \DeclareMathSymbol{\umu}{0}{UPM}{"16}
      \DeclareMathSymbol{\upartial}{0}{UPM}{"40}
      \DeclareMathSymbol{\leqslant}{3}{AMSa}{"36}
      \DeclareMathSymbol{\geqslant}{3}{AMSa}{"3E}

       \let\ge=\geqslant
    \fi
  \fi
\fi 

\ifCUPmtlplainloaded \else
  \ifAMStwofonts \else 
    \def\upi{\pi}
    \def\umu{\mu}
    \def\upartial{\partial}
  \fi
\fi

\title[Reconstruction of initial conditions]{Reconstruction of
cosmological initial conditions from galaxy redshift catalogues}

\author[P. Monaco \& G. Efstathiou]
       {Pierluigi Monaco$^{1,2}$ \& George Efstathiou$^1$\\
	$^1$Institute of Astronomy, Madingley Road, Cambridge CB3 0HA, GB\\
      $^2$Dipartimento di Astronomia, Via Tiepolo 11, 34131 Trieste -- Italy\\
	Email: monaco, gpe@ast.cam.ac.uk}
\date{}

\pagerange{\pageref{firstpage}--\pageref{lastpage}}
\pubyear{2000}

\begin{document}

\maketitle

\label{firstpage}

\begin{abstract}
We present and test a new method for the reconstruction of
cosmological initial conditions from a full-sky galaxy catalogue.
This method, called ZTRACE, is based on a self-consistent solution of
the growing mode of gravitational instabilities according to the
Zel'dovich approximation and higher order in Lagrangian perturbation
theory.  Given the evolved redshift-space density field, smoothed on
some scale, ZTRACE finds via an iterative procedure, an approximation
to the initial density field for any given set of cosmological
parameters; real-space densities and peculiar velocities are also
reconstructed.  The method is tested by applying it to N-body
simulations of an Einstein-de Sitter and an open cold dark matter
universe.  It is shown that errors in the estimate of the density
contrast dominate the noise of the reconstruction.  As a consequence,
the reconstruction of real space density and peculiar velocity fields
using non-linear algorithms is little improved over those based on
linear theory.  The use of a mass-preserving adaptive smoothing,
equivalent to a smoothing in Lagrangian space, allows an unbiased
(although noisy) reconstruction of initial conditions, as long as the
(linearly extrapolated) density contrast does not exceed unity.  The
probability distribution function of the initial conditions is
recovered to high precision, even for Gaussian smoothing scales of
$\sim 5$ \mpc, except for the tail at $\delta\ge 1$.  This result is
insensitive to the assumptions of the background cosmology.
\end{abstract}


\section{Introduction}

Large and homogeneous galaxy catalogues are defining the present-day
density field of galaxies to increasing precision.  Assuming that
galaxies trace the underlying matter field in some known way, it
should be possible to trace the matter density field back in time,
thus recovering the initial conditions of the local Universe.  The
initial conditions are of importance for at least two reasons:
firstly, it is interesting to examine the probability distribution
function (hereafter PDF) of the initial density field, which is
expected to be Gaussian in many theories of the origin of fluctuations
(see {\it e.g.} Linde, 1990); secondly, it is possible to use the
initial conditions derived from a galaxy redshift survey as the
starting point for an N-body simulation, to reproduce the non-linear
dynamics of our local Universe in some detail (see, for example,
Kolatt et al. 1996).

Although conceptually simple, the implementation of this idea to real
data presents some difficulties.  Firstly, it is not possible to
integrate the present-day density field back in time with an N-body
simulation: the familiar decaying mode of cosmological matter
perturbations will amplify noise in the backward integration.  This
can be avoided by using semi-analytical methods restricted to pure
growing-mode dynamics (see Nusser \& Dekel 1992).  Secondly, the
galaxy density field is measured in redshift space; this complicates
the analysis as redshift space is not isotropic, except around the
observer, and the map from initial to final positions is not
irrotational (see Section 2).  Moreover, ``multi-stream'' regions
around collapsing structures affect the mapping from the initial
conditions to both the real and redshift space galaxy distribution.
In redshift space, multi-stream regions are sometimes referred to as
``triple valued'' regions.  The existence of multi-stream regions
imposes a fundamental limit on the accuracy of reconstruction
algorithms.  In Section 2 we show that the effect of multi-streaming
is more severe in redshift space than in real space. Thus the
reconstruction of density fields using redshift surveys is limited to
modest (real-space) density contrasts ($\delta \rho/\rho \la 1$).
Thirdly, the galaxy density field is generally estimated from flux
limited galaxy catalogues, which become sparse at large distances; it
will be shown in Section 4 that the density estimate provides an
important source of noise in the reconstruction of initial density
fields.  Fourthly and most importantly, the relationship between
galaxies and the underlying mass density distribution is not well
known.  Thus progress can only be made if we adopt simple assumptions
concerning the nature of this relationship.

The most widely used analytical tool for the evolution of large-scale
structure is linear perturbation theory. The topic of redshift-space
distortion in linear theory has been reviewed by Hamilton (1998).
Linear theory has been used to recover the real-space density and
peculiar velocity fields from near all-sky galaxy catalogues, for
example, the QDOT (Kaiser et al. 1991), 1.2 Jy (Yahil et al. 1991;
Webster, Lahav \& Fisher 1997) and PSCz (Branchini et al. 1998) IRAS
catalogues, the optical catalogue compiled by Hudson (1994), the
Optical Redshift Catalogue (ORS; Baker et al. 1998), and the Abell
clusters catalogue (Branchini \& Plionis 1996).  On the other hand,
linear theory cannot be used to recover the initial conditions of our
local Universe. According to linear theory, the initial density field
is just a rescaled version of the final one (in real space), which is
strongly non-Gaussian as long as the mass variance is not much less
than unity.

The Zel'dovich (1970) approximation offers a useful tool for
reconstructing initial conditions, as it provides a description of the
growing mode of gravitational evolution in the mildly non-linear
regime when the variance of the density field is of order unity.
Nusser \& Dekel (1992) recast the Zel'dovich approximation in terms of
a Bernoulli-type evolution equation for the peculiar velocity
potential which can easily be integrated back in time once the
present-day peculiar velocity field is known.  In fact, another
reconstruction algorithm is needed to obtain the peculiar velocity
potential. Nusser, Dekel \& Yahil (1995) applied the formalism of
Nusser \& Dekel (1992) to the peculiar velocity potential found by
applying a non-linear approximation to the IRAS 1.2 Jy catalogue (see
also Yahil et al. 1991).  They found that the initial conditions of
the local Universe (smoothed with a Gaussian filter of width 10 \mpc)
are consistent with a Gaussian PDF for simple models of the relative
bias between the galaxy and mass distributions.  An alternative,
improved method, which is consistent with mass conservation, is given
by Gramann (1993a).  Gramann (1993b) and Susperregi \& Buchert (1997)
extended the formalism to second order in the Lagrangian perturbation
theory.  Taylor \& Rowan-Robinson (1993) developed a quasi-non-linear
reconstruction algorithm in which the dynamics is still described at a
linear level, but the mapping from real- to redshift-space is solved
exactly.

A simpler method was proposed by Weinberg (1992). If the PDF of
initial conditions is assumed to be Gaussian, then a better estimate
of the initial conditions is obtained by ``Gaussianizing'' the PDF
computed using linear theory. Assuming that non-linear dynamics
preserves the order of densities, the inferred initial densities can
be rescaled by an order-preserving transformation so that they have a
Gaussian PDF.  Methods of this kind were used by Kolatt et al. (1996)
and by Narayanan \& Weinberg (1998), where the Gaussianization
procedure was applied to the results of a Zel'dovich reconstruction.
The obvious problem with this method is that the Gaussianity of the
PDF is assumed instead of being inferred from the data. For example,
the initial PDF of the galaxy distribution may well not be Gaussian if
the relation between the galaxies and the mass distribution is
complex.  Moreover, the order-preserving condition may be a poor
approximation of the actual dynamics because of the non-locality of
gravitational evolution.

A different approach based on least action was proposed by Peebles
(1989, 1990). The initial positions of galaxies in a catalogue can be
found by searching for the displacements which satisfy a least action
constraint under some assumption regarding the mass distribution ({\it
e.g} that all of the mass is associated with point-like galaxies).  A
version of this method was tested by Branchini \& Carlberg (1994) who
found that it can lead to an underestimate of the cosmological density
$\Omega$, and applied to the groups of the local supercluster by
Shaya, Peebles \& Tully (1995).  Actually, as shown by Giavalisco et
al. (1993) and Croft \& Gatza\~naga (1997), the Zel'dovich
approximation is the first-order solution of the least-action
principle.  Croft \& Gatza\~naga (1997) used the least action
principle to set up a reconstruction method based on the Zel'dovich
approximation, called path-interchange Zel'dovich approximation
(hereafter PIZA). Given a set of points coming from a uniform
configuration, the map from initial to final positions is found by
minimizing the action constructed by the average square displacement.
Although different from the other reconstruction algorithms, the
least-action methods suffer from similar problems to the methods
described above: the mapping from redshift- to real-space presents
similar difficulties, multi-streaming gives multiple solutions for the
orbits, and some specific assumption (analogous to bias) is required
to relate the mass and galaxy distributions.

Narayanan \& Croft (1998) compared the performances of different
reconstruction algorithms to N-body simulations. The most accurate
initial conditions were recovered by the hybrid reconstruction scheme
of Narayanan \& Weinberg (1998) and by the PIZA scheme.  However, in
their tests the initial conditions were reconstructed from the full
output of the simulation in real space and the effects of
redshift-space distortions and of the galaxy selection function,
critical in the analysis of real galaxy catalogues, were ignored.

The Zel'dovich approximation can be used to obtain the initial density
field directly from the redshift-space density field, under some
assumption of bias.  This cannot be achieved with the Bernoulli
equation of Nusser \& Dekel (1992) or with the alternative one of
Gramann (1993a), as they require as input the peculiar velocity
potential, which is obtained (in a way that is not self-consistent)
from a linear theory analysis.  The PIZA method is in principle able
to give a self-consistent determination of initial condition.

The approaches of Nusser \& Dekel (1992) and Gramann (1993a) cannot be
extended beyond the Zel'dovich approximation without major changes to
the formalism (see Gramann 1993b; Susperregi \& Buchert 1997).  It is
well known that the Zel'dovich approximation is the first term of a
Taylor expansion of the Lagrangian map (see, e.g., Moutarde et
al. 1991; Buchert \& Ehlers 1993; Catelan 1995), and it would be
desirable to develop a reconstruction analysis that is easily
extendible to higher orders to check whether or not such contributions
are important.  The effects of higher order terms have been analyzed,
for instance, by Gramann (1993b), Hivon et al. (1995) and Chodorowsky
et al. (1998).

In this paper we present an iterative reconstruction algorithm based
on a self-consistent solution of the Zel'dovich (or higher-order)
approximation; this algorithm is able to find the initial density
field which evolves, with Zel'dovich (or 2nd-order Lagrangian
perturbation theory), into a given final density field in redshift
space.  In addition to the initial conditions, the real-space density
field and the line-of-sight (hereafter LOS) peculiar velocity field
can be recovered.  Here, we focus on the recovery of initial
conditions because, as will be shown in Section 5, when applied to
realistic simulated galaxy catalogues non-linear reconstruction
algorithms do not give greatly improved final density and velocity
fields over those inferred using linear theory (which is, of course,
much easier to implement).

The paper is organized as follows: Section 2 presents the mathematical
formalism and describes the iterative method.  Section 3 describes the
numerical simulations used to test the reconstruction algorithms.  In
Section 4 we describe the noise introduced by estimating the density
from a catalogue of discrete points.  The results of the
reconstruction algorithm and tests against N-body simulations are
presented in Section 5.  The conclusions are summarized in Section 6.

Throughout this paper all distances are given in \mpc, with
$h=H_0/(100\ km/s/Mpc)$.

\section{The Reconstruction Algorithm}

\subsection{Equations}

The evolution of a self-gravitating fluid can be described by
Lagrangian fluid-dynamics.  In this approach, the dynamical variable
is the displacement \S, which maps the point-mass particles from the
initial (comoving Lagrangian, \q) to the final (comoving Eulerian, \x)
position:

\be \mx(\mq,t)=\mq+\mS(\mq,t). \label{eq:map} \ee

The Euler-Poisson system of equations (see Peebles 1980) can be recast
in terms of a set of equations for the displacement \S; these
equations are given by Buchert (1989), Bouchet et al. (1995) and
Catelan (1995).  The notation used in this paper follows that of
Catelan (1995).  The system can be solved perturbatively for small
displacements.  Denoting by $D(t)$ the linear growing mode, the first
two terms of the perturbative series are:

\be \mS^{(1)}(\mq)=-D(t)\nabla\varphi(\mq); \label{eq:zel} \ee
\be \mS^{(2)}(\mq)=-\frac{3}{14}D(t)^2\nabla\varphi^{(2)}(\mq). \label{eq:2nd} \ee

\noindent
$\varphi(\mq)$ is simply a rescaled version of the initial peculiar
gravitational potential, such that:

\be \nabla^2\varphi(\mq)=\delta(\mq,t_i)/D(t_i), \label{eq:poi1} \ee

\noindent
where $\delta(t_i)$ is the initial density contrast ($\delta=(\varrho
-\bar{\varrho})/\bar{\varrho}$), and $D(t_i)$ is the growing mode at
the initial time $t_i$.  Note that the quantity $\delta_l(\mq)=
\delta(\mq,t)/D(t)$ is constant in linear theory. If the growing mode
is normalized to $D(t_0)=1$ at the final time $t_0$, $\delta_l$ is the
density contrast linearly extrapolated at $t=t_0$; this quantity will
be referred to as the linear density contrast in this paper.  Note
also that the map \S\ has been assumed to be irrotational because
rotational components do not contribute to the growing mode.

The second-order potential obeys another Poisson equation:

\be \nabla^2\varphi^{(2)}(\mq)=2\mu_2(\varphi_{,ab}), \label{eq:poi2} \ee

\noindent
where $\mu_2(\varphi_{,ab})=(\varphi_{,aa}\varphi_{,bb}-\varphi_{,ab}
\varphi_{,ab})/2$ is the second principal invariant of the tensor of
second derivatives of $\varphi$ (comma denotes derivative with respect
to the Lagrangian coordinate \q, summation over repeated indices is
assumed; see, e.g., Catelan 1995).

The peculiar velocity \v\ of the mass element \q\ is obtained from the
displacement \S\ as:

\be \mv(\mq,t)=a\frac{dD}{dt}\frac{d\mS}{dD} = \frac{dD}{dt}\frac{a}{D}
(\mS^{(1)}+2\mS^{(2)} +\ldots). \label{eq:pecv} \ee

\noindent
Here $a(t)$ is the scale factor (normalized to $a(t_0)=1$).  In a
reference frame centred on the observer, the redshift-space coordinate
\s\ is defined as:

\be
\ms(\mq,t)=\mx(\mq,t)+\frac{1}{aH}(\mv(\mq,t)\cdot\hat{\mx}(\mq,t))
\hat{\mx}(\mq,t).
\label{eq:redspace} \ee

\noindent
Here $H(t)$ is the Hubble parameter, while $\hat{\mx}$ is the versor
of \x.  It is possible to define a redshift-space map $\mS^{(s)}$ as
follows:

\be \ms(\mq,t)=\mq+\mS^{(s)}(\mq,t). \label{eq:ss_map} \ee

\noindent Then:

\bea \lefteqn{\mS^{(s)}(\mq,t)=}\label{eq:redmap} \\&&
\mS^{(1)}+\mS^{(2)}+\ldots + f(\Omega)[
(\mS^{(1)}+2\mS^{(2)}+\ldots)\cdot\hat{\mx}]\hat{\mx}. \nonumber
\eea

The function $f(\Omega)=d\ln D/d \ln a$ is usually approximated by
$f(\Omega)\simeq \Omega^{0.6}$ (Peebles 1980); if a cosmological term
is present, then $f(\Omega,\Omega_\Lambda) \simeq \Omega^{0.6}
+\Omega_\Lambda (1+\Omega/2)/70$ (Lahav et al. 1991).  It is
noteworthy that, while the real-space map \S\ is irrotational, the
redshift-space one $\mS^{(s)}$ is in general rotational (see, e.g.,
Nusser \& Davis 1994).

Given a map \S, the density contrast can be calculated through the
continuity equation:

\be 1+\delta(\mq)=\frac{1}{\det(\delta^K_{ab}+S_{a,b})}. \label{eq:delta} \ee

\noindent
Here $\delta^K_{ab}$ is the Kronecker tensor.  It is useful to recall
the identity:

\be \det(\delta^K_{ab}+S_{a,b}) = 1+\mu_1(\mS)+\mu_2(\mS)+\mu_3(\mS),
\label{eq:det} \ee

\noindent
where $\mu_i(\mS)$ are the principal invariants of the first
derivative tensor of \mS; $\mu_1=S_{a,a}$ is the divergence of the
displacement \S, $\mu_2(\mS)=(S_{a,a}S_{b,b}-S_{a,b} S_{a,b})/2$, and
$\mu_3(\mS)=\det(S_{a,b})$.

Inserting the redshift-space map $\mS^{(s)}$ (Eq.~\ref{eq:redmap})
into Eq.~\ref{eq:delta}, and noting that $\mu_1(\mS^{(1)})=-D\delta_l$
(Eqs.~\ref{eq:zel} and \ref{eq:poi1}), the following identity is
obtained:

\bea \lefteqn{\delta_l =}\label{eq:iter} 
\\ && \frac{1}{D}\left[\frac{\delta_s}{1+\delta_s}
-\mu_1(\mS^{(1)}) +\mu_1(\mS^{(s)})+\mu_2(\mS^{(s)})+\mu_3(\mS^{(s)})\right]. 
\nonumber \eea

\noindent
Here $\delta_s$ is the density contrast in redshift space, to be
distinguished from the real-space density contrast $\delta_x$.

The perturbative series breaks down when the map in Eq.~\ref{eq:map}
or Eq.~\ref{eq:ss_map} becomes multi-valued.  For the real-space map,
when this takes place the interested mass elements undergo pancake
collapse and go into the multi-stream regime. On the other hand,
redshift-space multi-streaming takes place when a perturbation
decouples from the Hubble flow; the infall velocity then becomes
comparable to the difference in redshift across the collapsing region
and the distance-redshift relation becomes triple-valued.  The
perturbative expansion will thus break down at lower real-space
density contrasts when applied to redshift space.

The linear theory limit can be derived easily from the equations
presented above.  Displacements are assumed to be infinitesimally
small, and the displacement from \q- to \x-space gives only a
second-order contribution to the density, so that the difference
between Eulerian and Lagrangian spaces can be neglected.  The
real-space density contrast is simply:

\be \delta_x = - \frac{\nabla\cdot\mv}{aHf(\Omega)}, \label{eq:lin_r} \ee

\noindent
while the redshift-space density is:

\be \delta_s = \delta_x + Df(\Omega)\nabla\cdot[(\nabla\varphi\cdot\hat{\mx}) 
\hat{\mx}]. \label{eq:lin_s} \ee

\subsection{The Method: Iteration and Convergence}

Given a smooth evolved density field evaluated on a regular grid in
redshift space, we want to obtain the linear density field $\delta_l$
which evolves into the input density field.  A direct inversion of the
equations presented in the previous section is intractable; however,
the identity given in Eq.~\ref{eq:iter} can be used to set up an
iterative method to reconstruct the initial conditions $\delta_l$.
The first guess for the linear density contrast can be simply given by
$\delta_s/(1+\delta_s)$, where $\delta_s$ is the evolved
redshift-space density contrast; from the first guess initial density
it is possible to obtain numerically the map \mS\ in real and redshift
space from Eqs~\ref{eq:poi1}, \ref{eq:zel} and \ref{eq:2nd}, and then,
using Eq.~\ref{eq:iter}, a new guess for the linear density can be
found. This cycle can be repeated until convergence is achieved.

The input redshift-space density field must be smoothed to eliminate
high density contrasts and regions of orbit-crossing, where we would
not expect the reconstruction method to work.

The final density field $\delta_s$ is given on a regular grid in
redshift \s-space, while the reconstruction is performed on a regular
grid in the Lagrangian \q-space; in other words, the quantity
$\delta_s$ in Eq.~\ref{eq:iter} is a function of \q, while the input
redshift-space field is given as a function of \s.  If the map
$\mS^{(s)}(\mq)$ is known (again on the regular grid in \q), one can
obtain the \q-space density from the \s-space density by interpolating
the density contrast $\delta_s$ at the positions in \s-space
corresponding to the regular grid in \q-space.  The term
$\delta_s(\mq)/(1+\delta_s(\mq))$ in Eq.~\ref{eq:iter} is then, in
practice, a function of \S, which changes at each iteration.

The first guess for the map is $\mS=0$, i.e. it is assumed that the
\q-space linear density is just given by the \s-space evolved density.
As a consequence, the first guess for the linear density is not
properly normalized.  In fact, the correct normalization of the
density field is never guaranteed, as the galaxy catalogue to which
the reconstruction algorithm is applied may not be a fair sample of
the Universe.

In summary, the iteration method has been implemented as follows:

\begin{enumerate}
\item The final, observed density field in redshift space, $\delta_s$,
is calculated on a cubic grid of points and provided as an input.
\item The first guess for the linear density is just $\delta_l(\mq)=
\delta_s(\ms)/(1+\delta_s(\ms))$, with $\mS^{(s)}=0$ (note that
$D(t_0)=1$: the initial density is just a simple function of the final
density.
\item The estimate of the linear density $\delta_l$ is used to
calculate the peculiar potential $\varphi$ (Eq.~\ref{eq:poi1}), the
\x-space map \S\ (Eq.~\ref{eq:zel}, and Eqs.~\ref{eq:poi2} and
\ref{eq:2nd} if 2nd order is used), the peculiar velocity \v\
(Eq.~\ref{eq:pecv}) and the \s-space map $\mS^{(s)}$
(Eq.~\ref{eq:redmap}).
\item Eq.~\ref{eq:iter} is used to obtain a new guess for the linear
density $\delta_l$.
\item Steps (iii) and (iv) are repeated until the mean square
difference between the old and new estimates of the linear density
field meets a specified criterion and the density field is convergent
at each grid point.
\end{enumerate}

Differentiation and the solution of the Poisson equations are done
with Fast Fourier Transforms (FFTs): FFTs provide fast and accurate
derivatives of the density and velocity fields.  However, periodic
boundary conditions do not hold in redshift space.  The natural
geometry of an all-sky galaxy catalogue is spherical, and the
imposition of periodic boundary conditions is clearly artificial.
Periodic boundary conditions are forced by padding the density field
outside the largest sphere inscribed within the cubic volume used for
the FFT.  The padding value is chosen so that the initial density
field has zero mean within the whole box; this procedure defines the
overall normalization of the density field at each iteration.

The iteration scheme is used to solve a complex non-linear and
non-local set of equations. The convergence of the iterative method is
not guaranteed and care is needed to achieve it.  The occurrence of
orbit crossing at a few points may indicate some problem with the
convergence, but Eq.~\ref{eq:iter} forces the system into the
single-stream regime, so it is possible to recover from orbit
crossings in many cases.  The most problematic points are (perhaps
counterintuitively at first sight) not large overdensities, but deep
underdensities; the final density $\delta_s$ enters Eq.~\ref{eq:iter}
as $\delta_s/ (1+\delta_s)$, so the contribution from overdensities
asymptotically approaches unity. Deep underdensities, however, give
large negative contributions (moreover, as shown by Sahni \& Shandarin
1996 and Monaco 1997, the perturbative Lagrangian series does not even
converge in deep underdensities).  Another problem with the
convergence is that there are discontinuities at the border of the
padded region, especially when no selection function is applied to the
density field.  Both deep underdensities and discontinuities can make
the solution ``explode'' at some points and this, because of the
non-locality of the system of equations, propagates over the whole
volume.  Finally, the system does not converge at the position of the
observer, which is a singular point in the \x--\s\ tranformation.

The following techniques have been used to help achieve convergence:

\begin{enumerate}
\item To avoid oscillations of the solution, at each iteration the new
guess for the linear density is given by a weighted mean of
Eq.~\ref{eq:iter} and the old guess.  This is equivalent to
introducing a numerical viscosity for damping the oscillations.  The
first ``old'' guess is assumed to be a null field.  The weights $w1$
(for Eq.~\ref{eq:iter}) and $w2=1-w1$ (for the old guess) are set to
0.2 and 0.8 at the starting time, so that the variance of the first
guess is small and no orbit crossing occurs from the start. The
weights are increased so as to take values 0.4 and 0.6 after 6
iterations.

\item To force the convergence at the center, both the guess for the
density field and the LOS peculiar velocity are smoothed with a
filter, $exp(-(q_{cut}/q)^2)$.  As a consequence, the initial density
within the central $\sim 2 q_{cut}$ grid points is not correctly
recovered.

\item To avoid problems due to discontinuities at the border of the
padded region, the guess for the linear density is linearly suppressed
(i.e. multiplied by a function linearly decreasing with radius) from a
given radius up to the padding radius (half the box size).  The
suppression, which is performed to smoothly join the field with the
constant padding value, depends on the padding value itself. In turn
the padding value depends on the average value of the field inside the
sphere, and hence on the suppression.  The padding value is thus found
iteratively and usually 3 iterations are sufficient to reach
convergence.  The distance at which the suppression starts depends on
whether a selection function is applied to the density: if we apply
the method to the full density field from a numerical simulation, then
the density is suppressed from 0.7 times half the box length; if a
simulated galaxy redshift survey is used, with the smoothing strategy
described in Section 5.2, the outer more sparsely sampled regions are
smoother than the inner regions and the suppression can be applied at
a radius of 0.9 times half the box length.

\item The deepest underdensities are truncated to a fixed limiting
value, set to $\delta_s/(1+\delta_s)=-5$ ($\delta_s=-0.83$) with
Zel'dovich and $-3$ with 2nd order ($\delta_s=-0.75$).  This does not
have a significant impact on the final results, as just a few grid
points are affected by the truncation. On the other hand, such points
can make the solution explode.
\end{enumerate}

The linear density is reliably reconstructed only in the central parts
of the computational volume.  This is a consequence of using an FFT
and a Cartesian grid for a problem with a more natural spherical
symmetry.  However, FFTs allow us to solve the complex system of
equations in an acceptable amount of computer time.  In principle, we
could use spherical harmonics in place of FFTs, but at the cost of a
great increase in the complexity of the equations and computer time.
It is worth noting that the the non-iterative method of Fisher et al
(1995), based on spherical harmonics, can be used only in linear
theory when the displacements from the \q- to the \x-space are
neglected; the transverse components of such displacements couple
different spherical harmonics and a direct inversion of the coupling
matrix rapidly becomes computationally intractable as the number of
spherical harmonics is increased.

We applied the iterative method to numerical simulations of cold dark
matter (CDM) models with a computational box length of 240 \mpc. The
models are described more fully in Section 3. Typically, we used a
grid of $64^3$ points to define the density and velocity fields, thus
the grid spacing is 3.75 \mpc.  For a smooth density field with
variance in the range from 0.1 to 0.7, the iteration method is able to
converge in 10-15 iterations; for a $64^3$ grid, convergence is
achieved in about 5 minutes of CPU time on a DEC Alpha 4100 5/300.
The solution is assumed to have converged when the mean square
difference between the old and new linear field is smaller than 1\% of
its variance; we have always checked that the largest difference,
which can be comparable to the variance of the field, is decreasing at
the convergence.  The convergence is slowest in overdensities, so that
the height of the highest peaks can be underestimated; on the other
hand, those peaks are presumably in multi-stream regime (or in
triple-valued regions), so they would be underestimated in any case.
The variance of the inferred linear field converges to a stable value.

The method described in this Section will be referred to as ZTRACE in
the rest of this paper.

\begin{figure*}
\centerline{
\psfig{figure=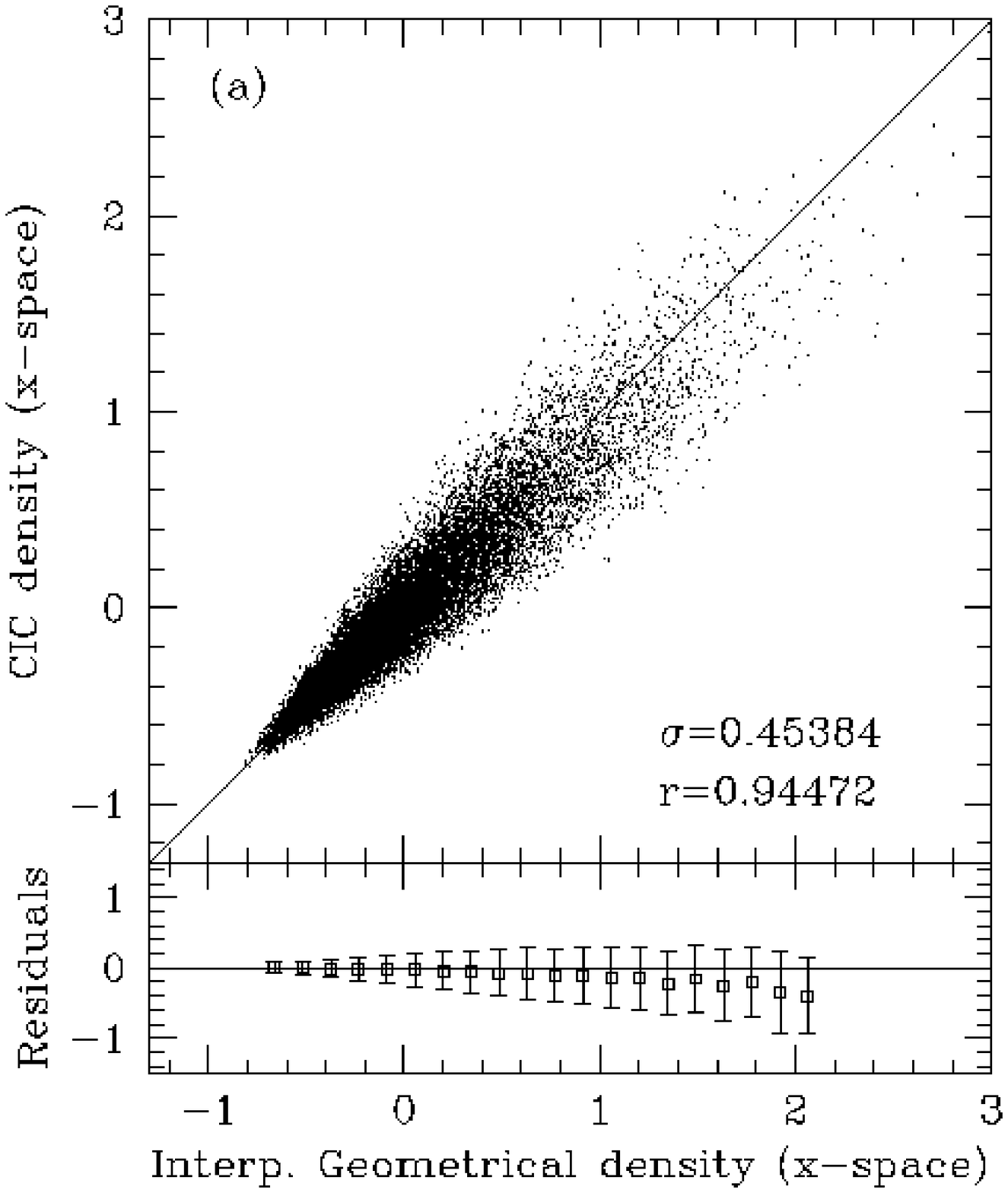,width=6cm}
\psfig{figure=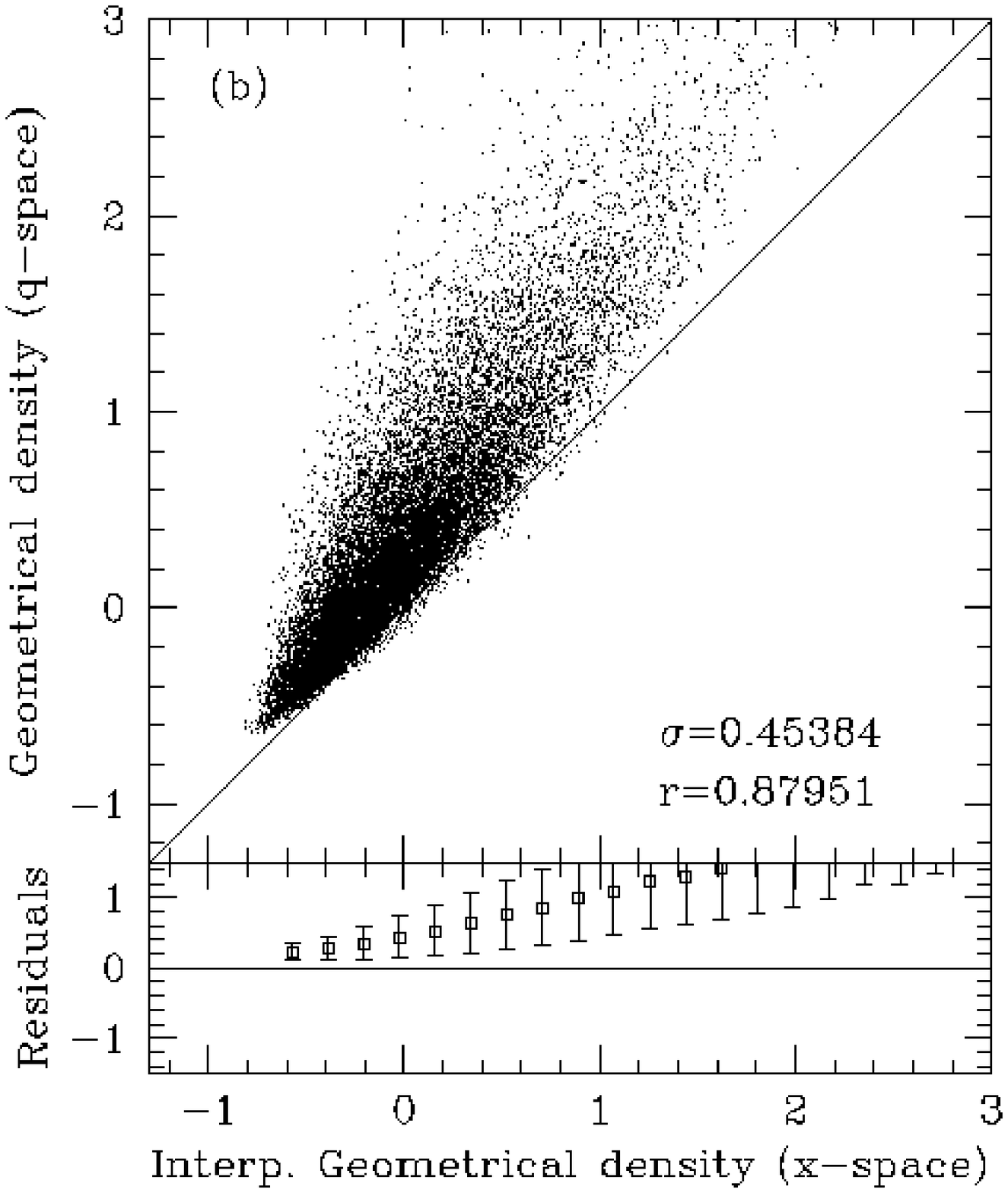,width=6cm}}
\centerline{
\psfig{figure=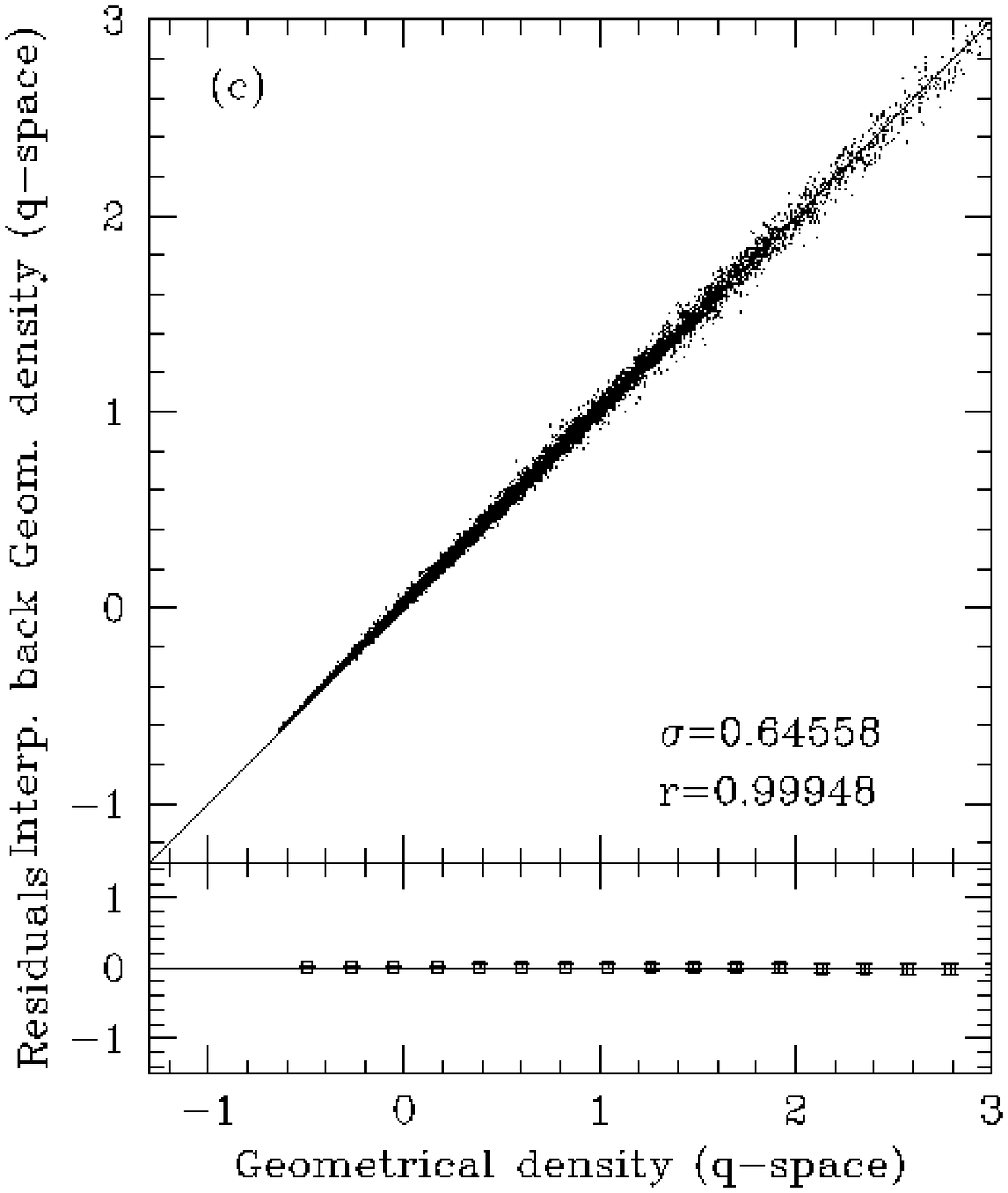,width=6cm}
\psfig{figure=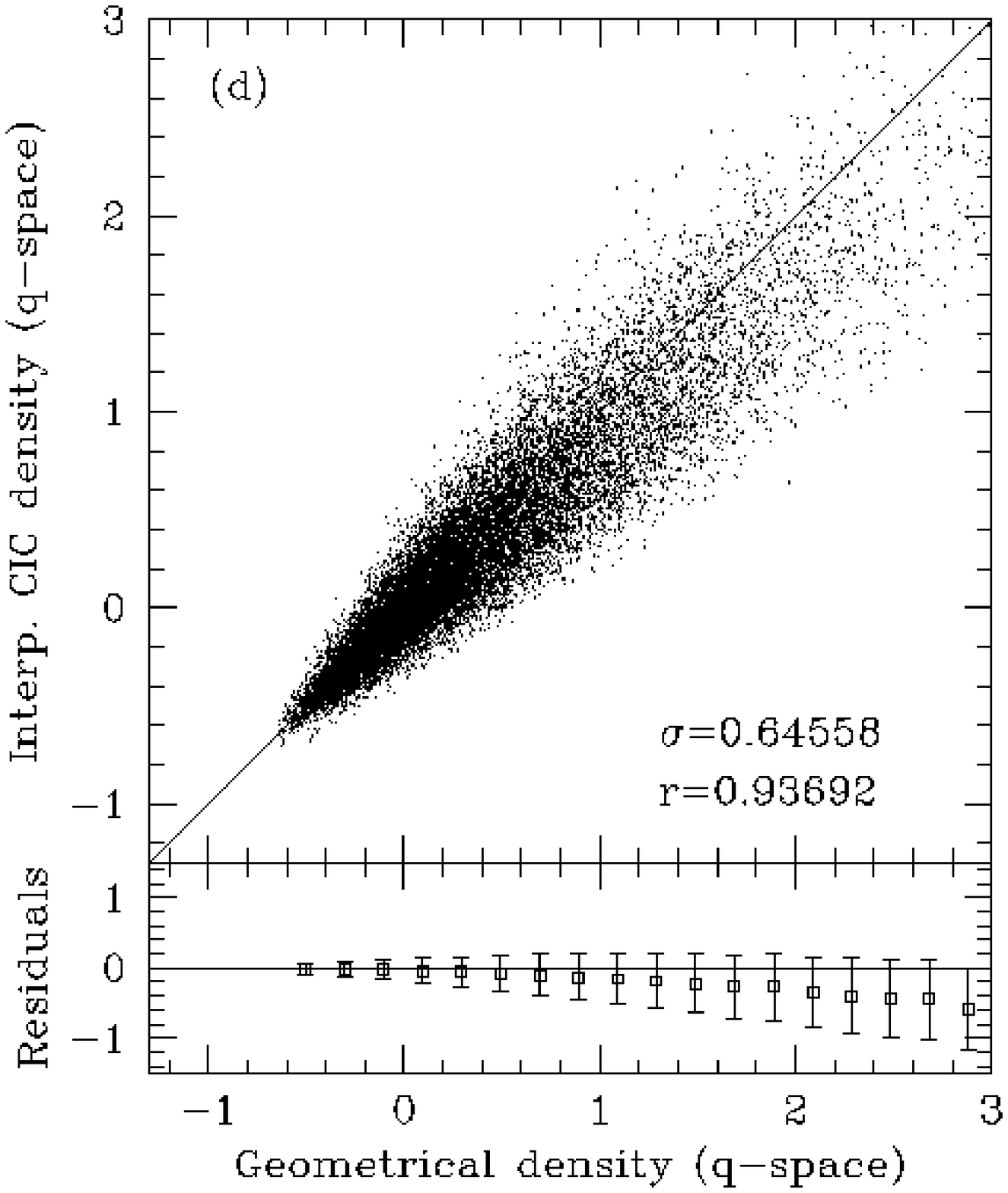,width=6cm}}
\caption{Comparison of the geometrical and CIC density estimates: (a)
in \x-space; (b) geometrical density in \q- and \x-space; (c) error
introduced by the double interpolation (\q-space); (d) error
introduced by the CIC estimate (\q-space).}
\end{figure*}

The same method has been adapted to two simpler cases, that of linear
theory and that in which the real-space density field is given as
input.  The implementation of linear theory requires minor changes to
the method: Eqs.~\ref{eq:lin_r} and \ref{eq:lin_s} are used to obtain
an estimate of the linear density from the redshift-space density
field, while Eq.~\ref{eq:pecv} (obviously with only the first order
term) is used to obtain the peculiar velocity.  The map from \q- to
\x-space has only a formal meaning, as the difference between the two
spaces is neglected (at first order in the density).  The padding
procedure is left unchanged, no limit is set to underdensities, the
central smoothing of the field is not performed and the weights $w1$
and $w2$ are set to (0.3,0.7) at the start, increasing to (0.5,0.5)
after three iterations.  The convergence is reached within 5
iterations.  This linear version of the reconstruction method will be
referred to as LTRACE.

To compare with the results of Narayanan \& Croft (1998), the method
has been adapted to recover initial conditions from the real-space
density field of a numerical simulation.  In this case, periodic
boundary conditions hold and neither the padding procedure not the
central smoothing are required.  The truncation of deep underdensities
is still performed, and the weights $w1$ and $w2$ are fixed to
(0.5,0.5).  The convergence is faster than the application to
redshift-space and is achieved in 8-10 iterations.  This version will
be referred to as XTRACE.

\section{N-body Simulations}

To test the reconstruction methods we have run N-body simulations with
version 2.0 of the Hydra code using pure dark matter (see Couchman,
Thomas \& Pearce 1995 for full details).  $128^3$ particles are
simulated in a box of length 240 \mpc\ using a $256^3$ base mesh with
adaptive refinements.  Two simulations have been run: one in an
Einstein-de Sitter (henceforth EdS) Universe with $h=0.5$, and one in
an open Universe with $\Omega=0.3$, $h=0.7$ and no cosmological
constant.  The two simulations have been run with the same random
number seed, so that they have the same phases.  To generate the
initial power spectra, we used the parameterization of Efstathiou,
Bond \& White (1992) for the power spectrum of an initially scale
invariant, adiabatic, CDM-dominated universe. For the power spectrum
shape parameter $\Gamma$, we adopted values of $\Gamma=0.5$ for the
EdS simulation and $\Gamma = 0.21$ for the open simulation.  In both
simulations the final dispersion of the mass fluctuations within
spheres of radius 8 \mpc\ was set to $\sigma_8=0.7$.

The box length has been chosen so that the radius of the largest
sphere contained within the box is 120 \mpc, so that we can reliably
simulate a galaxy catalogue out to at least 80 \mpc while having the
mass resolution to resolve non-linear structures.  The tests of the
reconstruction algorithms using the EdS and open simulations are
almost identical and so with the exception of Section 5.6, we present
comparisons exclusively with the EdS simulation.

\section{Noise in Density Estimates}

The largest source of noise in the reconstruction algorithms comes
from noise in the estimate of the input evolved density field.  This
can be demonstrated by using a true density field that is given by
Eq.~\ref{eq:delta} from a known displacement map \S, which is in the
single-stream regime at all points.  The map we use here was generated
on a $64^3$ grid by applying XTRACE to the output of the Einstein-de
Sitter simulation described in the previous section.  Even though the
displacement is from the \q- to the \x-space, all the conclusions of
this Section are valid for \s-space.

The true density, which will be called the {\it geometrical density}
in this paper, is found with Eq.~\ref{eq:delta} in Lagrangian space.
The corresponding Eulerian-space density can be obtained in two ways:
first, one can find by interpolation the \q-grid which according to
the map \S\ corresponds to a regular grid in \x-space (we use a linear
interpolation for this step); the \x-space density can then be found
by interpolating the geometrical density at the \q-points
corresponding to the \x-space grid (we use a quadratic interpolation
for this step).

\begin{figure*}
\centerline{
\psfig{figure=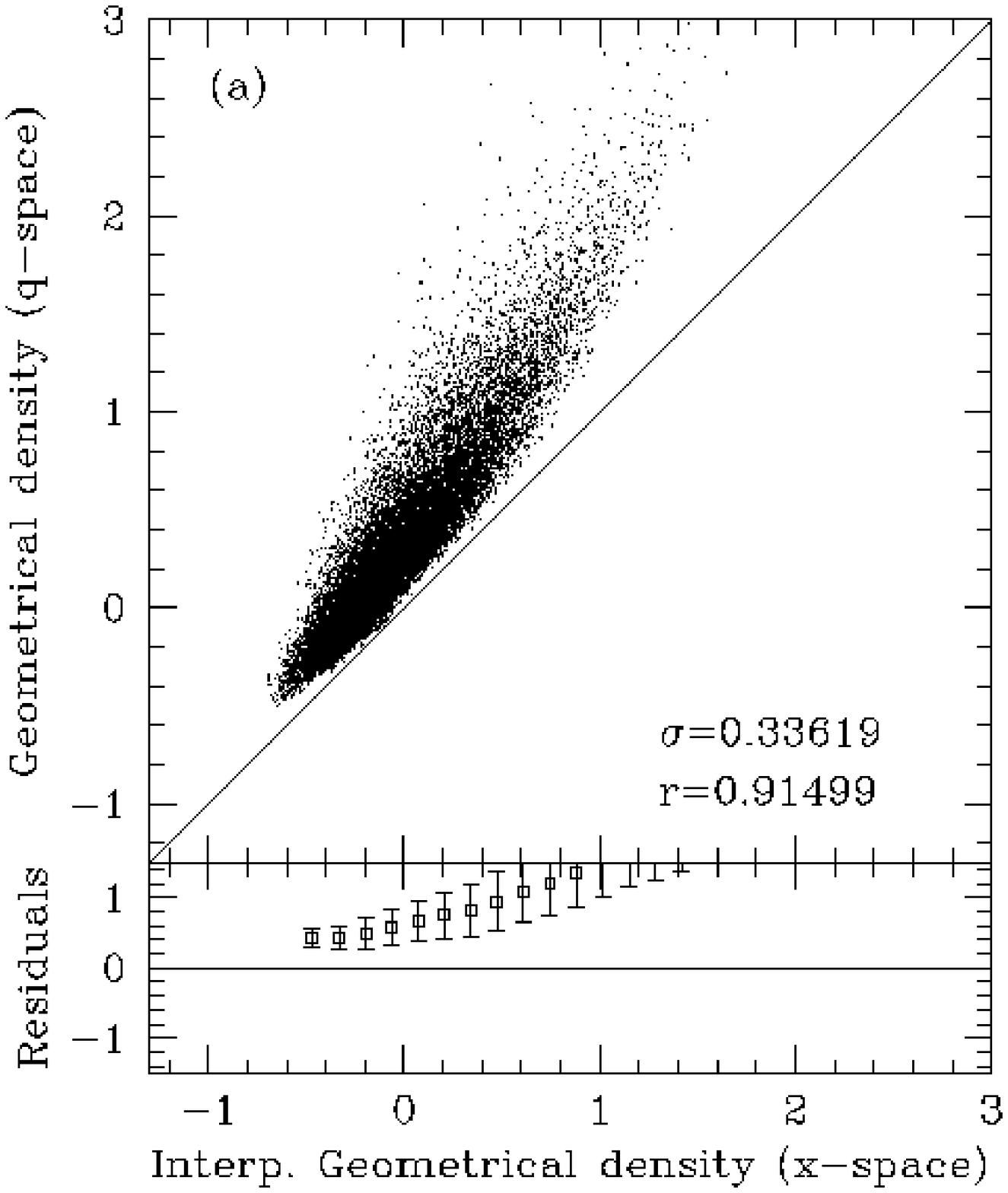,width=6cm}
\psfig{figure=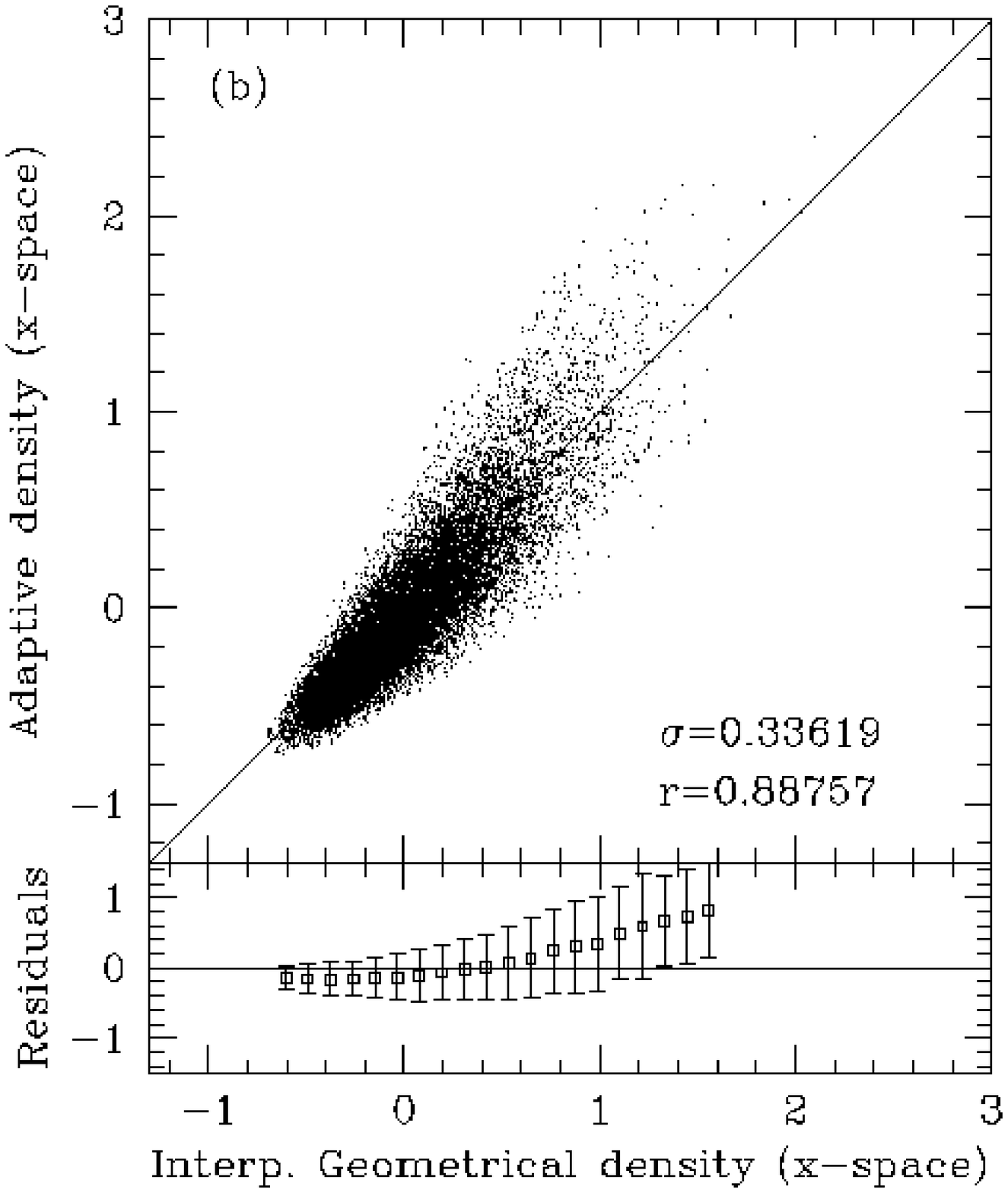,width=6cm}
\psfig{figure=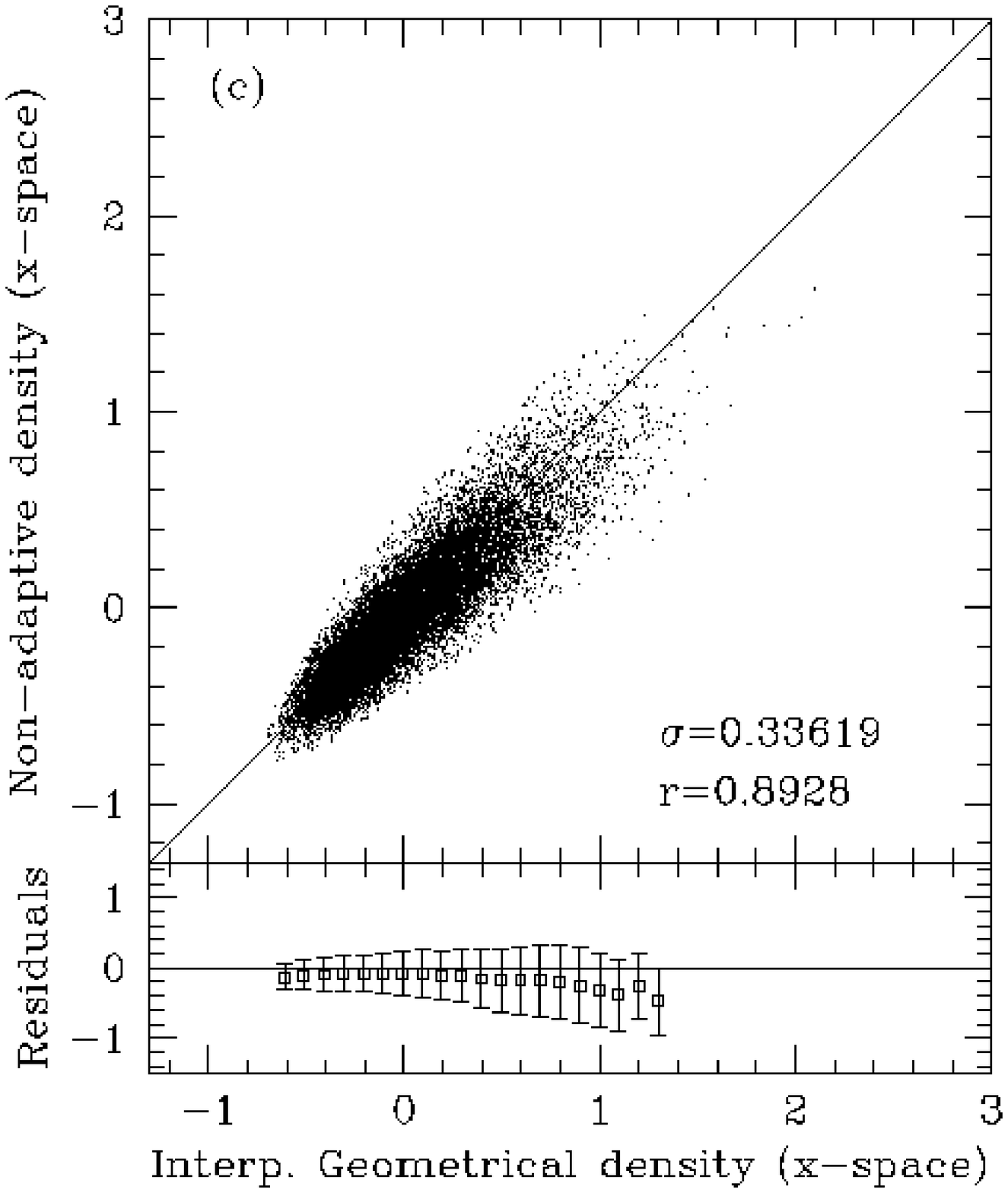,width=6cm}}
\caption{(a) Geometrical density in \x- and \q-space, smoothed over
1.64 grid points (6.15 \mpc), so that 10 particles in the simulated
catalogue lie within the Gaussian filter radius on average. (b) and
(c) show adaptive and non-adaptive smoothed density estimates of the
simulated catalogue compared with the smoothed geometrical density
(in \x-space).}
\end{figure*}

The second way of determining the \x-space density, which is valid
even if the map is not in the single-stream regime, is to perform a
standard cloud-in-cell (hereafter CIC) or similar interpolation on the
final \x\ positions of the grid points.  Fig. 1 shows the comparison
between CIC and geometrical densities, both in \x- and in \q-space.
The upper panels in each figure give point-to-point comparisons of the
two fields. The lower panels give the residuals around the 45 degree
bisector lines in units of the standard deviation of the field plotted
along the abscissa (which is listed as the quantity $\sigma$ in the
upper panels). The Spearman rank correlation coefficient, $r$, is also
listed, providing a non-parametric measure of the correlation between
the two fields.  The fields are smoothed over one grid size with a
Gaussian filter.

Fig. 1a shows the comparison between the CIC and interpolated
geometrical density. The two density fields are highly correlated,
though the CIC density underestimates slightly the height of the
highest density peaks. However, the scatter between the two density
estimates is large, and comparable to the scatter in the relation
between \q-space and interpolated \x-space density (which is however
very biased; Fig. 1b). This last relation shows the error which is
made in linear theory by assuming that the \q- and \x-spaces are the
same.

The noise in Fig. 1a is caused by both the CIC smoothing and by the
interpolation. To understand which source contributes more to the
noise, the densities have been interpolated back into the \q-space,
and compared to the geometrical density.  Figs.1c-d show that the main
contribution to the noise is given by the CIC smoothing procedure,
while the noise introduced by the interpolations is negligible (and,
furthermore, does not introduce any bias).

The reason why the CIC density estimates plotted in Fig. 1 are biased,
besides being noisy, is because the smoothing is performed in Eulerian
space.  Smoothing in Eulerian space is a problem for the Lagrangian
schemes, as the smoothing does not commute with dynamics and mixes
different scales (see, e.g., Bernardeau \& Kofman 1995). For instance,
in an overdense region, mass is gathering from a Lagrangian patch
which is larger than the actual size of the structure, while in
underdensities the opposite happens.  As a consequence, smoothing with
a fixed radius in Eulerian space mixes the larger Lagrangian scales
associated with overdensities with the smaller ones associated with
underdensities.

On the other hand, smoothing is required for at least two reasons,
first to reduce the variance of the input density field by erasing
small-scale, highly non-linear, structure, and second to construct a
continuous density field from a set of discrete points.  In the more
realistic case of estimating a density field from a catalogue of
galaxies, the density can be constructed by assigning a (Gaussian)
smoothing kernel to each galaxy and evaluating the resulting density
on a grid in the \x-space (or \s-space).  In this case, it is useful
to adaptively change the radii of the smoothing kernels associated to
each galaxy in a mass-preserving way: galaxies in overdensities or
underdensities will have smaller or larger smoothing radii, so as to
keep constant the mass inside the filter.  This adaptive smoothing
procedure is equivalent to smoothing in Lagrangian space, as the
Lagrangian mass elements contain the same mass by construction.  In
the applications described below, we have used an adaptive smoothing
algorithm based on the code described by Springel et al. (1998), and
used in Canavezes et al. (1998) to describe the topology of the IRAS
PSCz catalogue.  In this code the smoothing radius associated with
each galaxy is chosen so that, starting from a reference radius, the
actual number of galaxies inside the adapted filter volume is equal to
the average number of galaxies in the reference filter volume.

It will be shown in Section 5.4 that the use of adaptive smoothing
greatly improves the performances of the ZTRACE code.  However,
adaptive smoothing is slow, and in practice can be applied only to
catalogues with $\la 10^4$ galaxies (for which it takes about 30
minutes of CPU time to find the smooth field on a $64^3$ grid).  The
true density field in the simulations is, however, specified by $\ga
10^5$ points, thus we apply a constant average smoothing to the full
density field in the simulations.  As a consequence the adaptively
smoothed field from a galaxy catalogue constructed from the
simulations is slightly different from the smoothed true density
field, underdensities being more underdense and overdensities being
more overdense.

To quantify the noise in the density estimate derived from a limited
set of points, we have selected $\sim 20000$ particles at random
within a sphere of diameter $240$ \mpc\ from the final output time of
the EdS simulation described in Section 3.\footnote{The average
particle density in this sphere is close to the mean density of the
PSCz IRAS redshift survey at a radius of $\sim 70$ \mpc\ (see Section
5.1 for more details of this Survey).} A density field from these
points was computed with a fixed smoothing radius and with the
adaptive smoothing procedure described above.  The fixed smoothing
radius was set to $\sim 1.64$ grid points ($\sim 6.15$ \mpc), so that
there are on average 10 points within the filter.  Figs. 2b and 2c
show the comparison of these density estimates with the true
(interpolated geometrical) density computed from the full set of
points in the simulation after smoothing with a Gaussian filter of
radius 1.64 grid points.  Fig. 2a shows, for reference, the
scatterplot of the geometrical density for the full N-body simulation
computed in \q-space and interpolated in \x-space smoothed with
filters of width 1.64 grid-cells.

The non-adaptive smoothing gives an almost unbiased but noisy estimate
of the true density. The adaptive smoothing estimate is biased, as
explained above, and is noisier than the non-adaptive density
estimate as the adaptive refinements introduce further noise.  In
both cases, the noise is comparable to  that in the
\q-space -- \x-space scatterplot (Fig. 2a).

The code for adaptive smoothing allows one to change the shape of the
smoothing kernel into a triaxial ellipsoid, which can be oriented to
follow the inertia tensor of matter inside the filter.  This could in
principle improve the results compared to simple adaptivity of a
spherical smoothing radius.  However, we have verified that shape
adaptivity introduces further noise in the density estimate, but no
appreciable improvement for the applications described in this paper.

In summary, density estimates give an important (usually dominant)
source of noise in the reconstruction algorithm.  For a sparse
catalogue, the noise introduced by the density estimate is at least as
large as that introduced by not distinguishing between \q- and
\x-space.  The latter provides a quantitative measure of the noise
introduced by non-linearities that are not included in linear
reconstruction algorithms. Thus, complex non-linear reconstruction
algorithms for the final real-space density and peculiar velocity
fields are unlikely to perform very much better than a simple linear
algorithm, since shot noise usually dominates in the density
estimates. This is not, however, true of reconstructions of the
initial density field.  As we will show in Section 5, the non-linear
reconstruction algorithm described here performs very much better than
linear theory in recovering the initial conditions from a highly
evolved density field.

Note that in analysing the numerical simulations, the final density
fields in both real in redshift space, have been estimated through a
CIC interpolation, as it is not possible to construct a geometrical
density directly if the map \S\ is in the multi-stream regime.
However, the ZTRACE method gives, as an output, the real-space
geometrical density and the LOS peculiar velocity fields in Lagrangian
\q-space.  As the recovered map is in the single-stream regime by
construction, it is possible to obtain the \x-space density and
velocity fields by interpolation, so as to minimize the noise in the
transformation.

\section{Tests with N-body Simulations}

To test their performances, the LTRACE, XTRACE and ZTRACE
reconstruction algorithms have been applied to the final outputs of
the N-body simulations (described in Section 3) in real (XTRACE) or
redshift (LTRACE, ZTRACE) space; the reconstructions of initial
conditions, real-space density and LOS peculiar velocity are described
in this Section.

\subsection{Simulated galaxy catalogues}

From the final output times of the the N-body simulations, simulated
catalogues of galaxies have been extracted that roughly mimic the
IRAS PSCz survey of Saunders et al. (1994). This survey consists of
redshifts for a near all-sky survey of about $15,000$ galaxies with
$60\mu$m fluxes $> 0.6$Jy. The selection function for the survey is
approximated by the functional form 

\be
\Phi(r)=\Phi_*\frac{(r/r_*)^{1-\alpha}}{(1+(r/r_*)^\gamma)^{\beta/\gamma}},
\label{eq:selfun} \ee

\noindent
where $\Phi_*=0.0132\ h^3Mpc^{-3}$, $r_*=64.6$ \mpc, $\alpha=1.61$,
$\beta=3.90$, $\gamma=1.64$ and $r = cz/H_0$. The parameters in this
expression are close to those given by Sutherland et al. (1999)
for the PSCz survey. Point masses in the simulation are selected
at random with this form of the selection function. Thus, galaxies
in our simulated catalogues are unbiased tracers of the mass. Other
bias schemes can be implemented easily, but our purpose in this paper
is to provide a test of the reconstruction methods rather than to
test the effects of more complex bias schemes.

\begin{figure*}
\centerline{
\psfig{figure=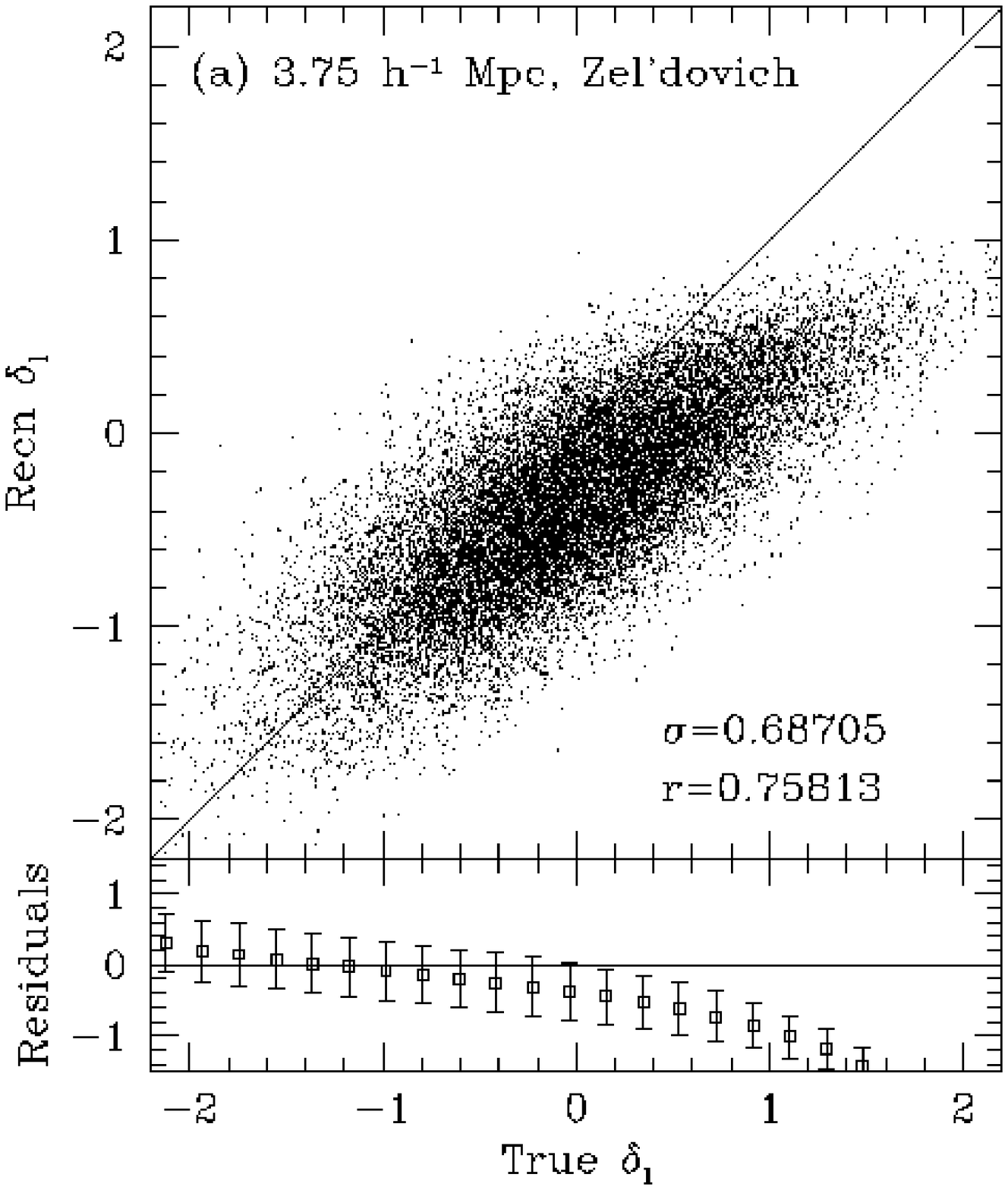,width=6cm}
\psfig{figure=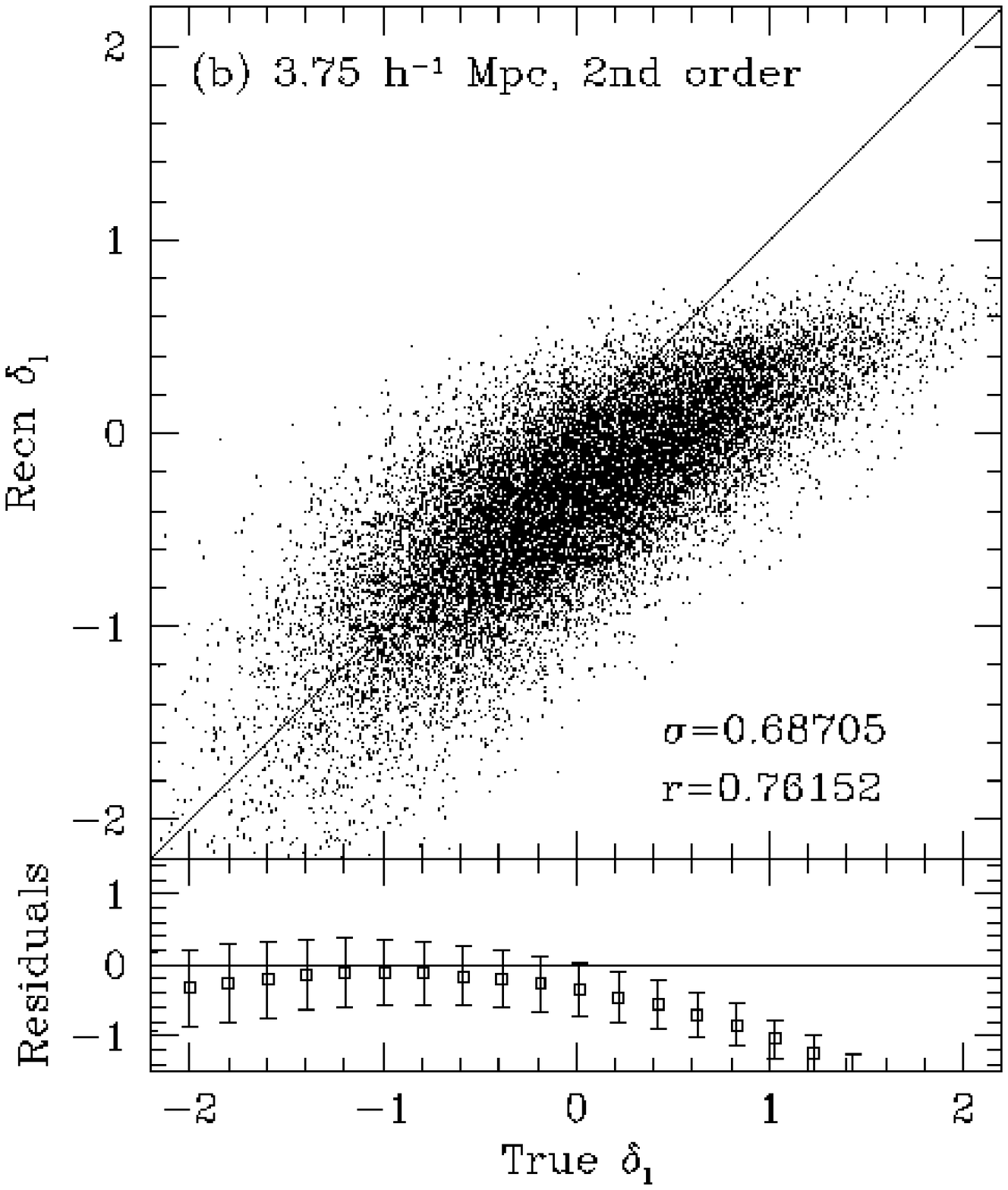,width=6cm}}
\centerline{
\psfig{figure=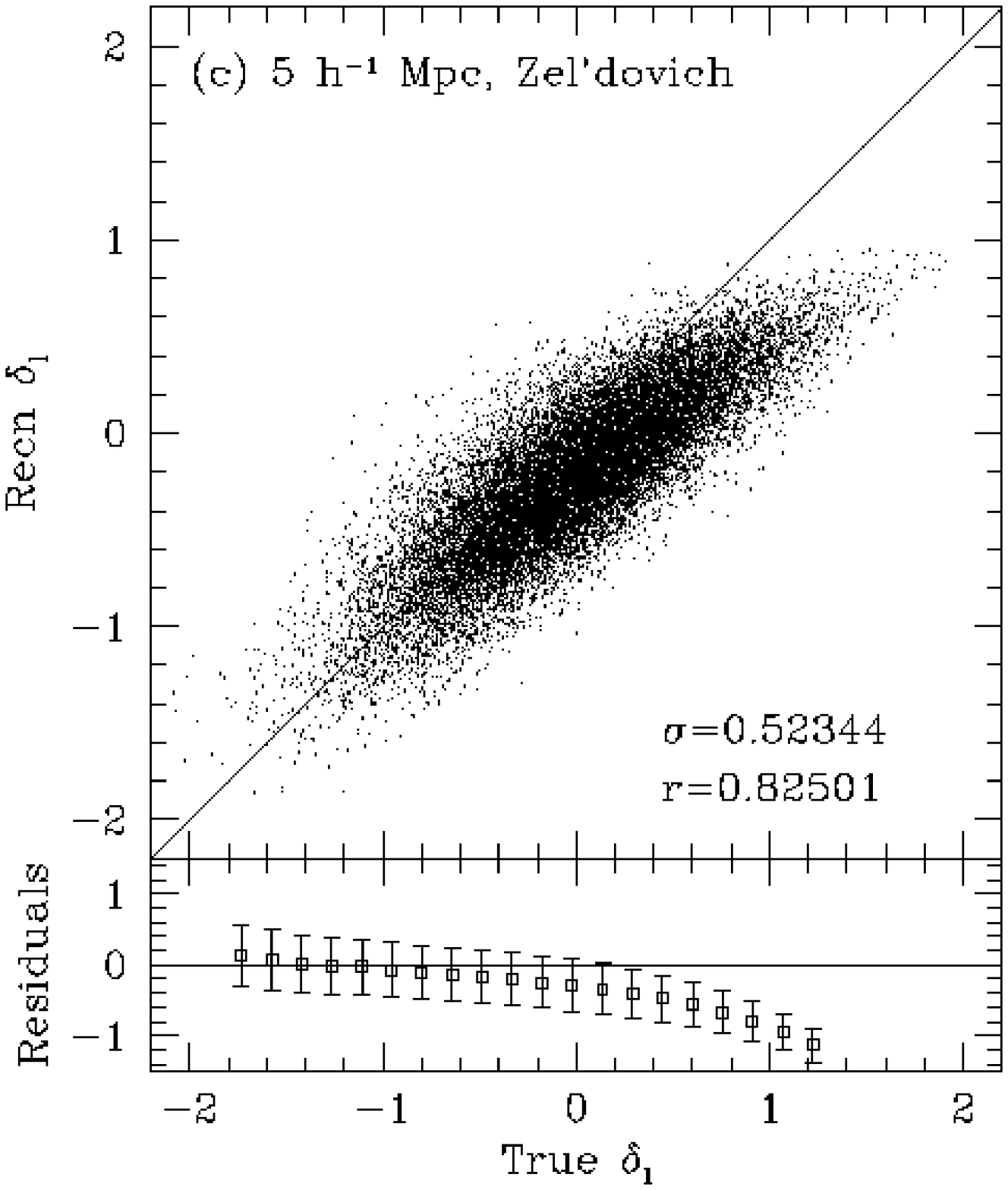,width=6cm}
\psfig{figure=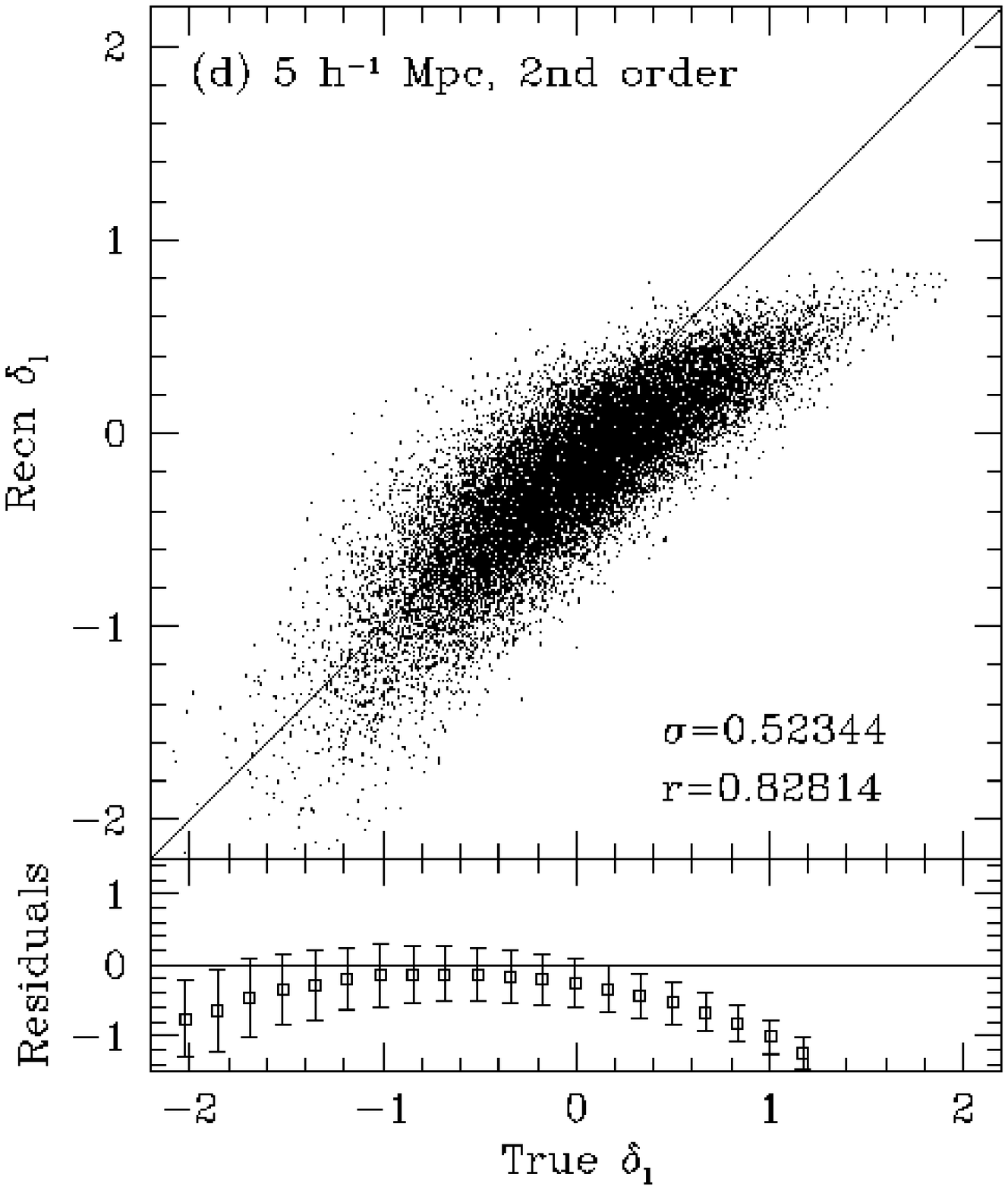,width=6cm}}
\caption{Reconstruction of initial conditions with XTRACE.  Upper
panels with 3.75 \mpc\ smoothing;  lower panels with 5 \mpc\ smoothing. 
Panels on the left show the  Zel'dovich approximation and panels on
the right show  2nd order Lagrangian perturbation theory.}
\end{figure*}
 
To suppress errors arising from the highly non-linear dynamics inside
relaxed structures, we have collapsed the ``fingers of God'' in
redshift-space.  Following Gramann, Cen \& Gott (1994), groups have
been found with a friends-of-friends algorithm, with radial and
tangential linking lengths respectively of 3 and 0.5 \mpc.  The radial
component of the distance of each galaxy from the centroid of the
group is rescaled so that the groups become spherical on average. This
procedure was applied to typically 50 or so prominent groups and
clusters in the simulated galaxy catalogues. In this way, the high
peaks of the redshift-space density are slightly more enhanced.  On
the scales tested, the collapse of the fingers of God produces a
slight improvement of the results.

\subsection{Smoothing strategies}

As already discussed in Section 4, smoothing is necessary to 
derive a continuous density field from the simulated catalogues and
to suppress highly non-linear structures.
When we use the full evolved density field from a simulation as an
input to the reconstruction algorithms, a moderate smoothing in
Eulerian space is performed. As described in Section 4, adaptive
smoothing is too slow to be applied to $\ga 100,000$  point
masses.

The simulated galaxy catalogues are smoothed with the adaptive
smoothing algorithm described in Section 4. The {\it reference}
smoothing radius is held fixed within a characteristic distance of
either $R_{*} = 50$ or $80$ \mpc, so that at this distance 10 galaxies
are on average contained in the filter volume (defined as $V_G=(2\pi)
^{1.5} R_{sm}^3$, where $R_{sm}$ is the smoothing radius).  At larger
radii, the smoothing radius is increased to compensate for  the selection
function, so that there are 10 galaxies on average inside the
filter volume.  The reference smoothing radii are 5.15 or
7.65 \mpc, when the characteristic distance $R_{*}$ is 50 or 80 \mpc.
The density field within $R_*$ is well sampled and is reconstructed
with a constant reference smoothing radius, while the outer parts are
used just to give the external tides to the inner parts of the simulated
catalgues.  The fact that
the density field is smoother in the outer parts is an advantage for
the reconstruction method, as it reduces the sensitivity to the
boundary conditions (see Section 2.2).

It is worth noting that the smoothing requires knowledge of the
selection function to determine the density contrast. The same
selection function is applied to the N-body simulations to select the
galaxies in real space and to estimate the density contrast in
redshift space.  This is not strictly correct, as the selection
functions in real and redshift space will be slightly
different. However, as pointed out by Hamilton (1998), these
differences will be small for an all-sky catalogue.  We have checked
that the differences between the real-space and redshift-space
selection functions for our simulated catalogues are indeed
negligible.

\begin{figure*}
\centerline{
\psfig{figure=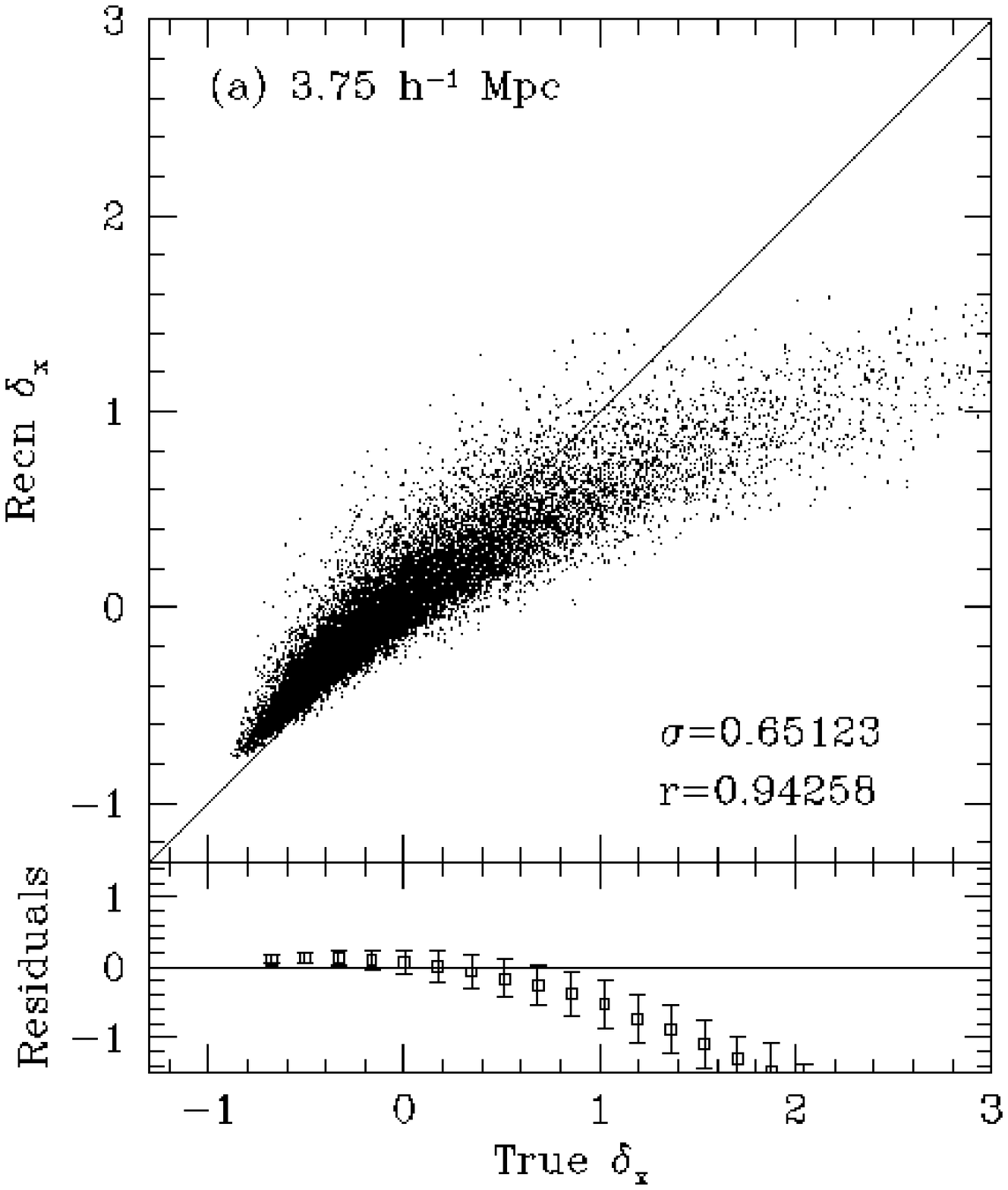,width=6cm}
\psfig{figure=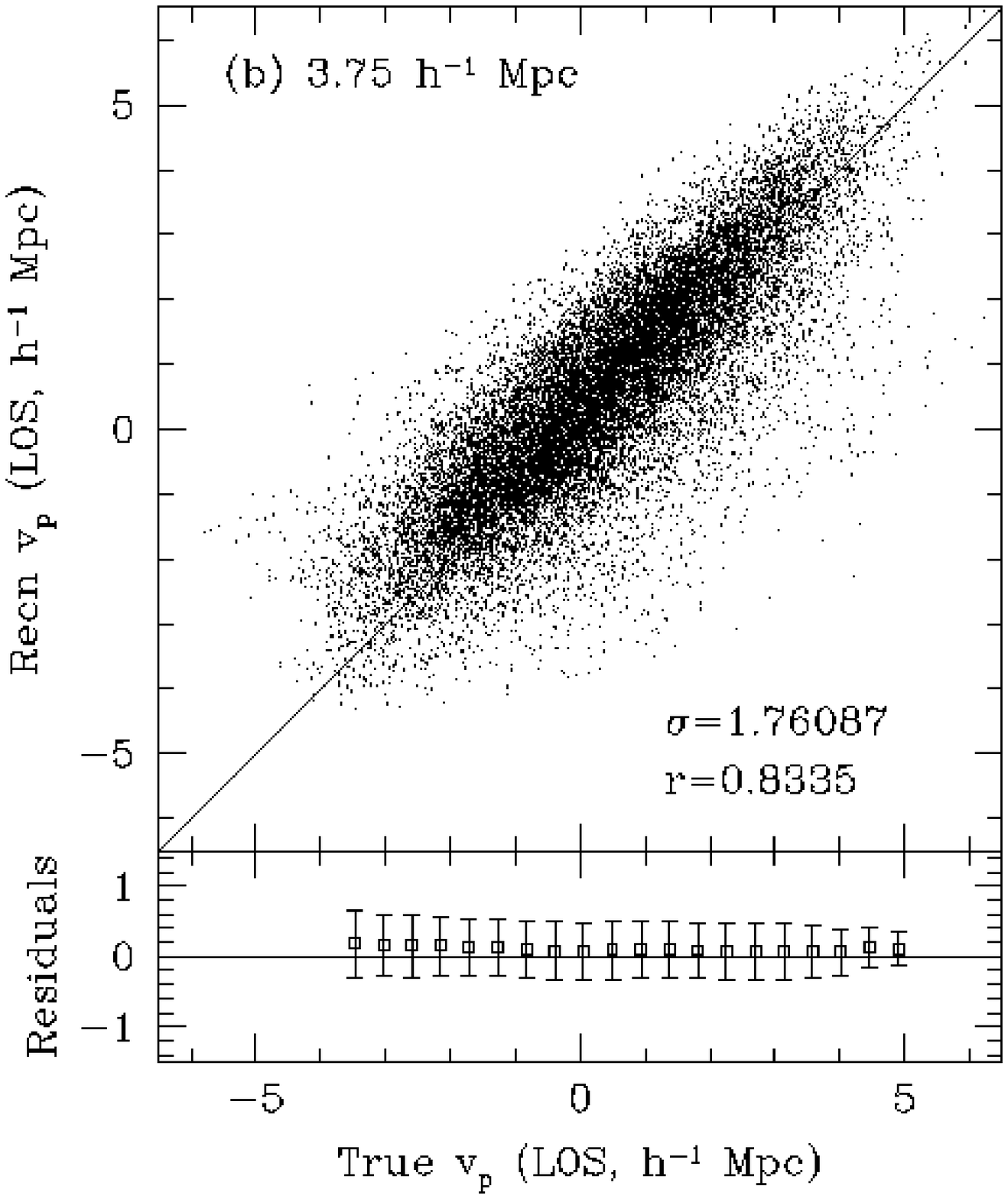,width=6cm}
\psfig{figure=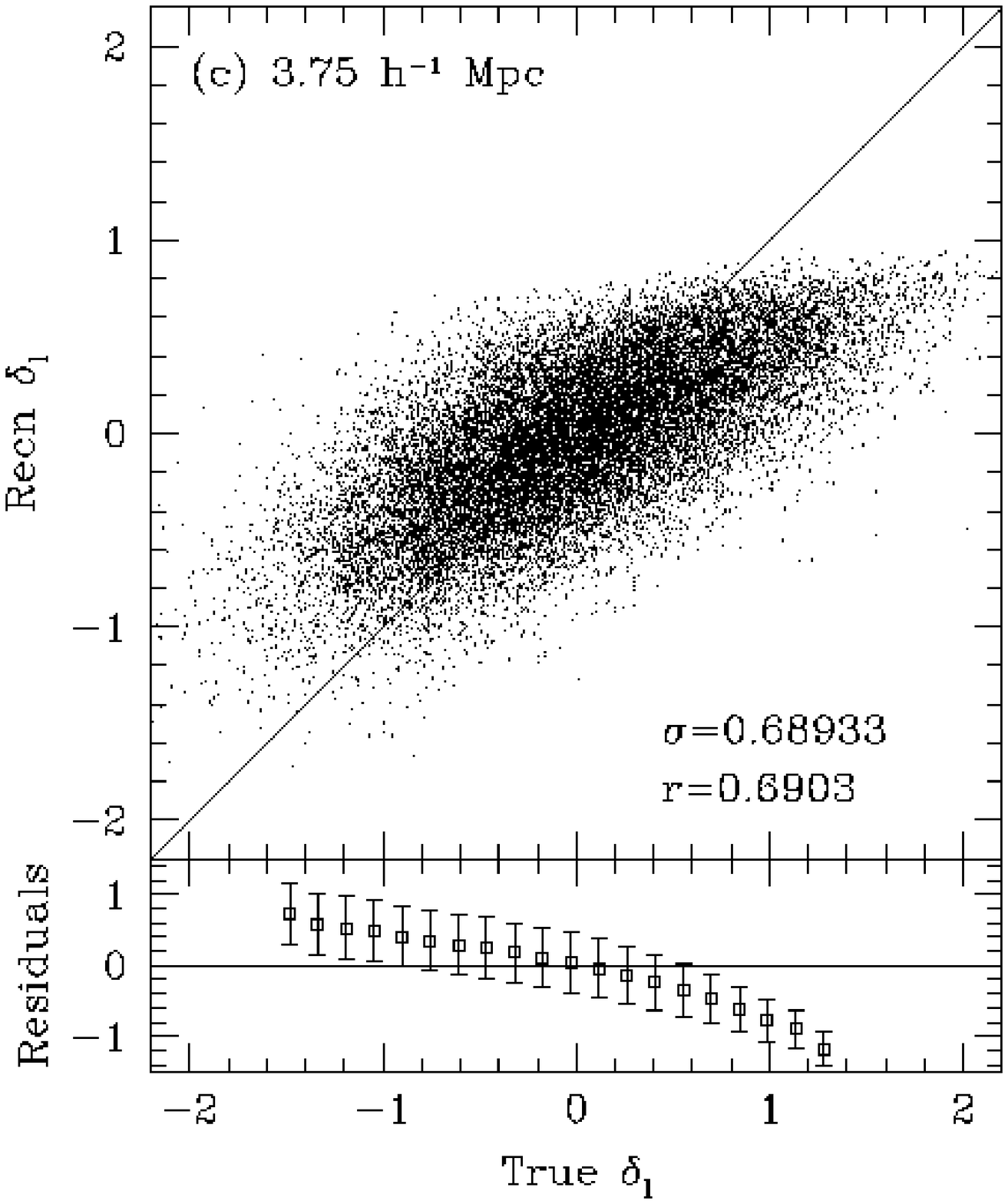,width=6cm}}
\centerline{
\psfig{figure=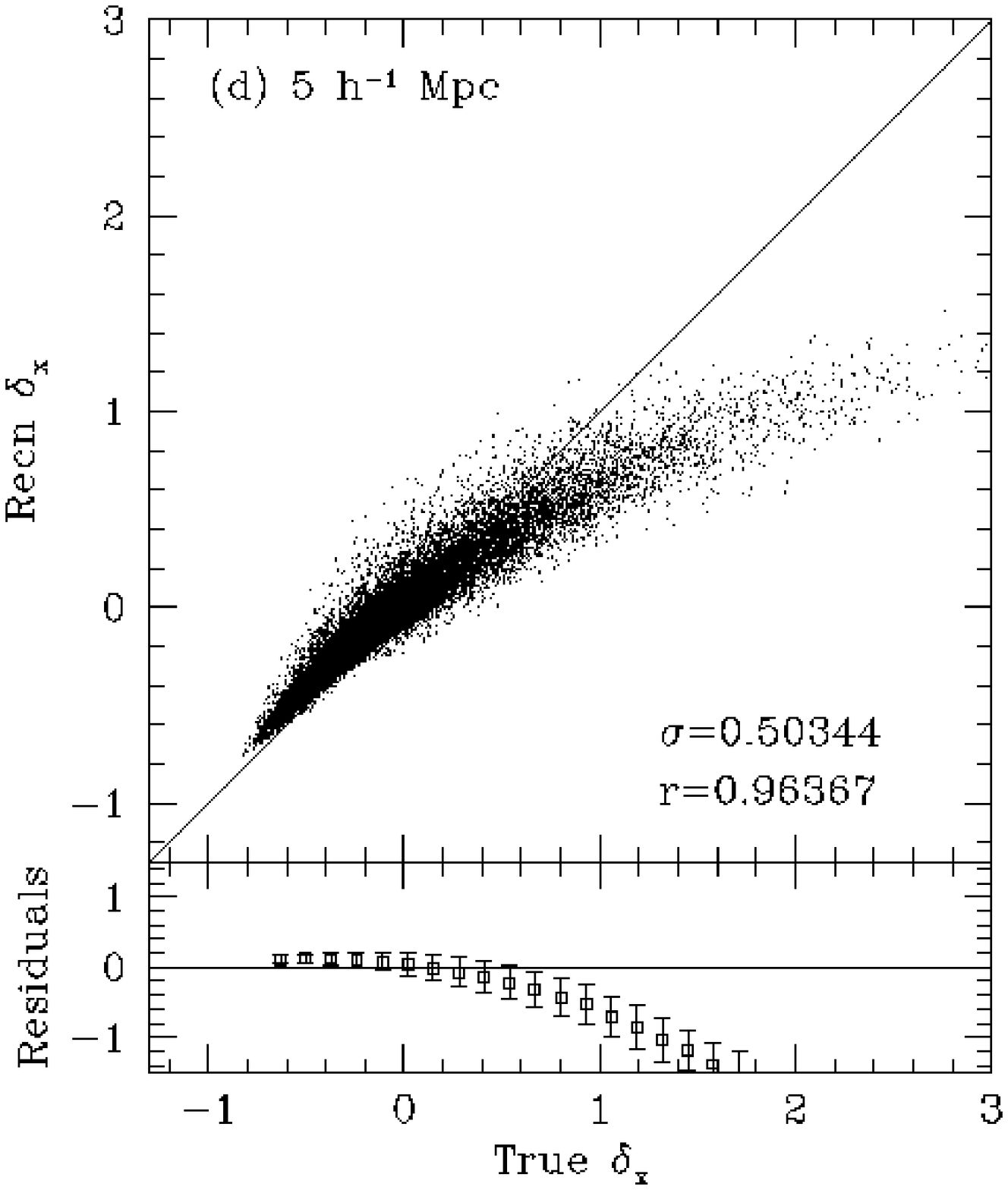,width=6cm}
\psfig{figure=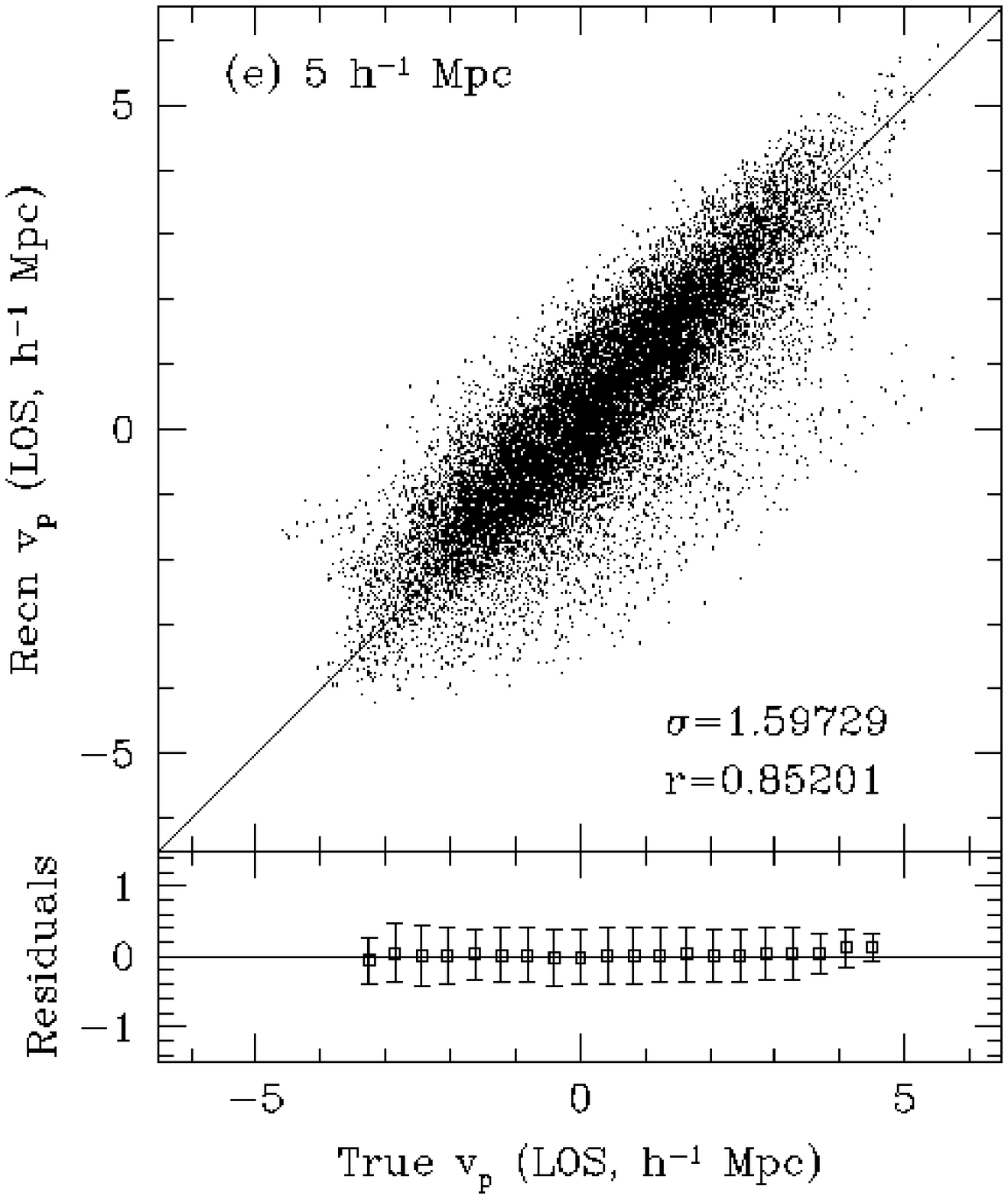,width=6cm}
\psfig{figure=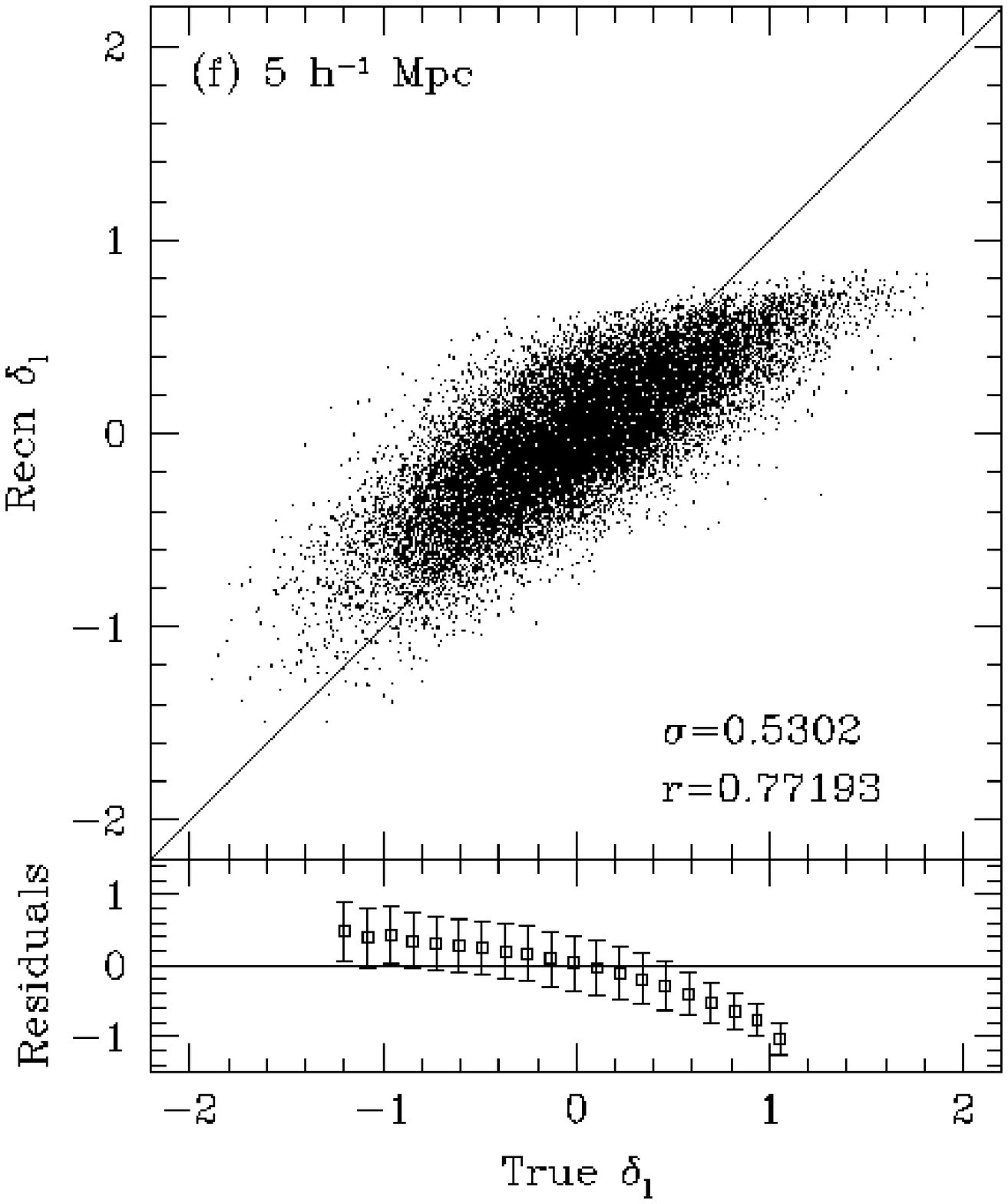,width=6cm}}
\caption{Reconstruction with ZTRACE with smoothing of 3.75 \mpc\ (upper
panels) and  5 \mpc\ (lower panels).}
\end{figure*}

\subsection{Results with XTRACE}

The complete real-space density field from the EdS simulation
has been given as an input to XTRACE.  This allows us to compare our
results with those of Narayanan \& Croft (1998). The input real space
density field is smoothed with Gaussian filters of widths 3.75 \mpc\
and 5 \mpc.  Fig. 3 shows the scatterplots of true and reconstructed
linear density fields. In these comparisons, the true linear
density field has been smoothed with the same Gaussian filters. The
reconstruction algorithm introduces some noise at the level of the grid cell,
which is difficult to subtract without removing true power present
at small scales.  To suppress this noise, a Gaussian smoothing of 2.5
\mpc\ has been applied to the reconstructed fields.  Finally, both the
Zel'dovich and the 2nd order reconstructions are shown.

The following points can be noted from Fig. 3:

\begin{enumerate}
\item
The reconstructed linear density rarely exceeds the value of unity.
This is a true physical effect as the high peaks in the evolved
density fields are in the multi-stream regime but are assumed to be
in the the mildly non-linear phase in the reconstruction. As a
consequence, they are reconstructed as though they evolved from
shallower peaks with linear density $\la 1$.  (In fact, in a
general pancake-like collapse, the real space density contrast is
$\delta_r \approx 4$ when the extrapolated linear density contrast is
$\delta_l\simeq 0.8$.)
\item
With larger smoothing, the reconstructed linear density field is less
noisy, with no additional  bias.  This is due to the noise introduced
at small scales by highly non-linear dynamics. Some smoothing is
required to suppress this noise.
\item
The 2nd order dynamics tends to improve the correlation with the true
field slightly, but induces some further bias in recovering
underdensities; as discussed in Section 2, this is because 2nd order
perturbation theory is inaccurate for deep underdensities.  As a
consequence, going to 2nd order does not improve the reconstruction
scheme significantly.
\end{enumerate}

\begin{figure*}
\centerline{
\psfig{figure=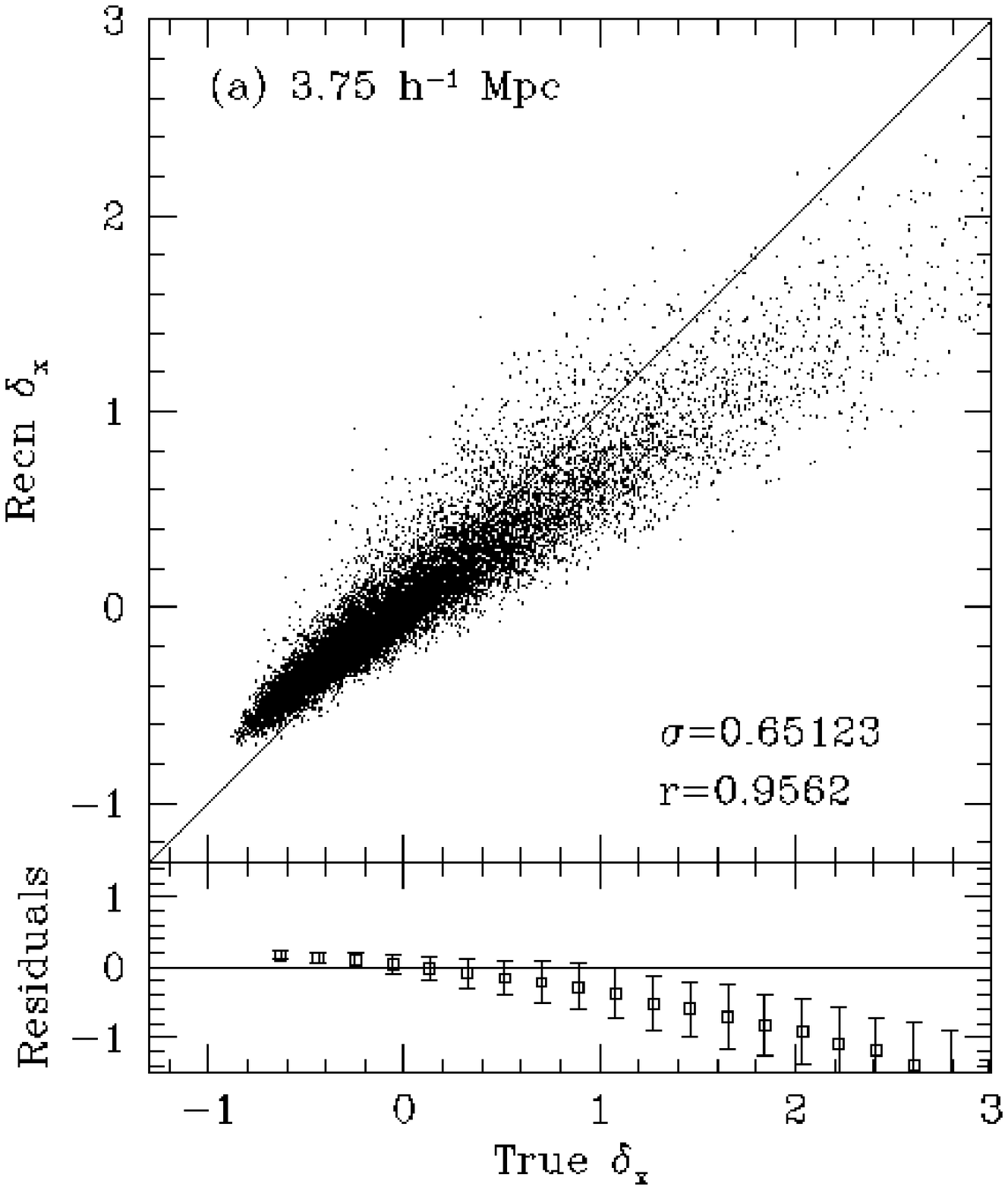,width=6cm}
\psfig{figure=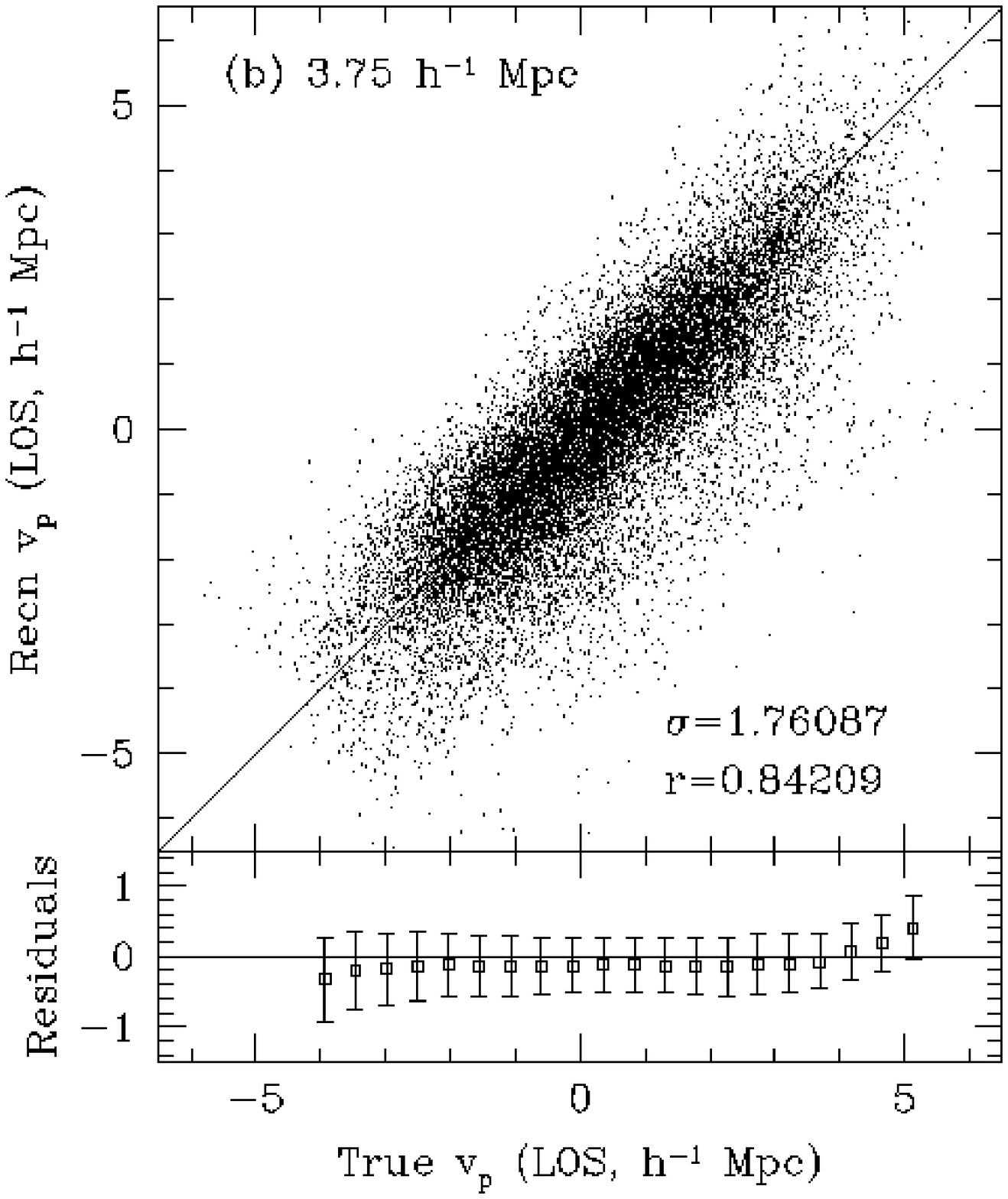,width=6cm}
\psfig{figure=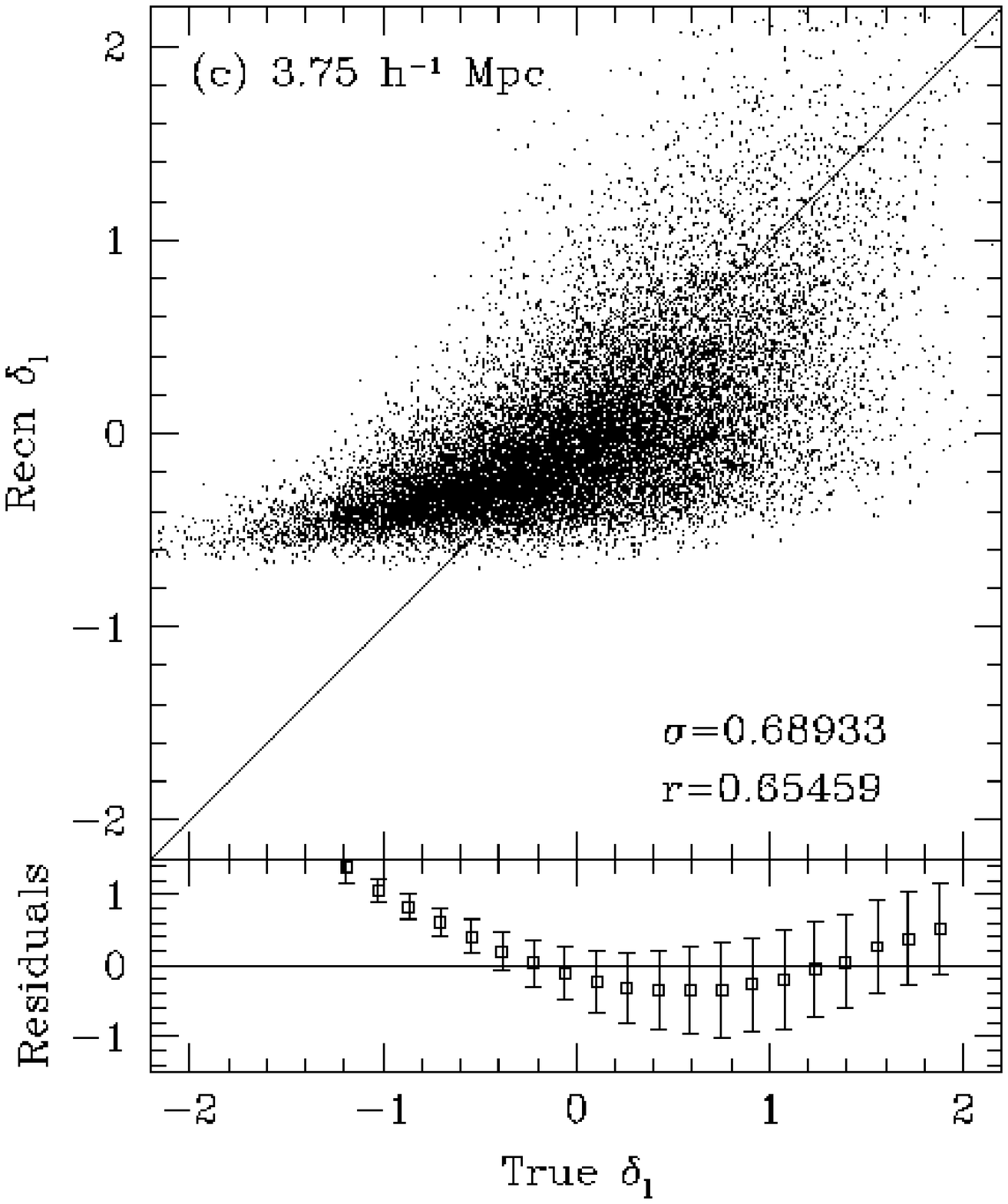,width=6cm}}
\centerline{
\psfig{figure=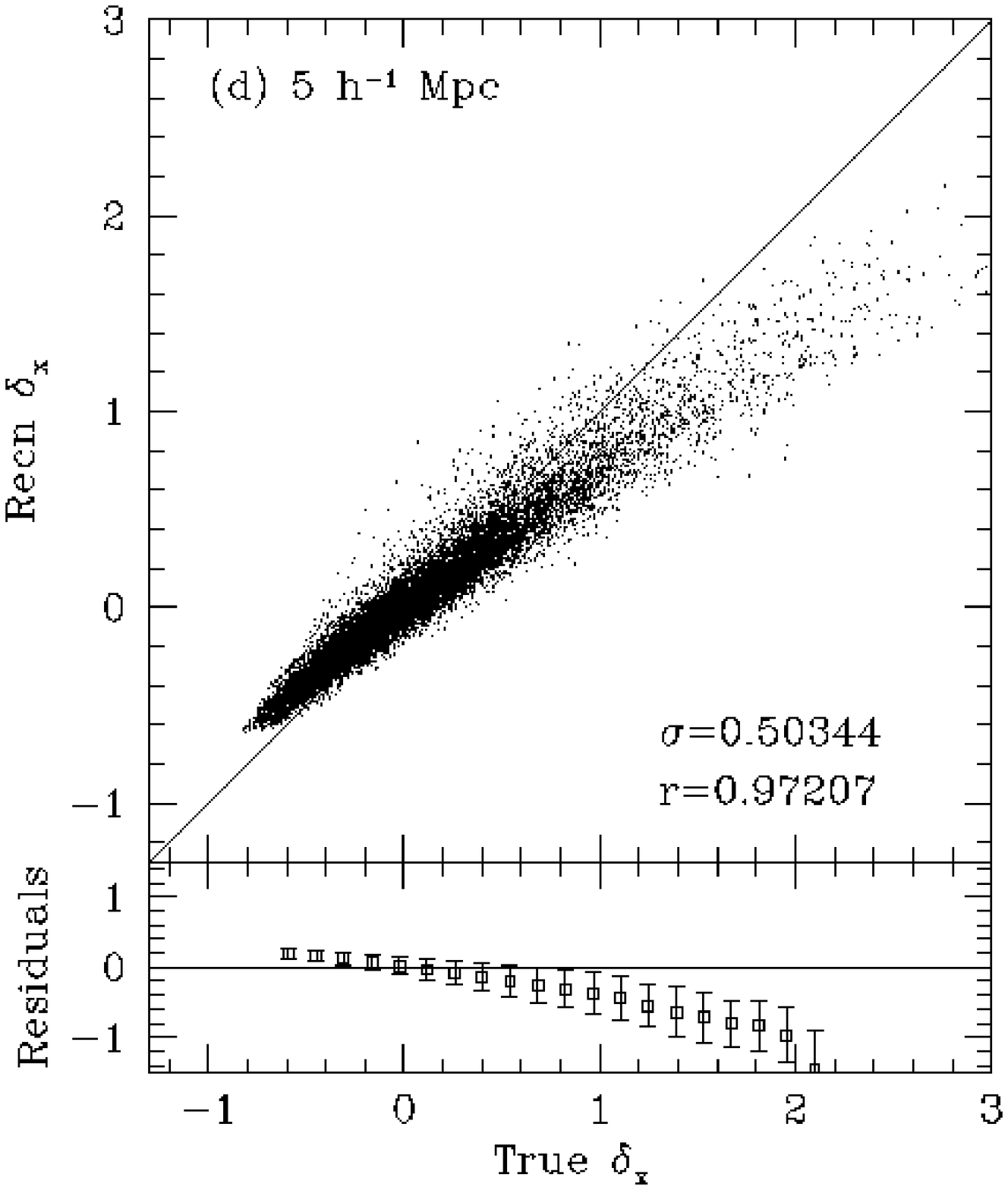,width=6cm}
\psfig{figure=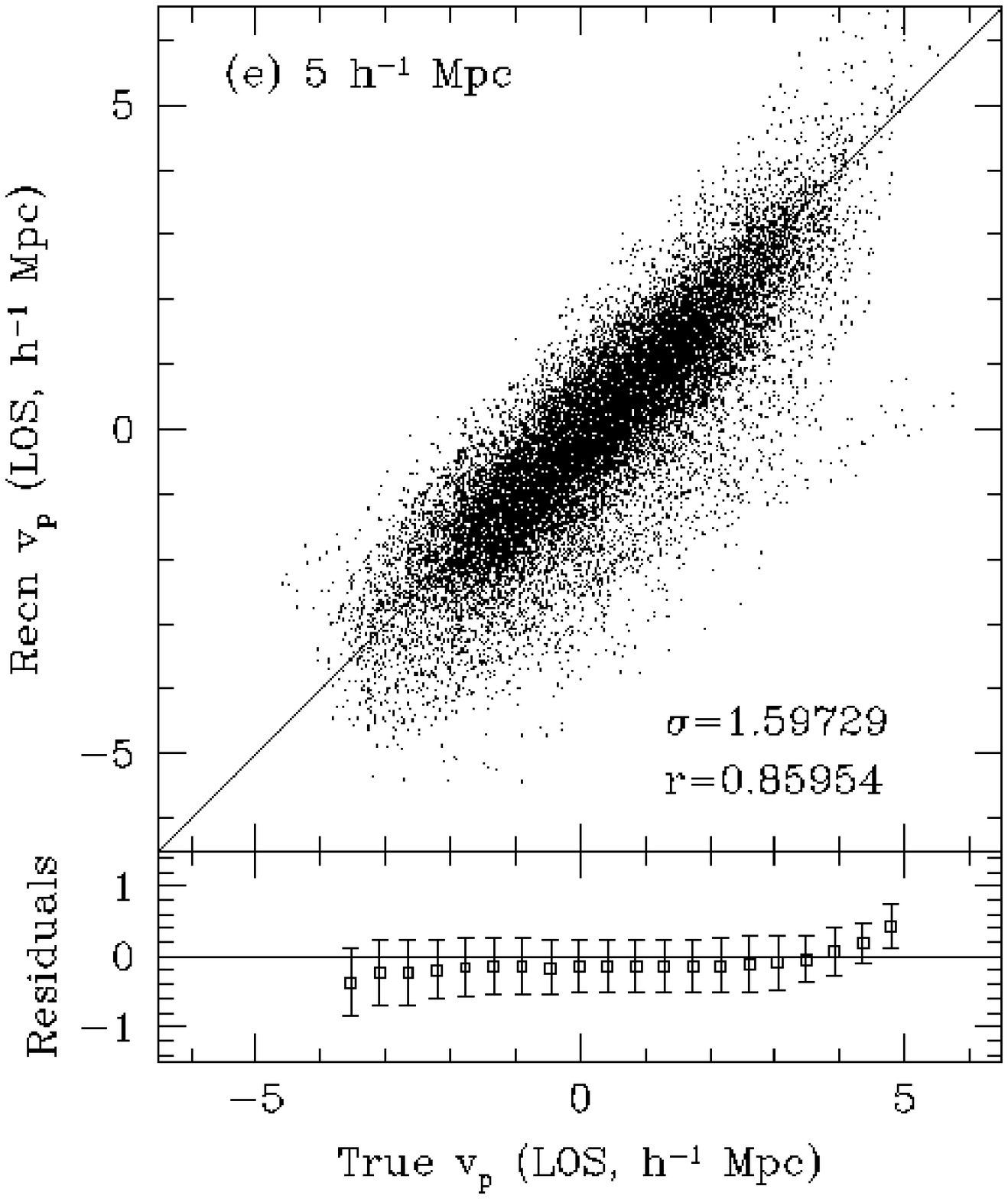,width=6cm}
\psfig{figure=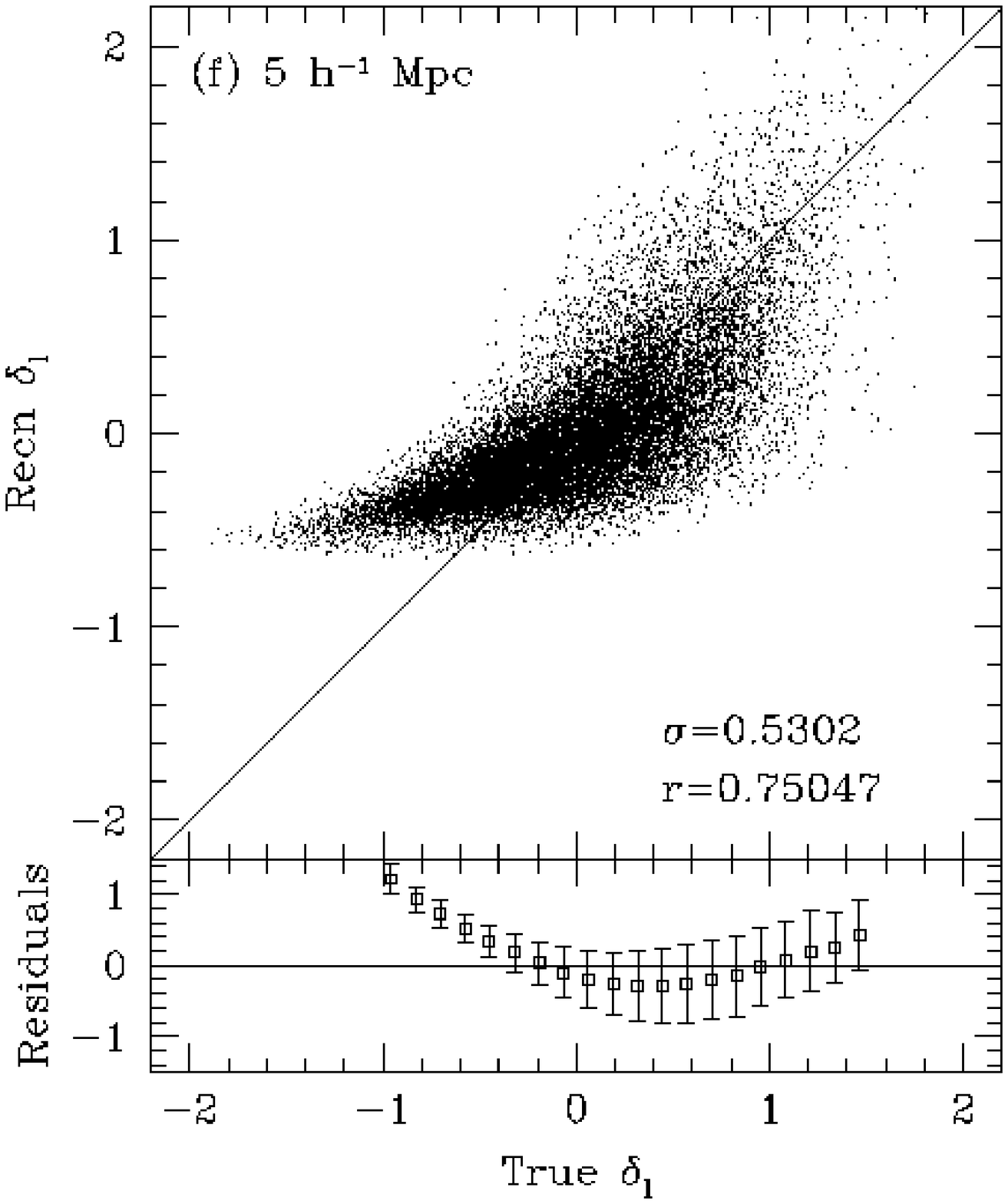,width=6cm}}
\caption{Reconstruction with LTRACE, with smoothing of 3.75 \mpc\ (upper
panels) and 5 \mpc\ (lower panels).}
\end{figure*}

The comparison with the results of Narayanan \& Croft (1998) are not
completely straightforward, as they use simulations with a smaller box
(200 \mpc) normalized to have $\sigma_8=1$.  Nonetheless, it appears
that the performance of XTRACE is comparable or even better than other
Zel'dovich-based approximations, especially for underdensities. The
XTRACE algorithm performs similarly to the hybrid Gaussianization
model of Narayanan \& Weinberg (1998), in which the PDF of initial
conditions is forced to be Gaussian.  The PIZA method of Croft \&
Gatza\~naga (1997) appears to perform better than both XTRACE and the
Narayanan \& Weinberg algorithm, recovering linear density contrasts
greater than unity, though underdensities are recovered with a
slightly biased.  The smaller scatter in the PIZA reconstruction
arises because the method uses the final positions of the N-body
particles and hence there is no need to smooth the evolved density
field in Eulerian space.  It would be interesting to investigate the
performance of PIZA on simulated galaxy redshift surveys which include
a realistic selection function.

\subsection{Results with ZTRACE}

As a first step, the full redshift-space density field of the evolved
simulation is given as input to ZTRACE and LTRACE.  The density field
is suppressed, as described in Section 2.2, at radii greater than 84
\mpc\ from the origin.  As in the test of XTRACE, we smooth the input
redshift-space density field with Gaussian filters of widths 3.75
\mpc\ and 5 \mpc\ and smooth the output density and velocity fields on
a scale of 2.5 \mpc\ to suppress the noise introduced by the
reconstruction algorithm.  The true fields from the simulations are
smoothed with the same filters as the input redshift-space density
field.  The comparisons are shown in Figure 4 for radii less than 84
\mpc\ and larger than 10 \mpc\ (excluding  the central region, which
is not recovered accurately as described in Section 2.2).  Note that
no attempt has been made in these comparisons to collapse ``fingers of
God'' in the input redshift-space density field.  Fig.4 shows the
results from ZTRACE for the real-space density field, the peculiar
velocity and the linear density.  Fig. 5 shows the same results
obtained with LTRACE.\footnote{ The LOS peculiar velocity in these
figures is given as an apparent displacement in \mpc.}

The following points can be noted from Figs. 4 and 5:

\begin{enumerate}
\item
Smoothing on a larger scale decreases the scatter in the Figures
without introducing a larger bias. 
\item
The highest peaks in the real-space density are not correctly
recovered.  As in the application of XTRACE, this is because 
smoothing in  Eulerian space introduces bias  and because the
reconstruction algorithm does not correctly model
multi-stream regions.
\item
The peculiar velocity is recovered in an almost unbiased way.
\item
The reconstruction of the linear density field is noisy; again, the density
peaks with $\delta_l>1$ are considerably suppressed.  Underdensities
are recovered in a biased way due to the smoothing in
Eulerian space.
\item
ZTRACE always performs better than LTRACE; however,
both the real-space density field and the LOS peculiar velocities are
reconstructed fairly well by linear theory.  The linear density
field recovered by LTRACE is, however, very
different from the true one.
\end{enumerate}

\begin{figure*}
\centerline{
\psfig{figure=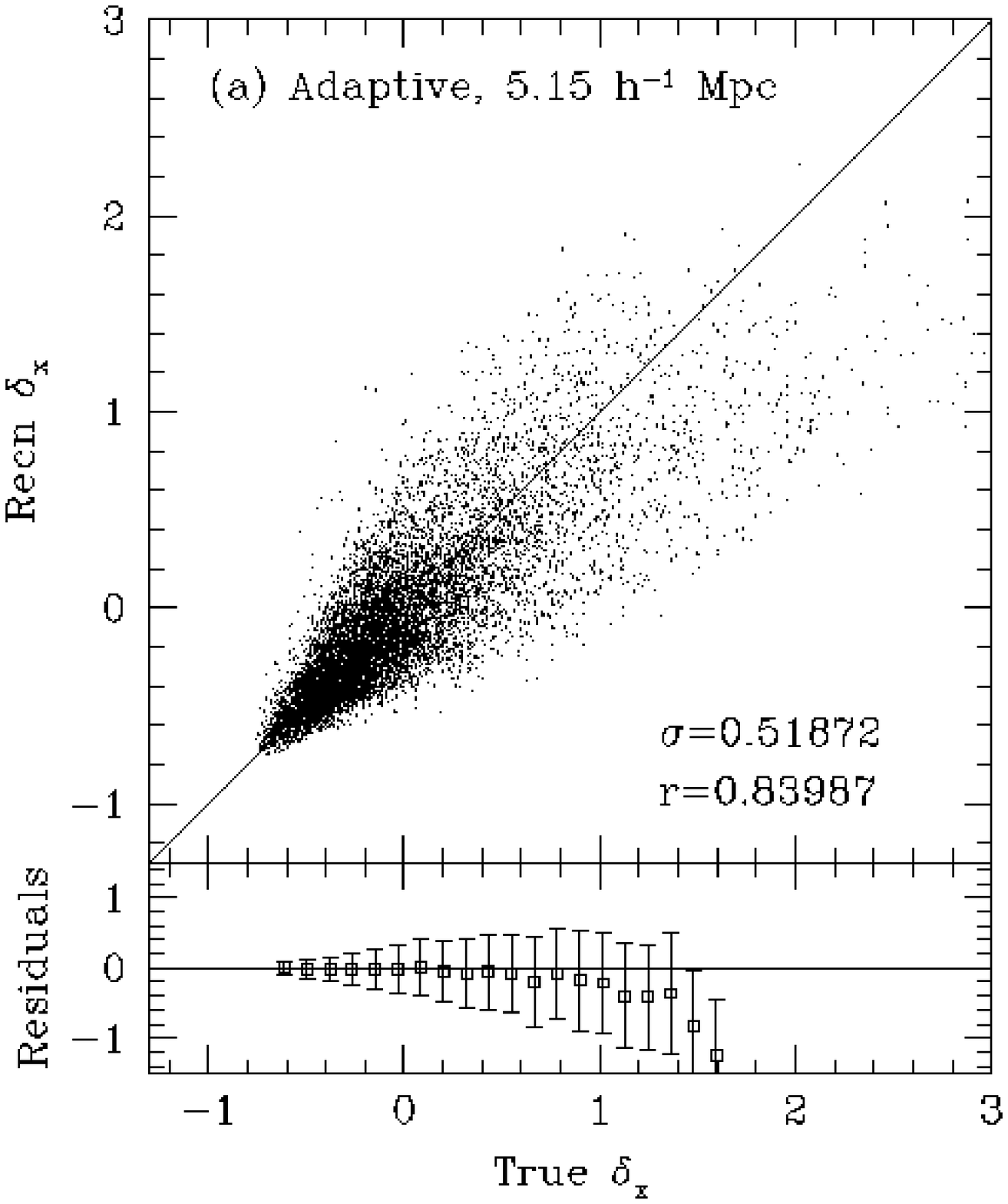,width=6cm}
\psfig{figure=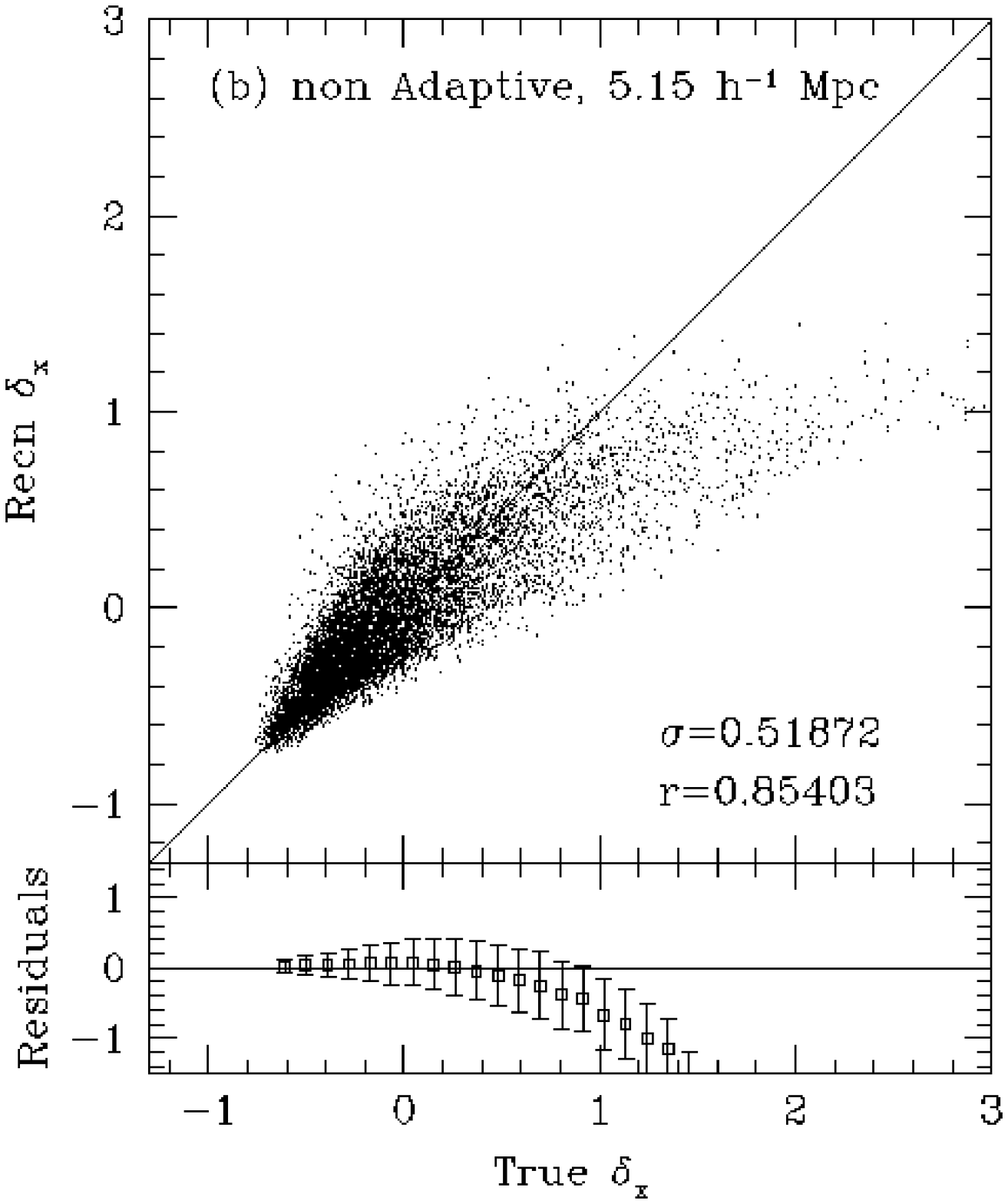,width=6cm}
\psfig{figure=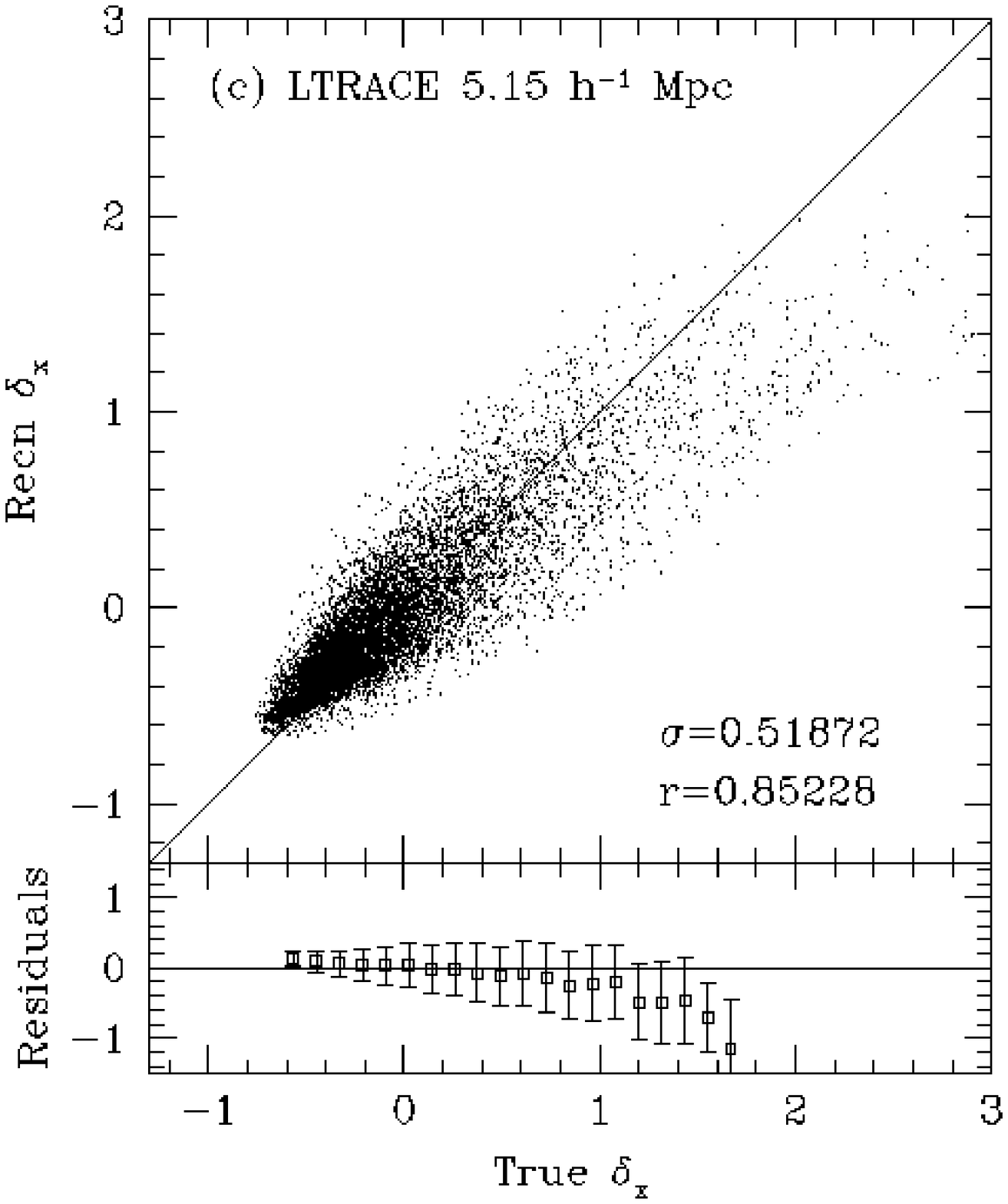,width=6cm}}
\centerline{
\psfig{figure=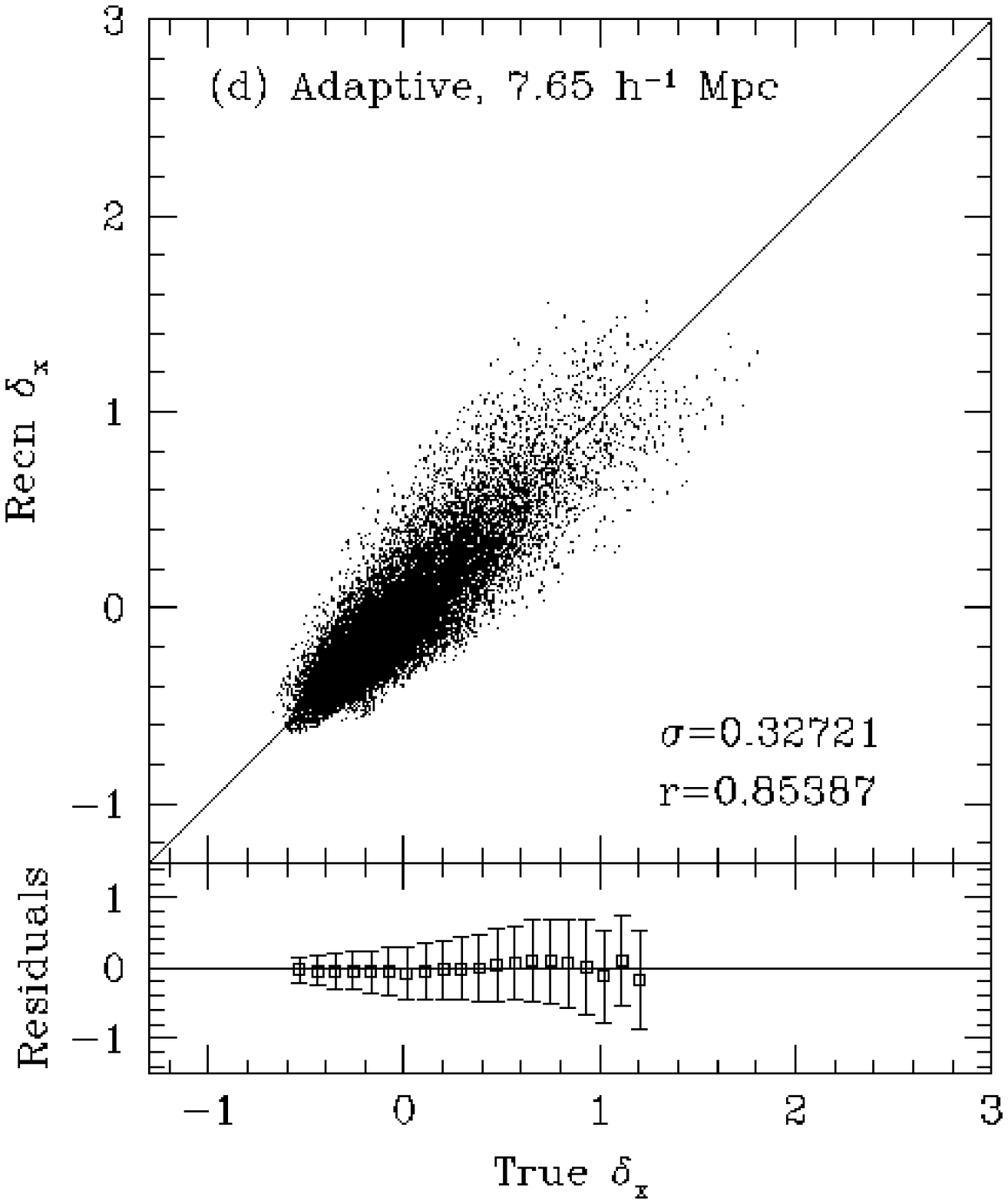,width=6cm}
\psfig{figure=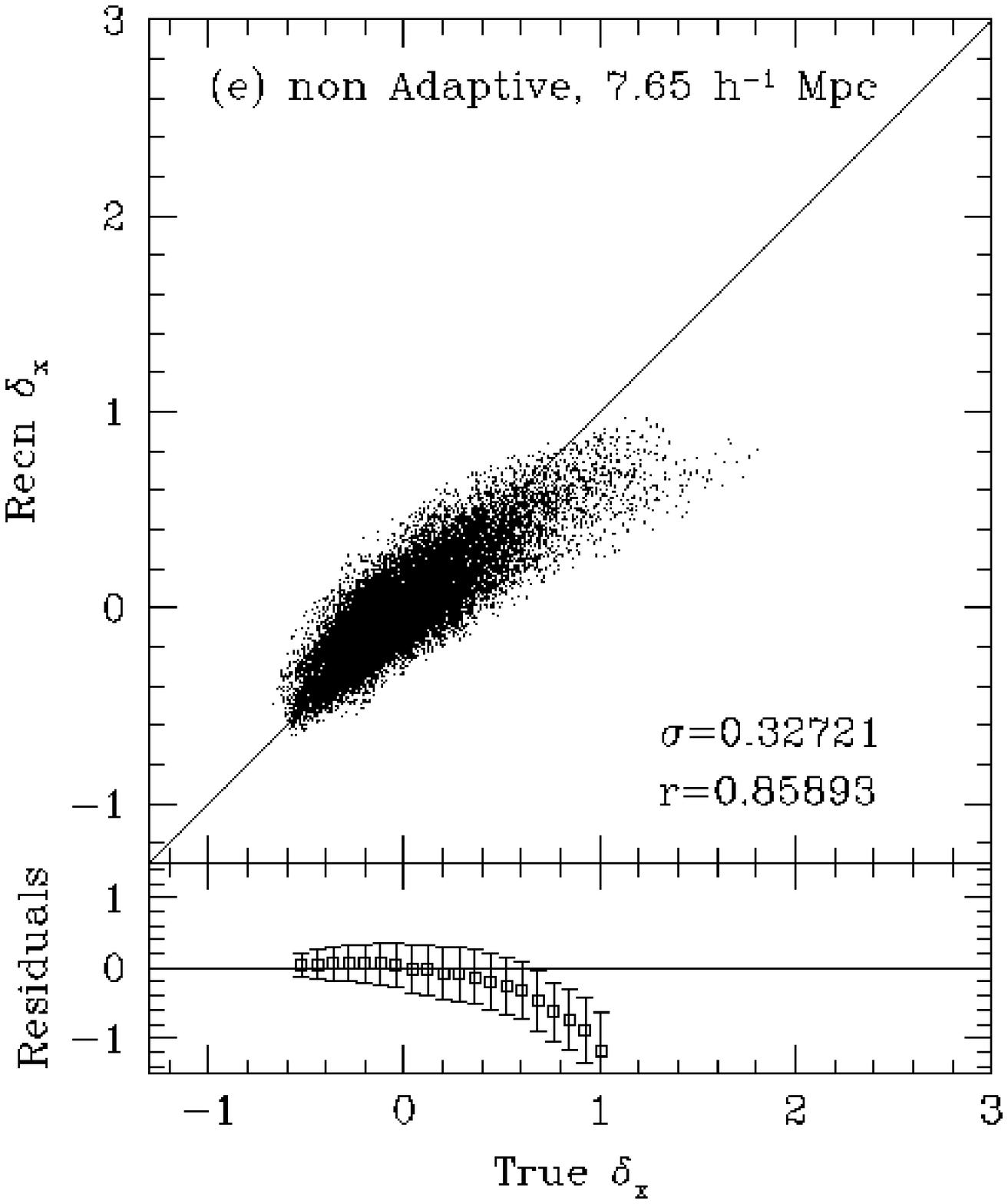,width=6cm}
\psfig{figure=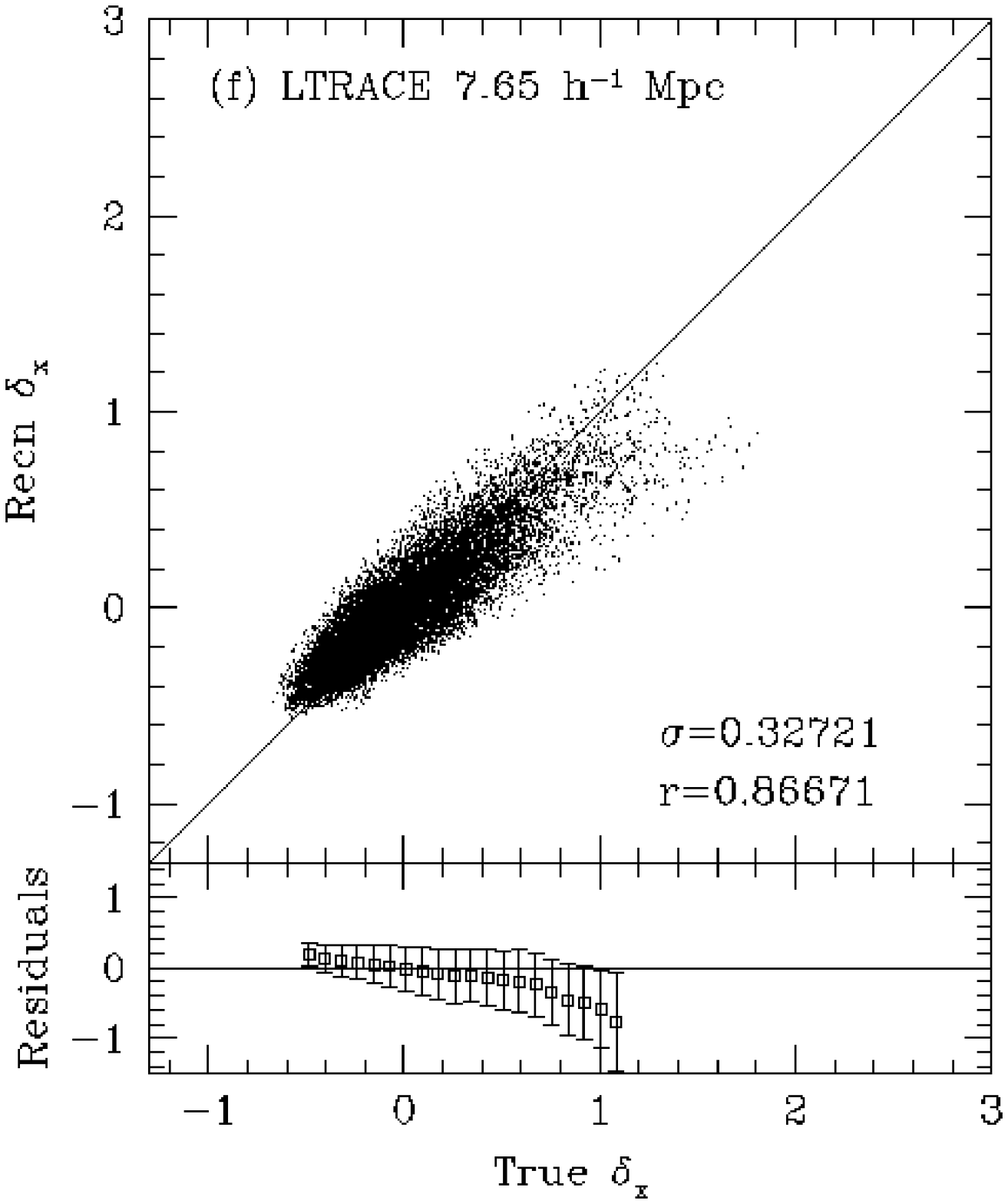,width=6cm}}
\caption{Reconstruction of the real-space density field with ZTRACE
and LTRACE within 50 \mpc\ (upper panels, smoothing of 5.15 \mpc) and 80
\mpc\ (lower panels, smoothing of 7.65 \mpc).}
\end{figure*}

This analysis shows that the ZTRACE procedure improves over linear
theory, especially in recovering the initial density field. We will
show below that the use of adaptive smoothing in ZTRACE gives a
further improvement making it a useful tool for the analysis of real
galaxy redshift surveys.

\begin{figure*}
\centerline{
\psfig{figure=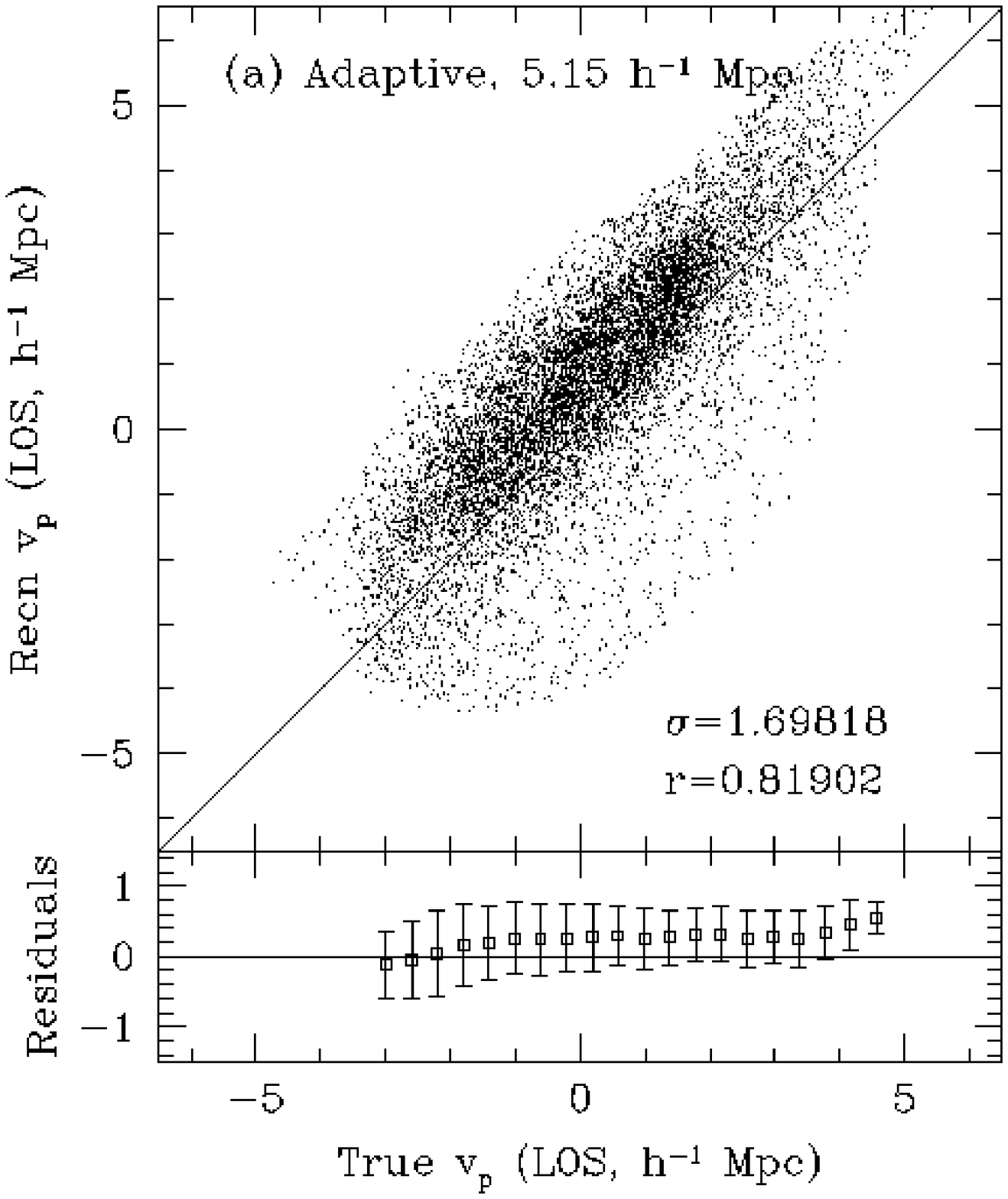,width=6cm}
\psfig{figure=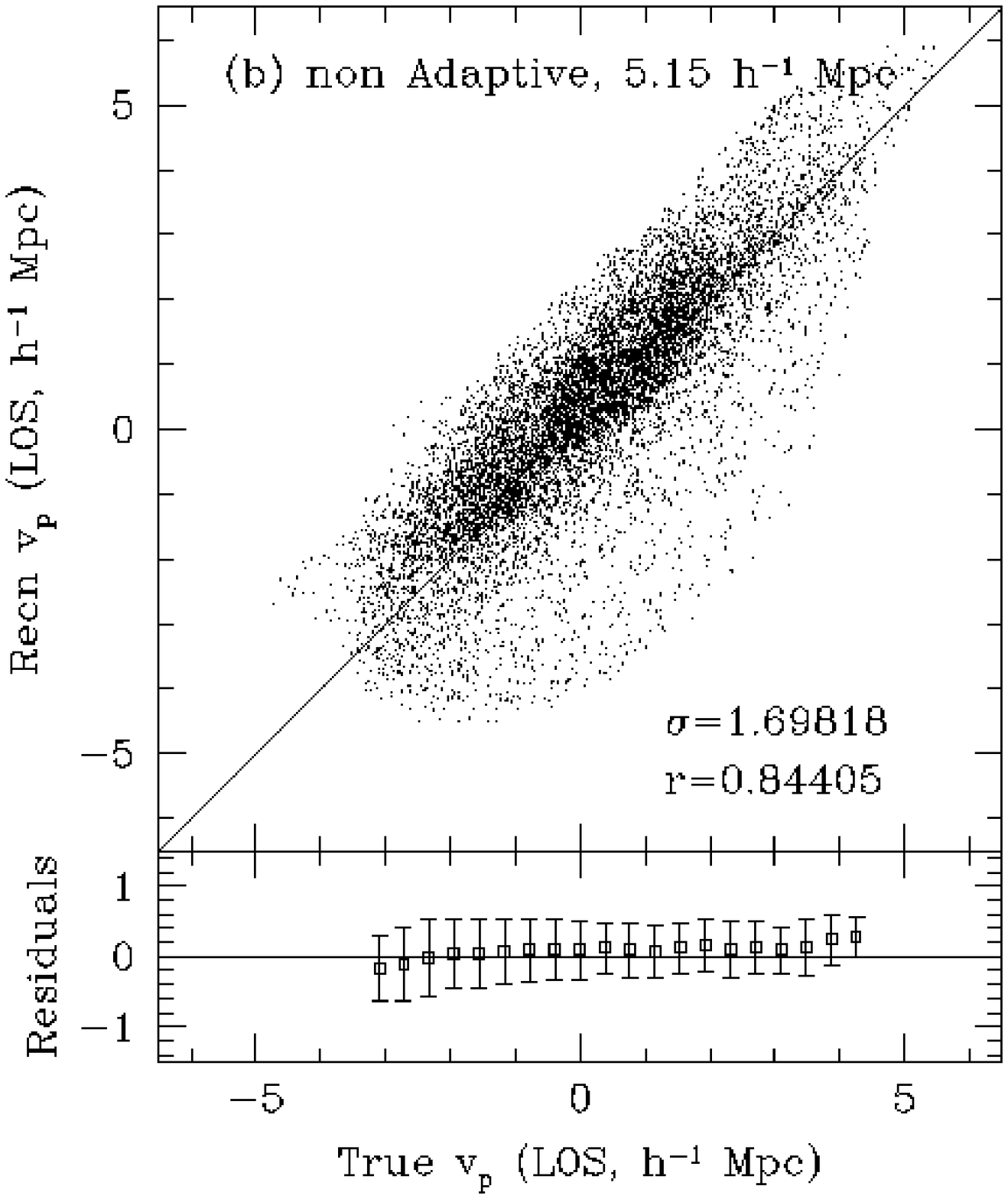,width=6cm}
\psfig{figure=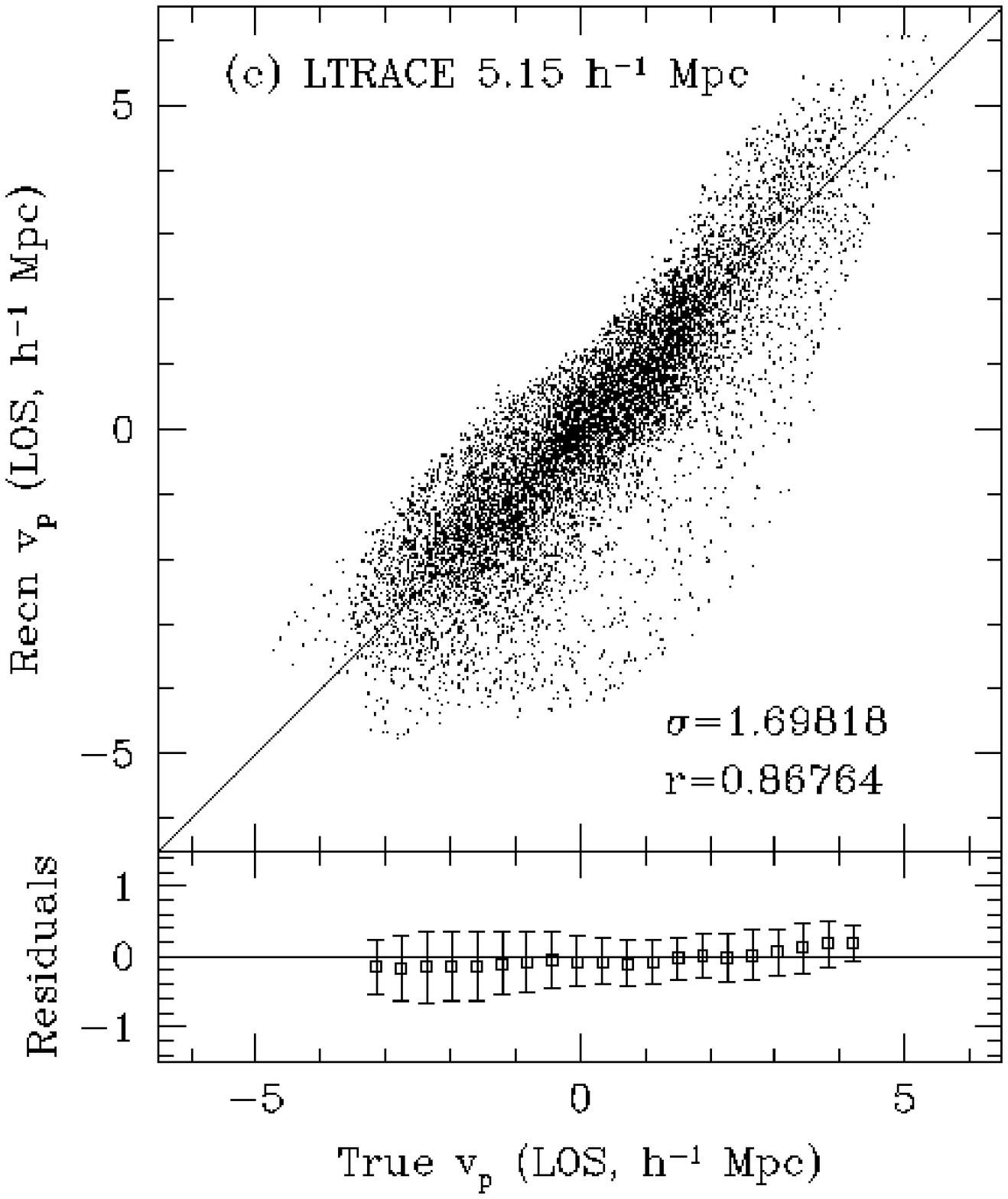,width=6cm}}
\centerline{
\psfig{figure=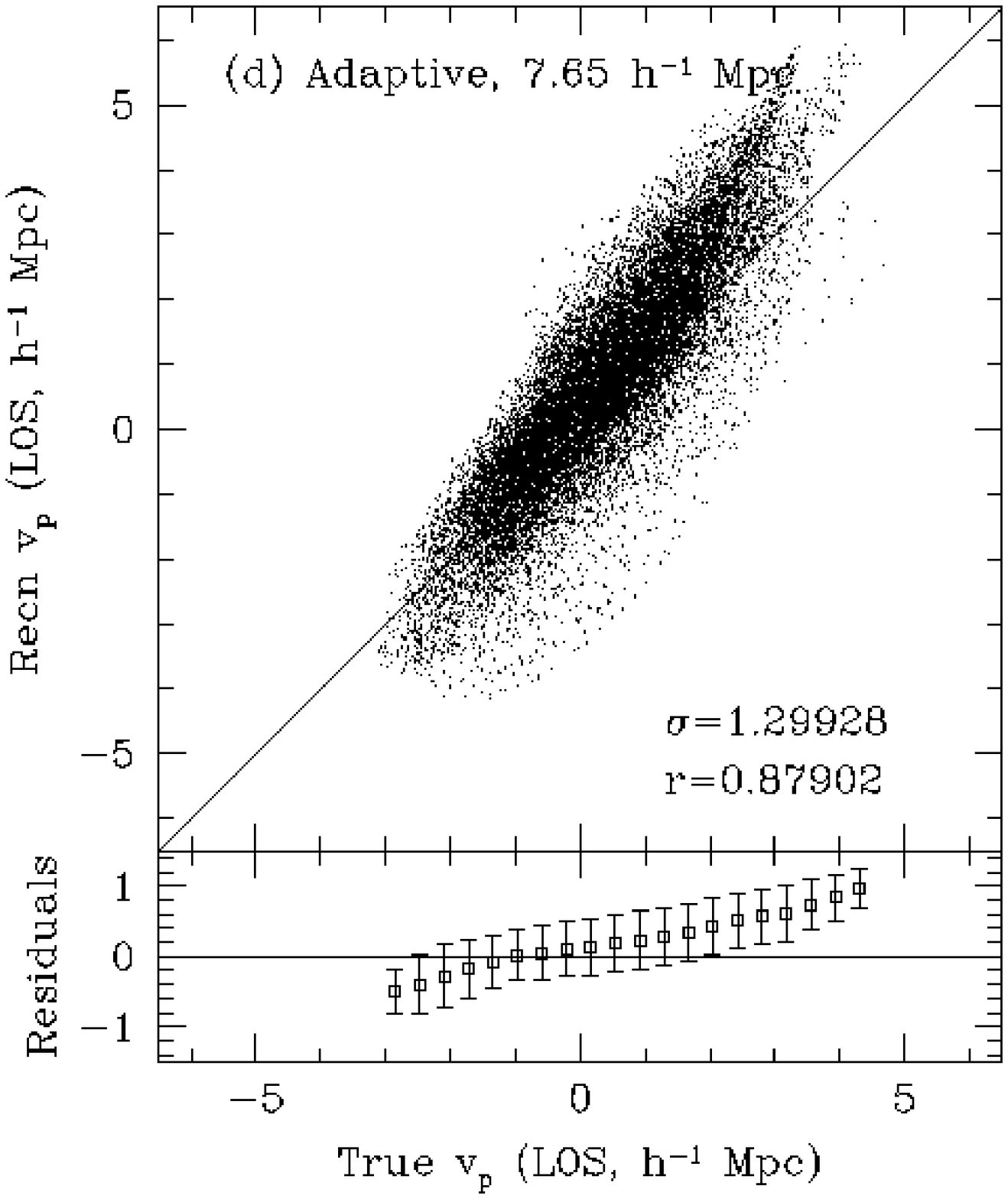,width=6cm}
\psfig{figure=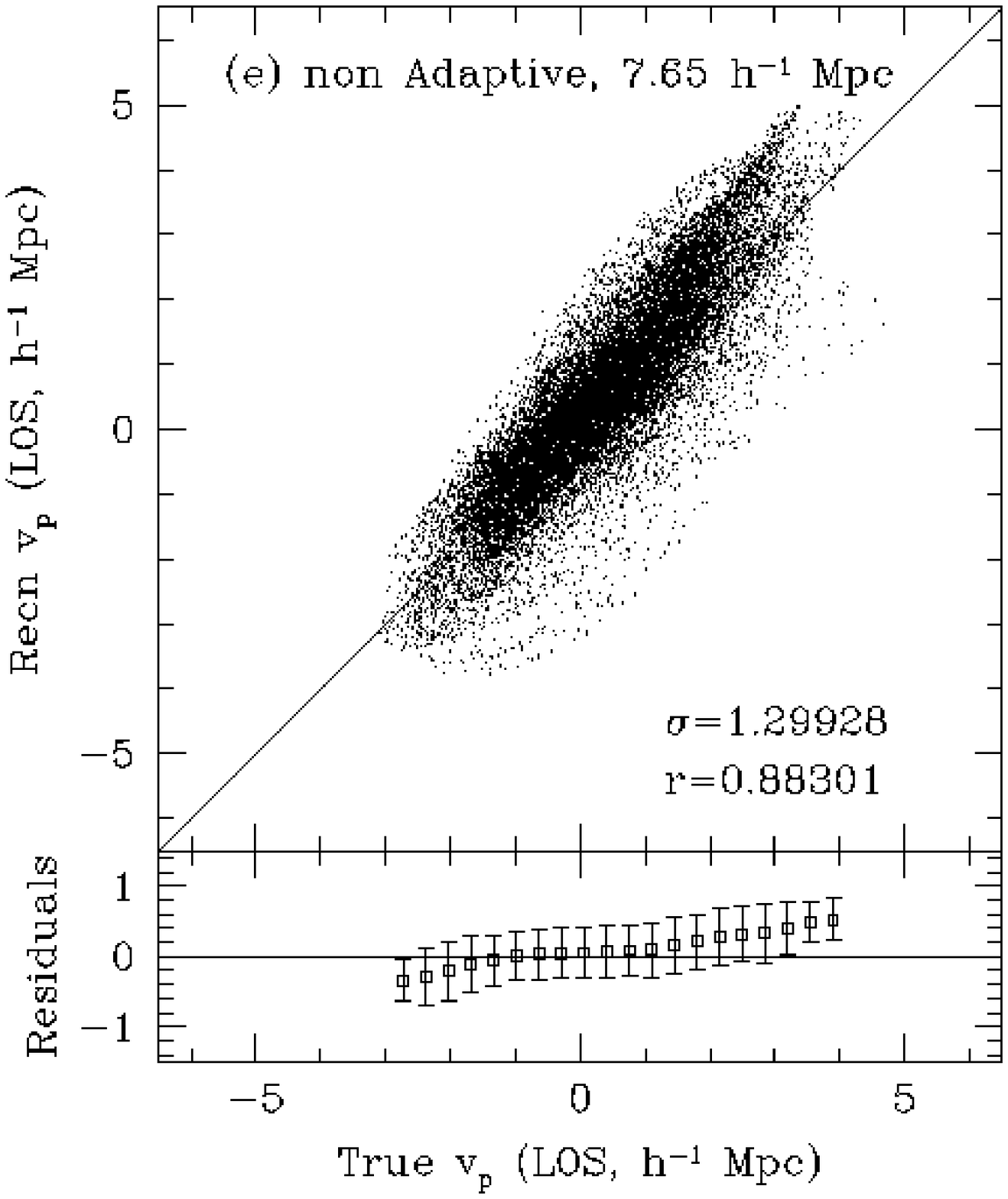,width=6cm}
\psfig{figure=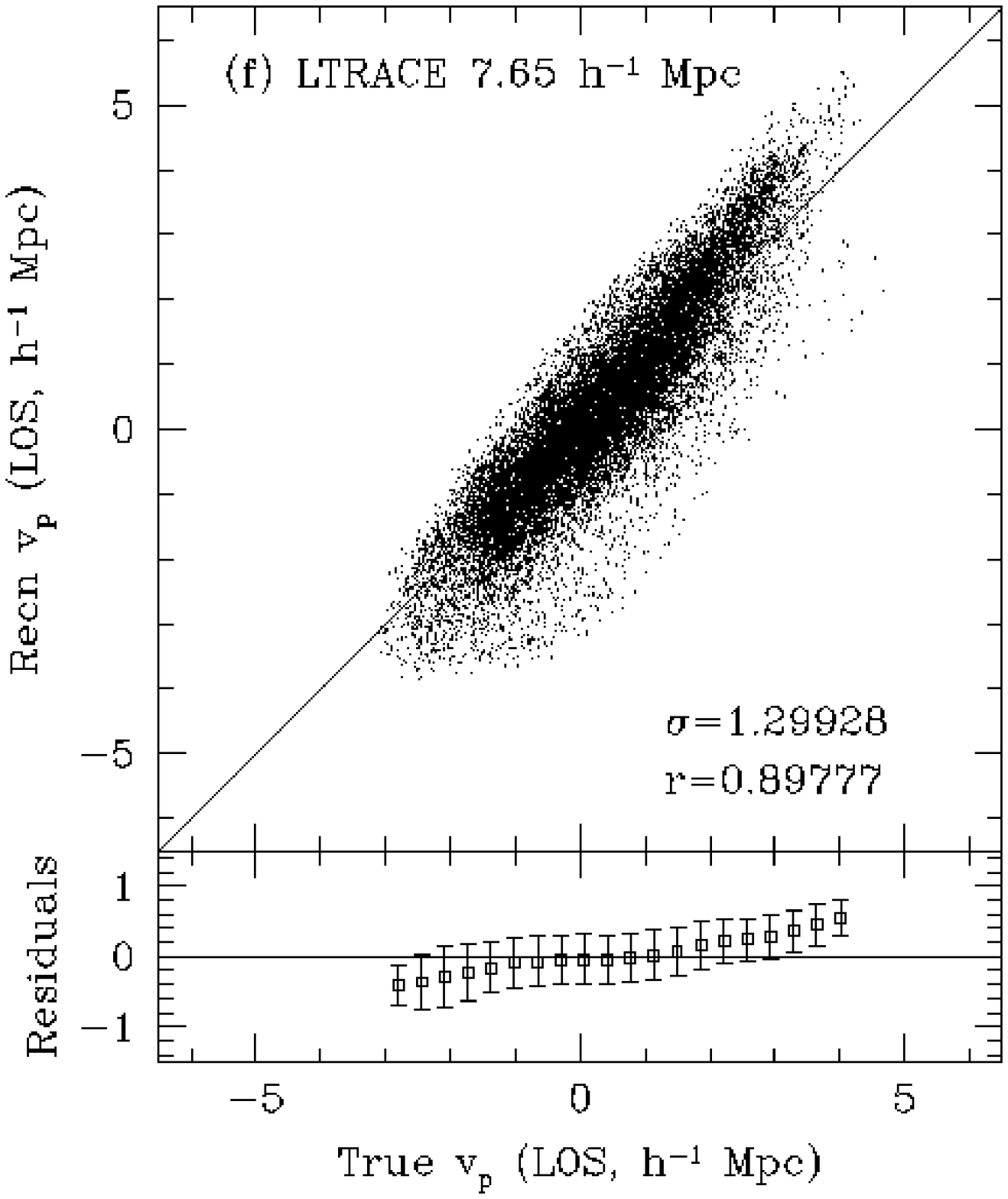,width=6cm}}
\caption{Reconstruction of the LOS peculiar velocity (expressed as an
apparent displacement) with ZTRACE and LTRACE, within 50 \mpc\ (upper
panels, smoothing of 5.15 \mpc) and 80 \mpc\ (lower panels, smoothing
of  7.65 \mpc).}
\end{figure*}

Figs. 6 to 8 show the performance of ZTRACE and LTRACE in
reconstructing the real-space density, LOS velocity and initial
density, from the simulated PSCz catalogue described in Section 5.1.
We provide the redshift-space density field as input smoothed with
adaptive and non-adaptive Gaussian filters as described in Section
5.2.  The figures show results for the ZTRACE algorithm for smoothing
radii of 5.15 and 7.65 \mpc\ (for adaptive smoothing these numbers
refer to the values of the reference radii). The upper panels show the
reconstruction within 50 \mpc\ of the origin and the lower panels show
results within 80 \mpc.  To suppress the grid-level noise introduced
by the reconstruction, the recovered fields are smoothed on a scale of
2.5 \mpc\ in the 50 \mpc\ case, and on a scale of 3.75 \mpc\ in the 80
\mpc\ case. The results for the LTRACE algorithm with non-adaptive
smoothing are also shown in the figures.

\begin{figure*}
\centerline{
\psfig{figure=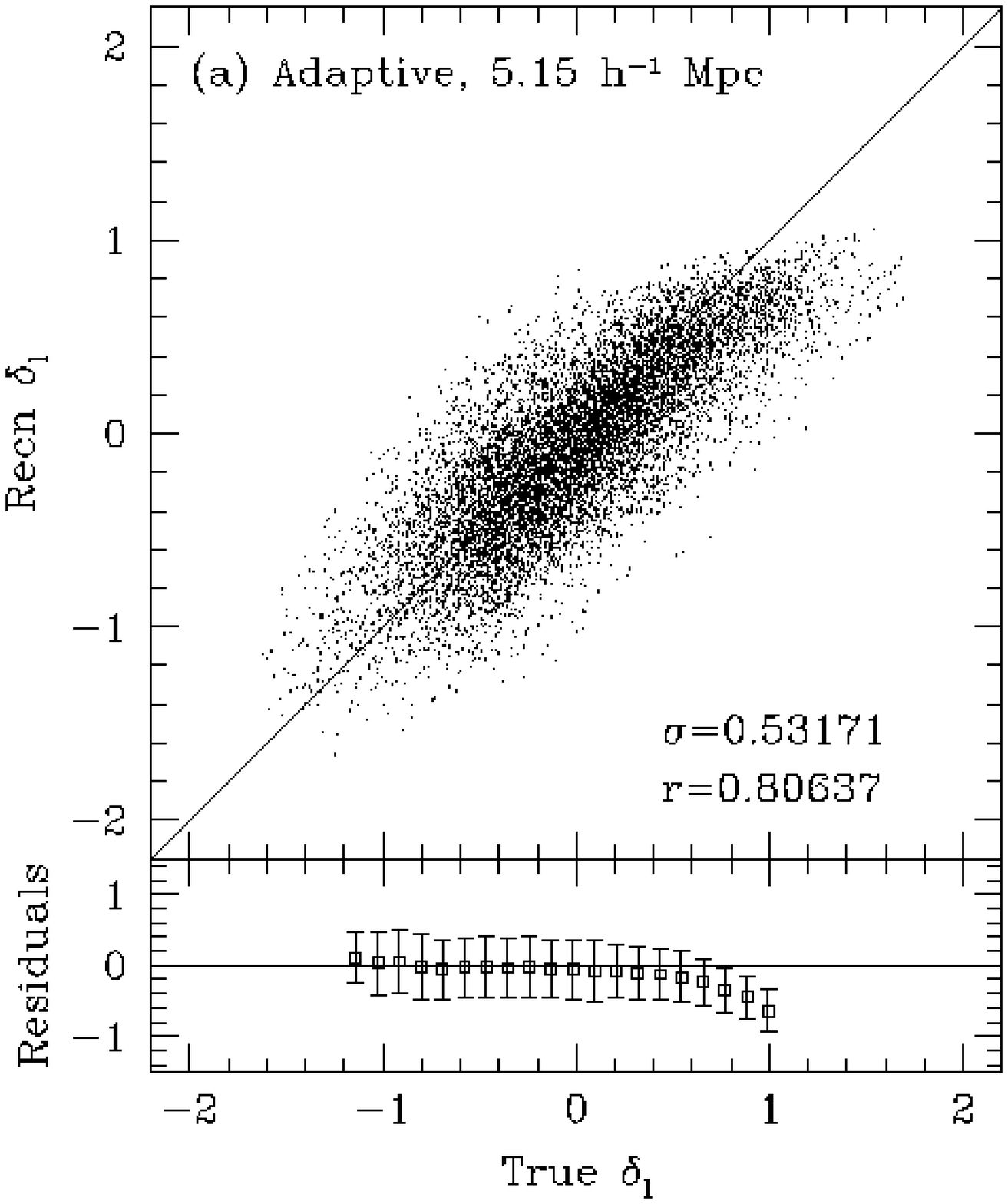,width=6cm}
\psfig{figure=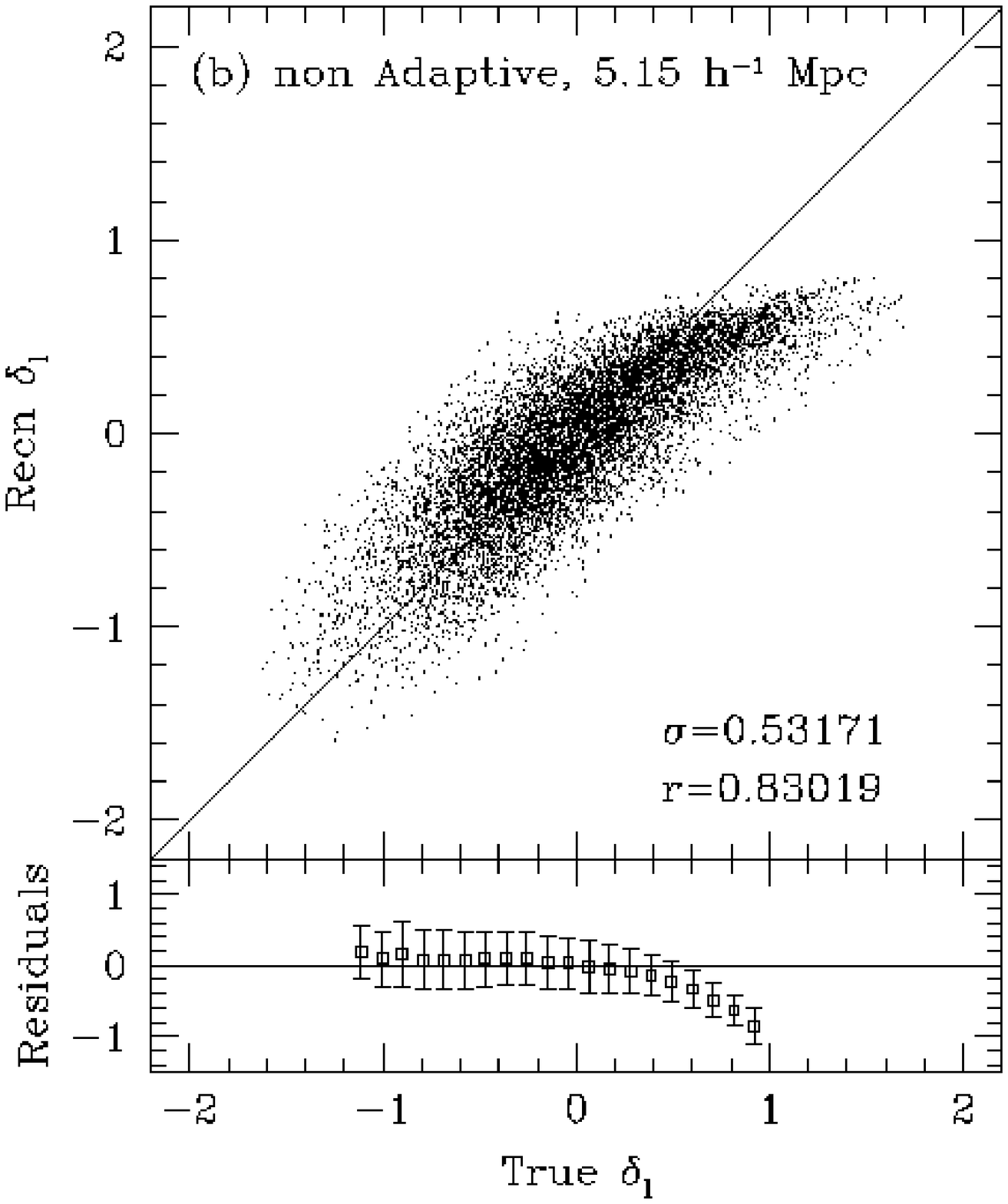,width=6cm}
\psfig{figure=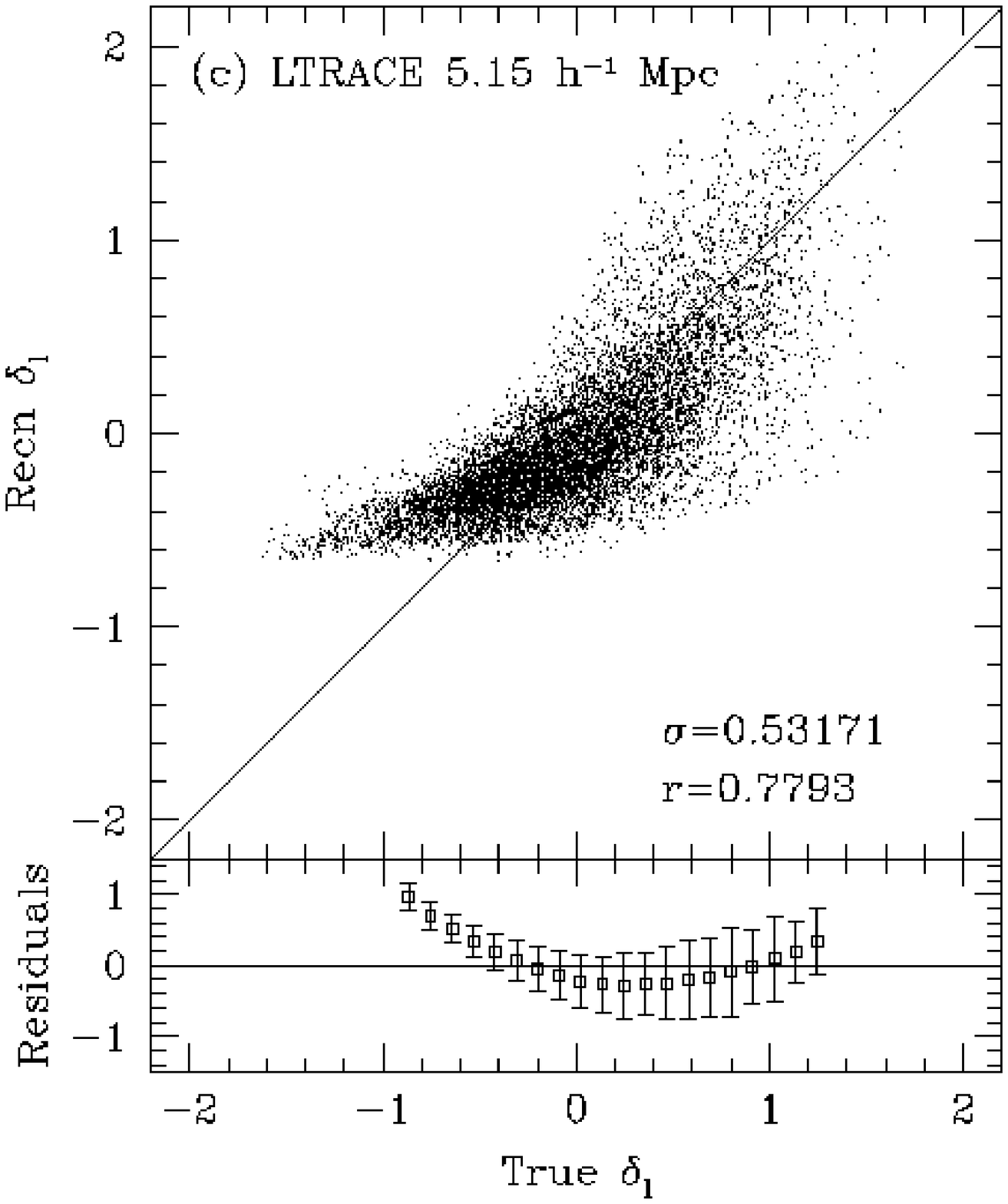,width=6cm}}
\centerline{
\psfig{figure=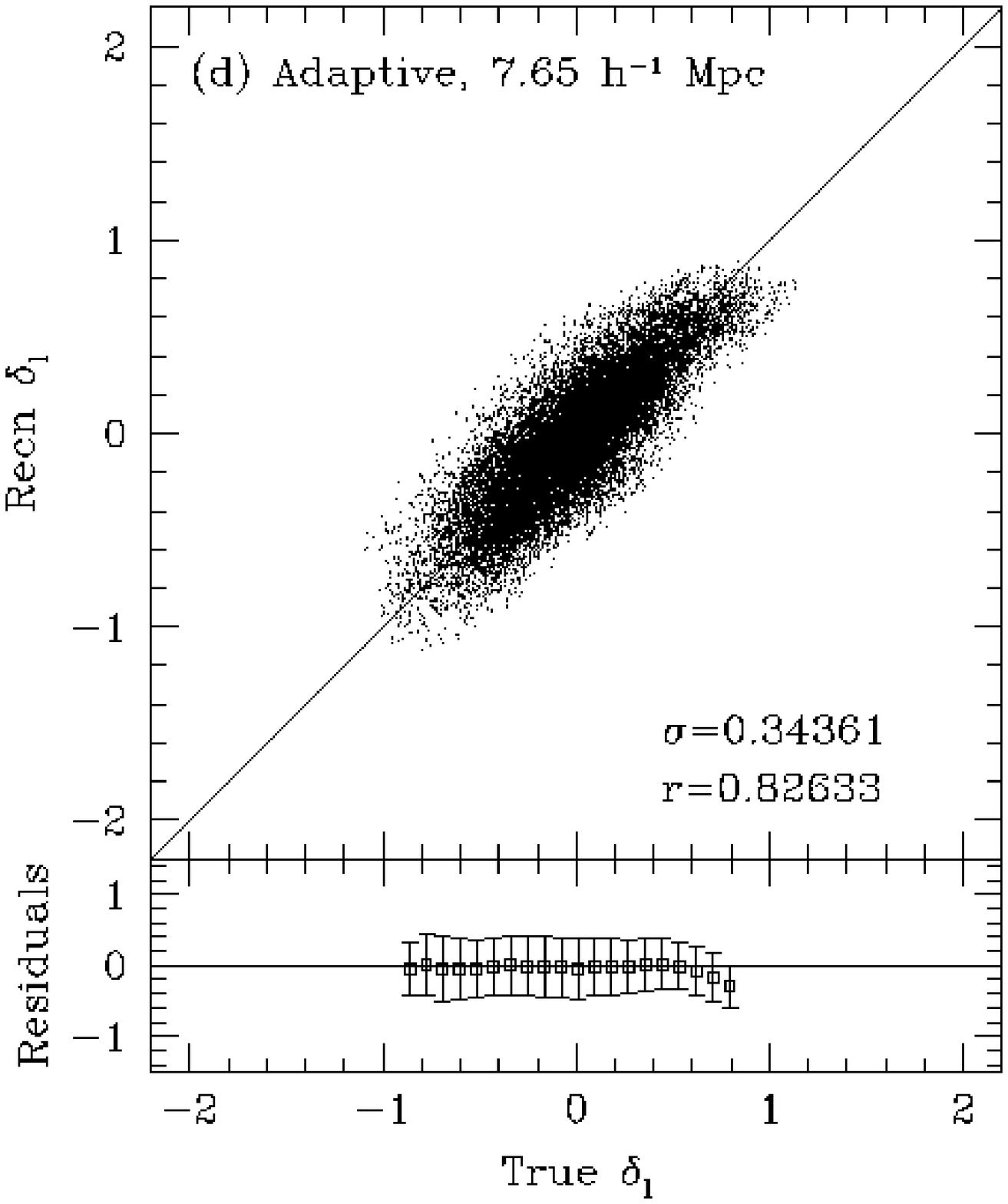,width=6cm}
\psfig{figure=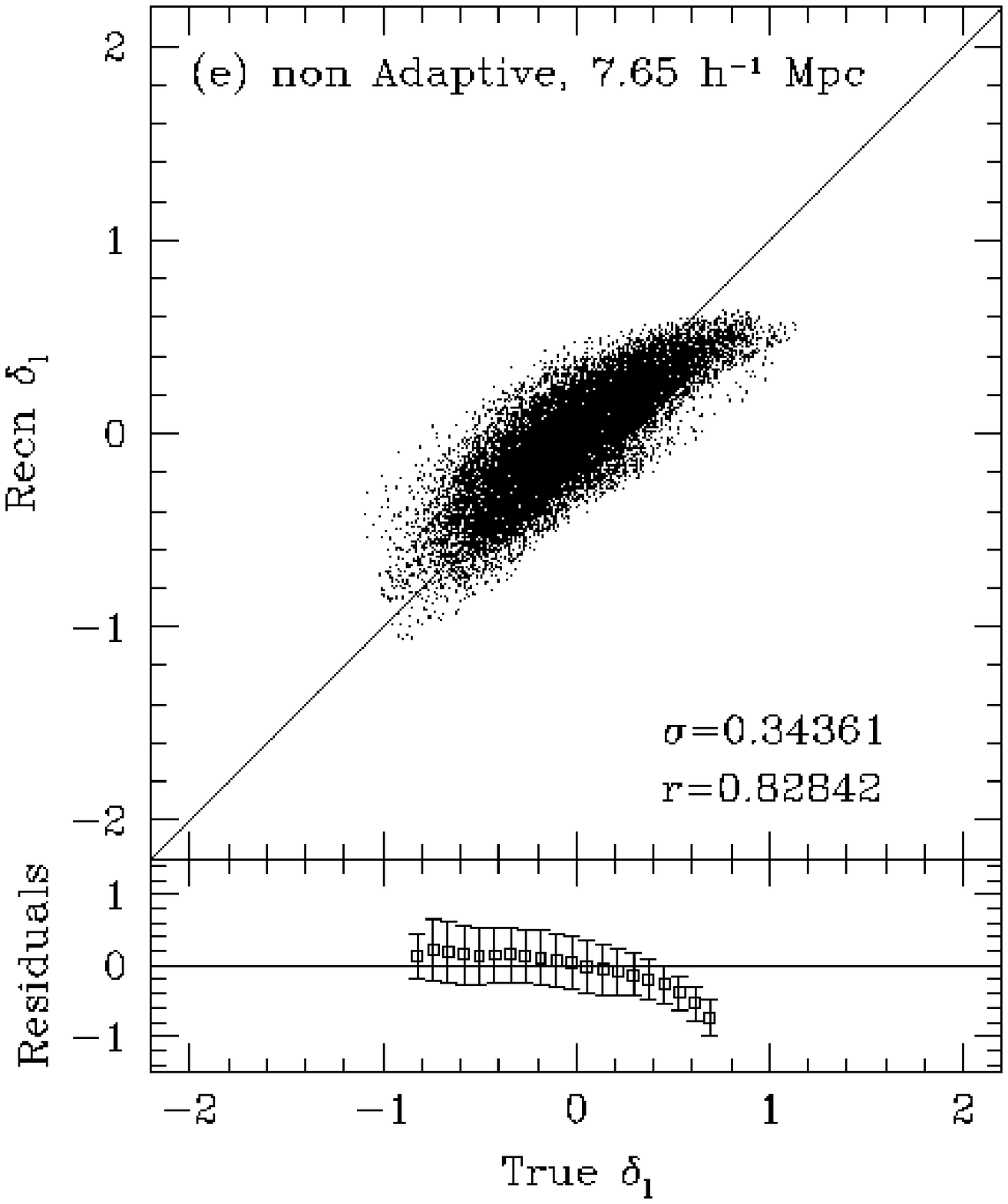,width=6cm}
\psfig{figure=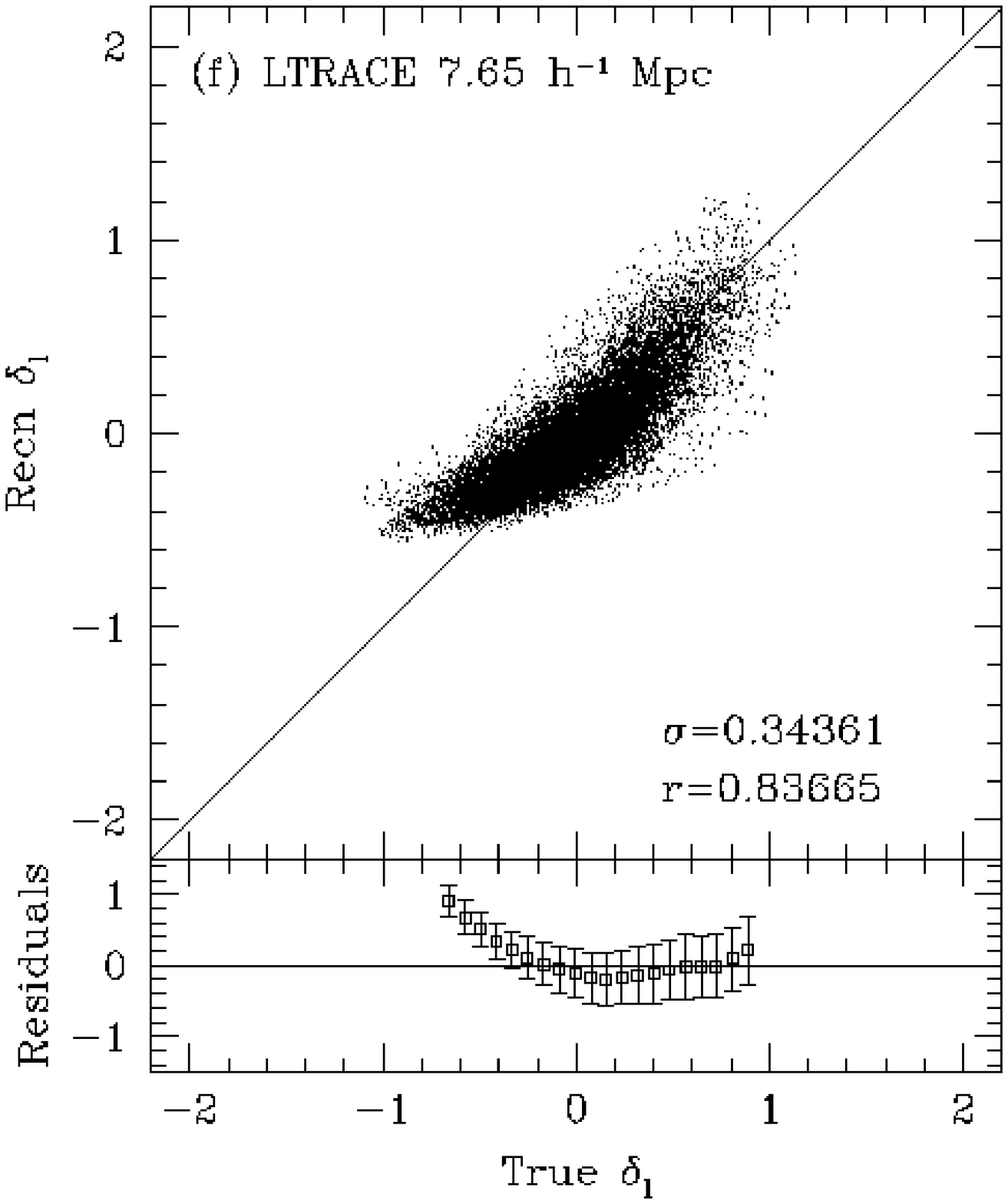,width=6cm}}
\caption{Reconstruction of the linear density with ZTRACE and LTRACE,
within 50 \mpc\ (upper panels, smoothing of 5.15 \mpc) and 80 \mpc\ (lower
panels, smoothing of 7.65 \mpc).}
\end{figure*}

For reference, Fig. 9 shows  a comparison between the redshift-space
density field, as estimated from the simulated PSCz catalogue, and the true
redshift-space density of the full simulation smoothed with constant
smoothing radii of 5.15 and 7.65 \mpc. The dominant source of noise in
this figure arises from the estimation of the density field from the
sparsely sampled redshift catalogue as we have discussed in Section 4. 
The bias at high densities seen
when adaptive smoothing is applied to the redshift catalogue is an
artifact of comparing the adaptively smoothed field with the
non-adaptively smoothed density estimates from the N-body simulation.

Fig. 10 shows the reconstructed initial densities from ZTRACE and
LTRACE with 7.65 \mpc\ smoothing, compared with the true initial
density field of the simulation along a slice of the computational
volume centred on the observer. The ZTRACE algorithm gives a good
representation of the initial conditions even at the edge of the
simulated redshift survey.  In contrast, the initial conditions
recovered by LTRACE resemble a heavily smoothed (and biased) version
of the initial conditions. Fig. 8 provides a more quantitative
comparison of the initial densities recovered by these algorithms.

\begin{figure*}
\centerline{
\psfig{figure=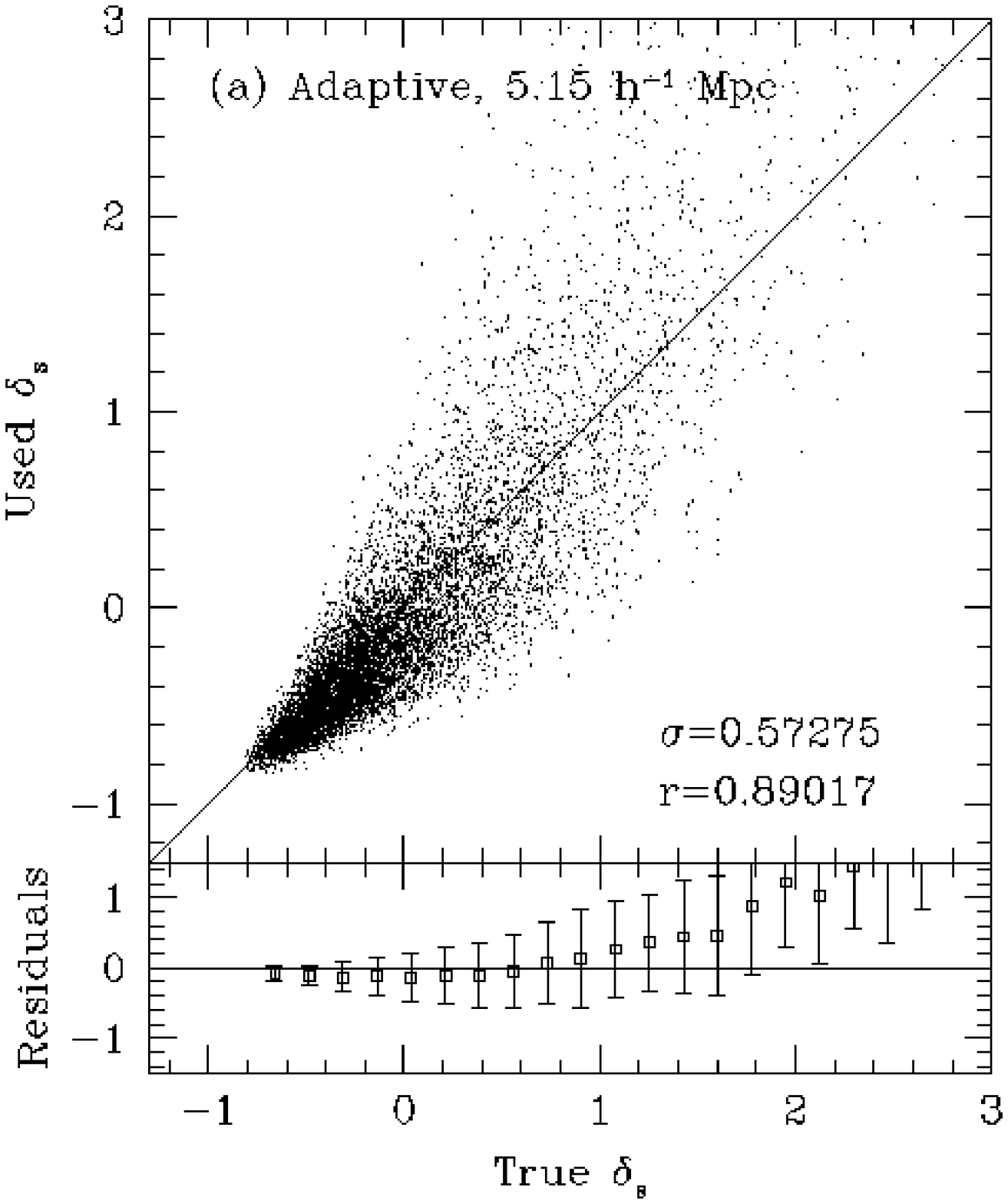,width=6cm}
\psfig{figure=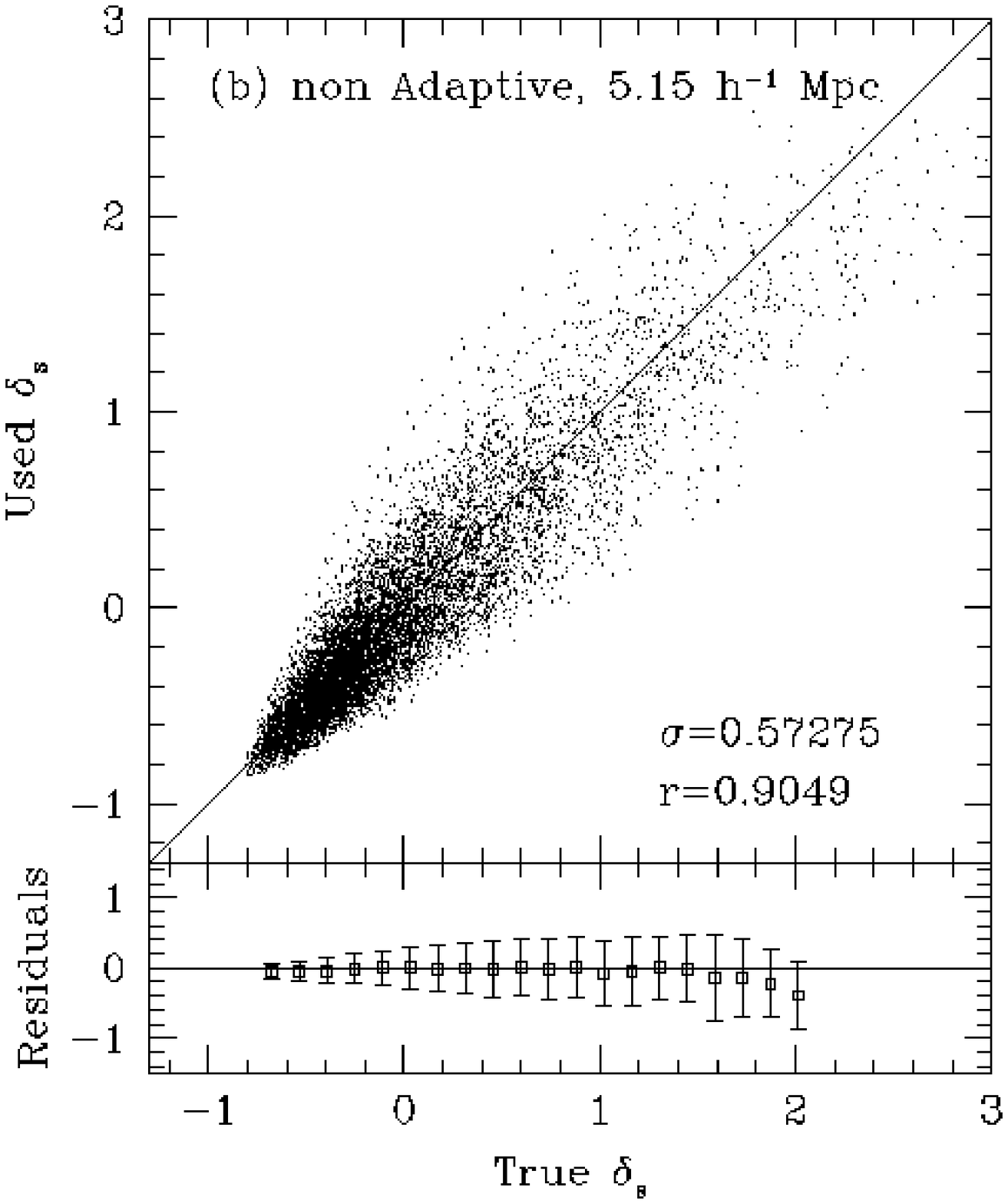,width=6cm}}
\centerline{
\psfig{figure=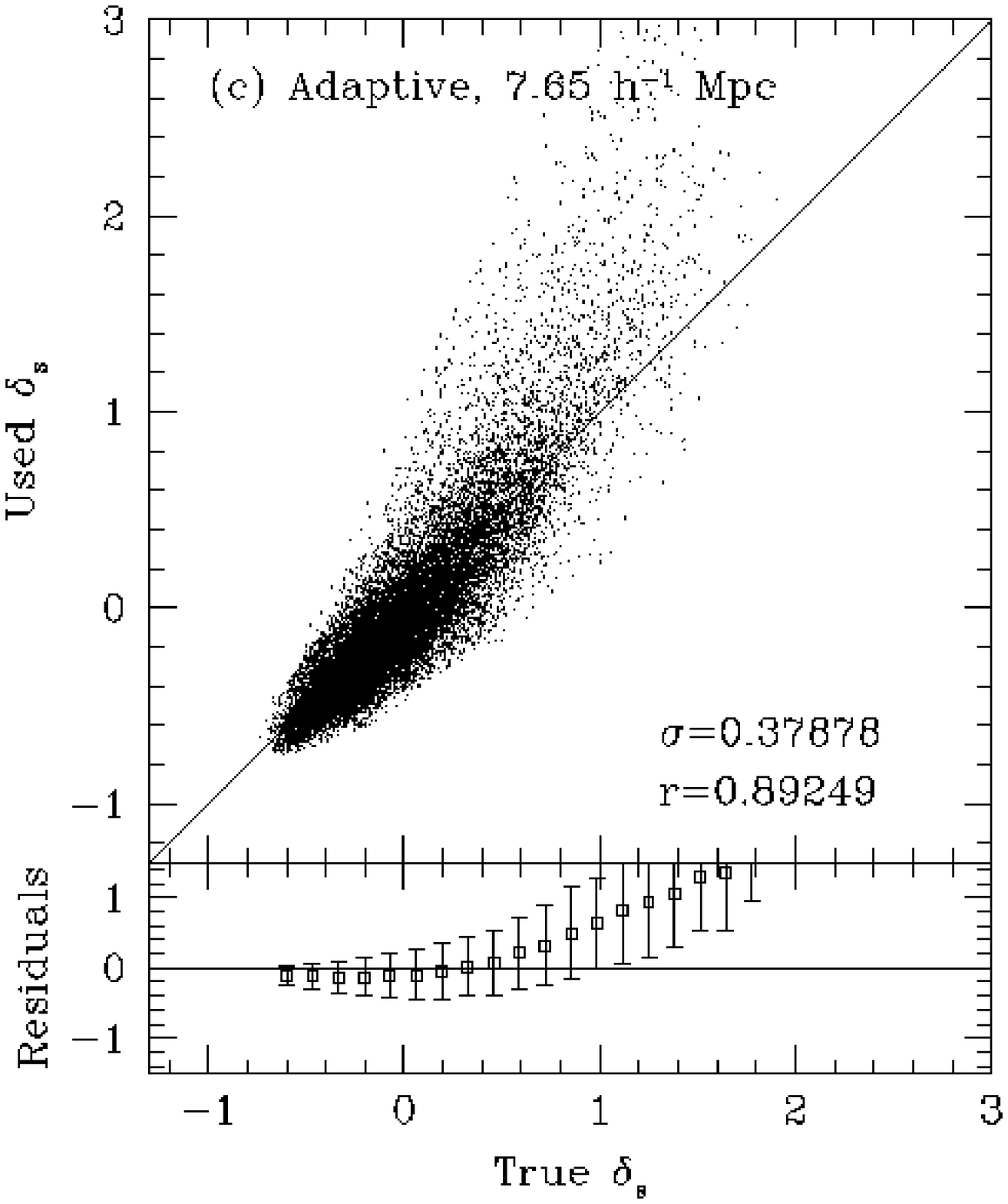,width=6cm}
\psfig{figure=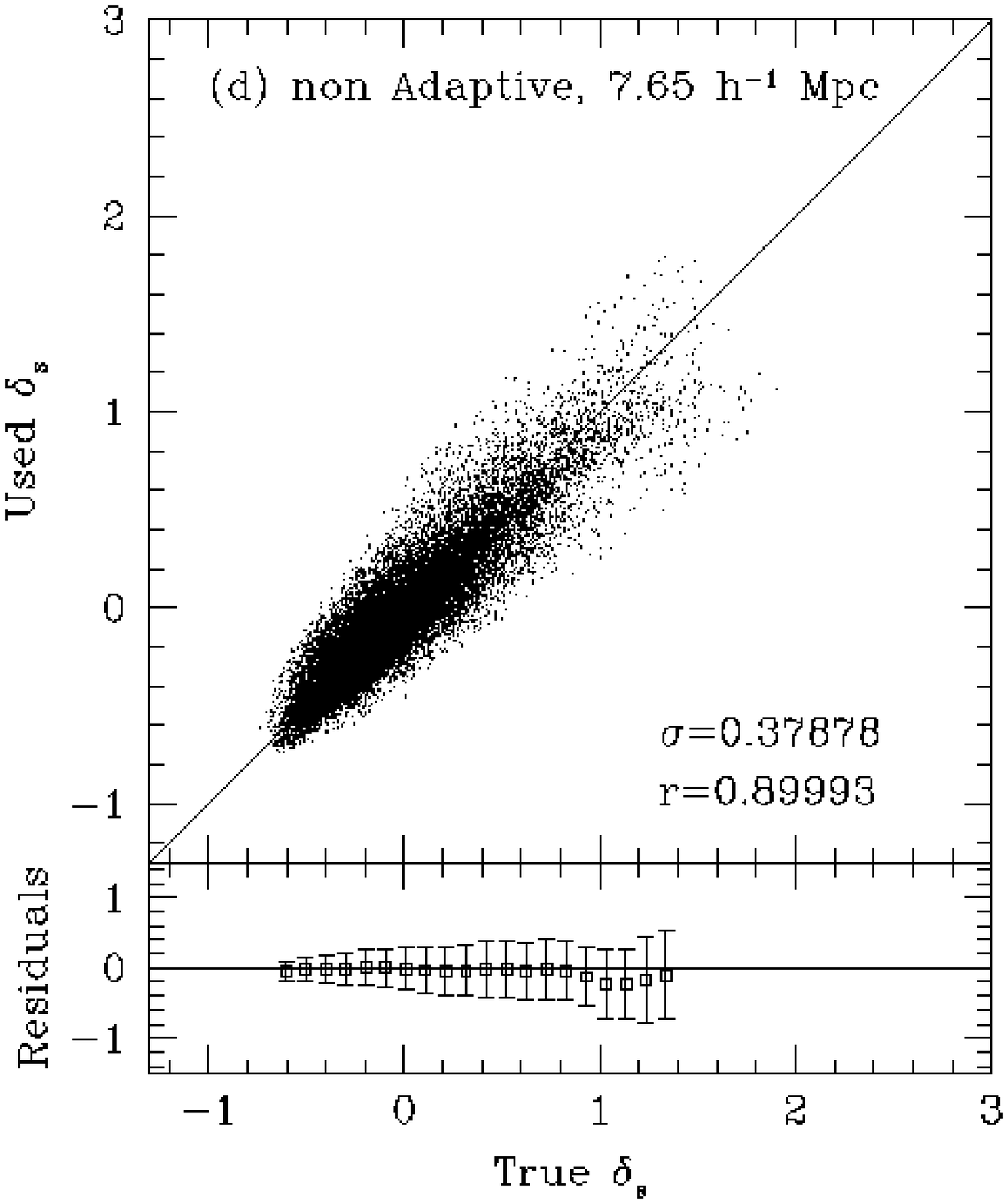,width=6cm}}
\caption{Reconstruction with ZTRACE, within 50 \mpc\ (upper panels,
smoothing of 5.15 \mpc) or 80 \mpc\ (lower panels, smoothing 7.65 \mpc).
Redshift-space densities used as an input to the ZTRACE algorithm
compared  with the
smoothed true density from the full output of the simulation.
(a) and (c) shows adaptive smoothing with a reference radii of 
5.15 \mpc\ and 7.65 \mpc; (b) and (d) show non-adaptive smoothing
with radii of 5.15 \mpc\ and 7.65 \mpc.}
\end{figure*}

From this analysis we conclude the following:

\begin{enumerate}
\item
From Fig. 9, we see that the scatter in the density estimate is large,
as expected from the analysis of Section 4.  Adaptive smoothing
increases the noise slightly.
\item
Adaptive smoothing leads to an accurate reconstruction of the real-space
density field (Fig. 6). Even  high-density peaks with $\delta \sim 1$ 
are  recovered with little bias. In contrast, the real space 
densities recovered with non-adaptive smoothing are 
strongly biased at density contrasts approaching unity.

\item
All of the methods recover the LOS peculiar velocities to comparable
accuracy (Fig. 7). The peculiar velocity field is more strongly
correlated than the density field and the number of effectively
independent volumes in our $50$ and $80$ \mpc\ spheres is relatively
small.  The velocity fields thus show structure arising from distinct
objects (especially in the lower panels of Fig. 7).

\item
Adaptive smoothing significantly improves the accuracy of the 
linear density field reconstruction from ZTRACE algorithm (Figs. 8 and 10).
The recovered linear density field is almost unbiased, even to density
contrasts approaching unity. Furthermore, the adaptive ZTRACE
linear density field has the correct variance.

\end{enumerate}

\begin{figure*}
\centerline{\psfig{figure=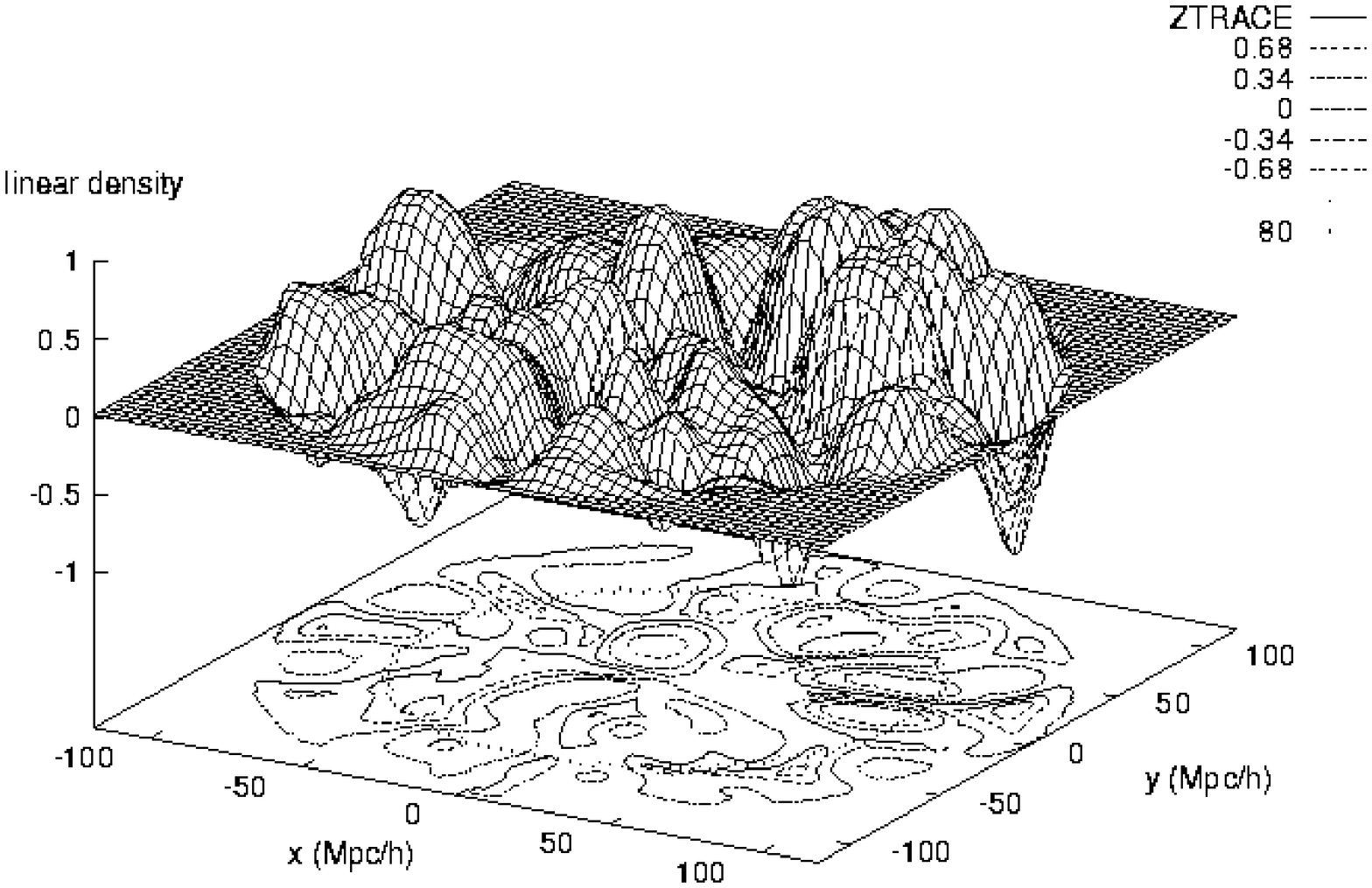,angle=-90,width=10cm}}
\centerline{\psfig{figure=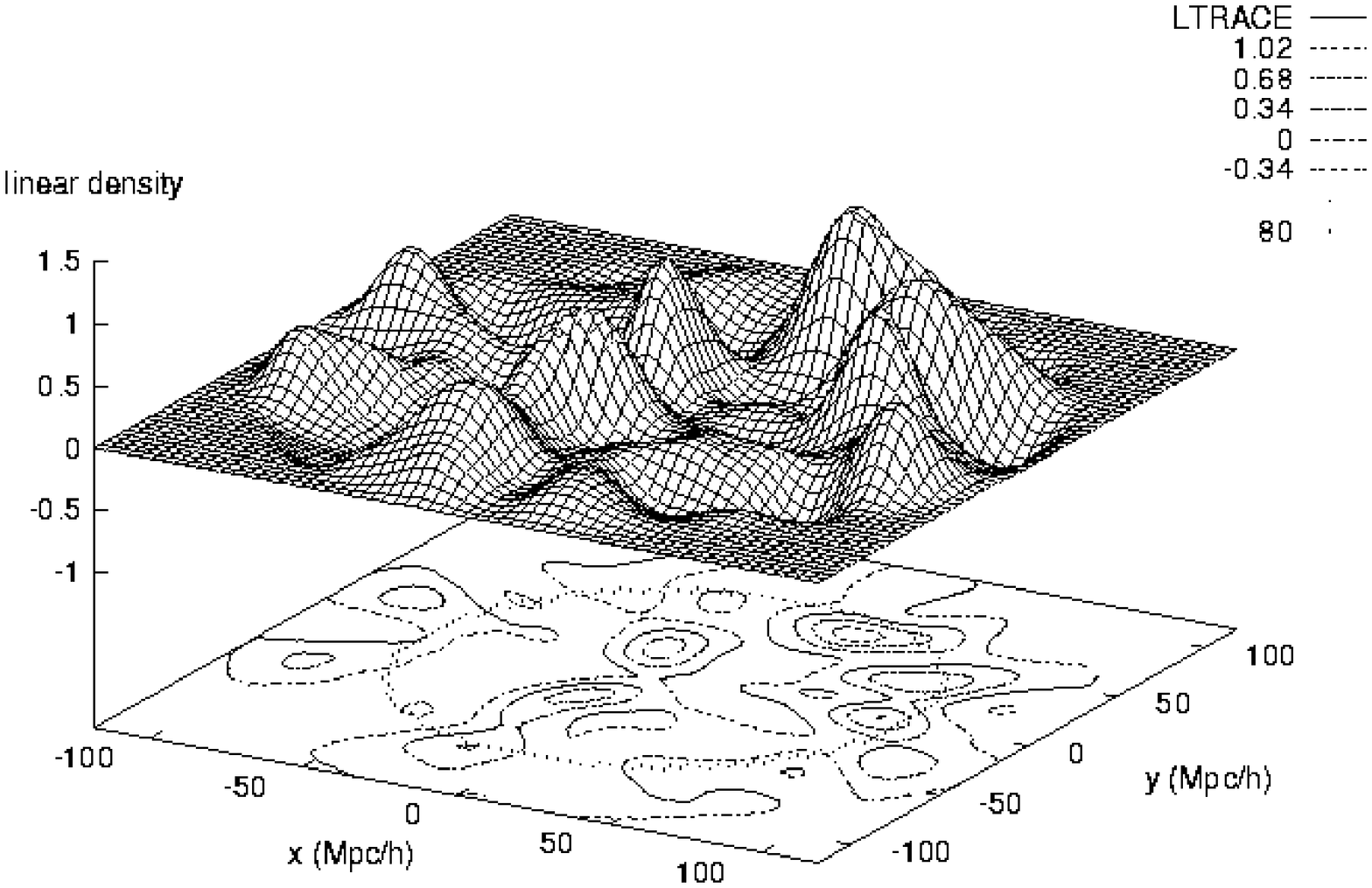,angle=-90,width=10cm}}
\centerline{\psfig{figure=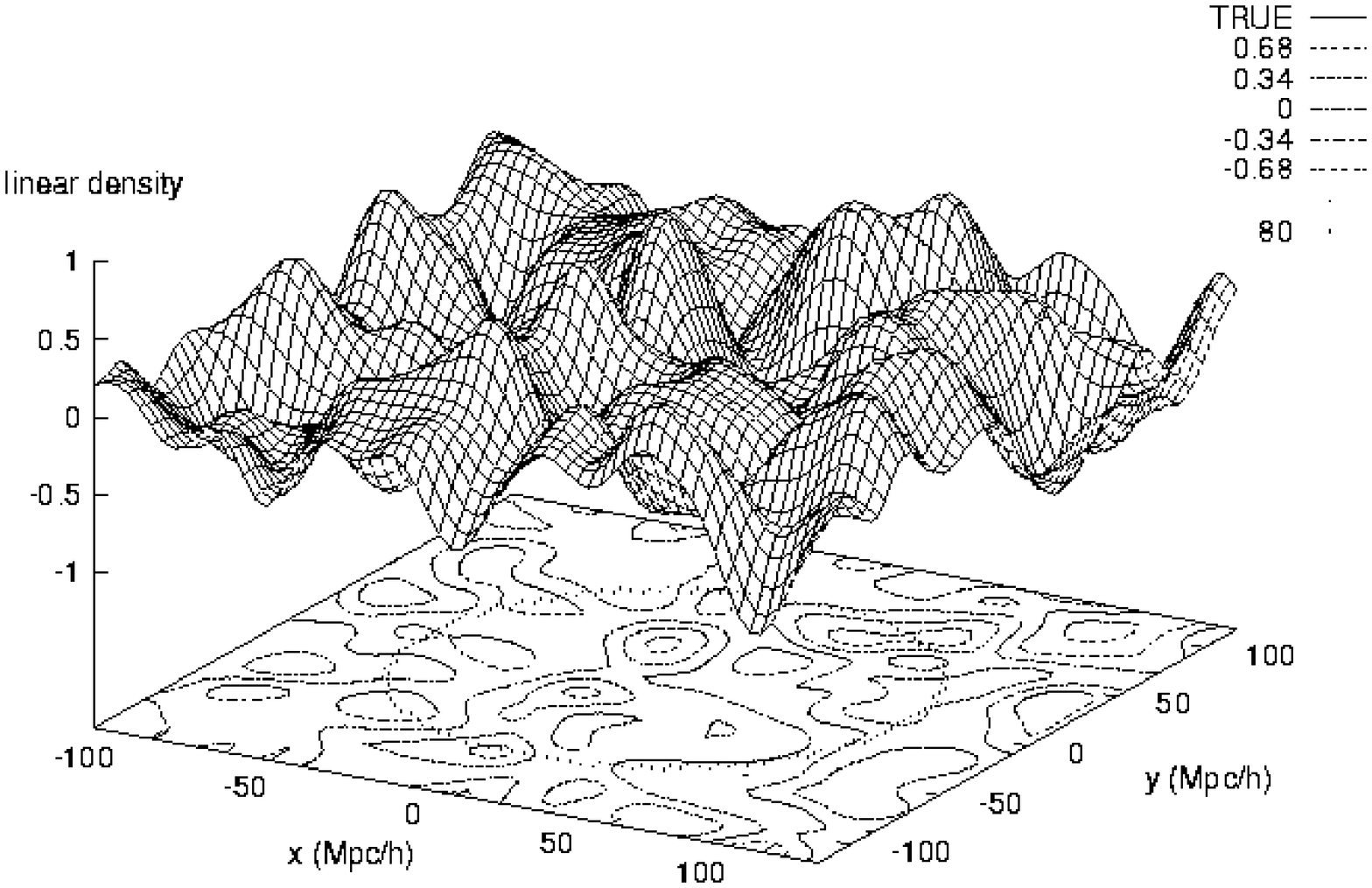,angle=-90,width=10cm}}
\caption{Reconstruction of the linear density from a simulated PSCz
redshift survey along a thin slice centered on the observer. The upper
two figures show the results from ZTRACE and LTRACE, within 80 \mpc\
(smoothing of 7.65 \mpc), compared to the true linear density field
plotted at the bottom.  The contours show one, two and three $\sigma$
levels and the radius of 80 \mpc\ is shown by the dotted lines.}
\end{figure*}

\subsection{The PDF of the Reconstructed Linear Density}

One of the problems with many of the reconstruction methods described
in the Introduction is that they are not able to reproduce the correct
PDF of the initial conditions.  This problem can be avoided
heuristically by ``Gaussianizing'' the reconstructed field as
described by Weinberg (1992) and Narayanan and Weinberg (1998).
Clearly, it is important to develop techniques for recovering the
initial conditions without making any specific assumptions concerning
the PDF. Fig. 11 shows the 1-point PDF of the reconstructed linear
density computed from ZTRACE compared to the Gaussian PDF of the true
initial conditions.  The adaptive ZTRACE algorithm reproduces the true
PDF very accurately, except for the tail at $\delta_l>1$ which is
severely truncated.  The PDF of underdense regions, even the tail at
large negative amplitudes, is reproduced extremely well by the adaptive 
ZTRACE  algorithm. 

When applied to real redshift surveys, this could provide an
interesting test of the Gaussianity of the initial conditions. More
realistically, the PDF of the initial conditions recovered by
ZTRACE could be useful in constraining models of bias.

\begin{figure*}
\centerline{
\psfig{figure=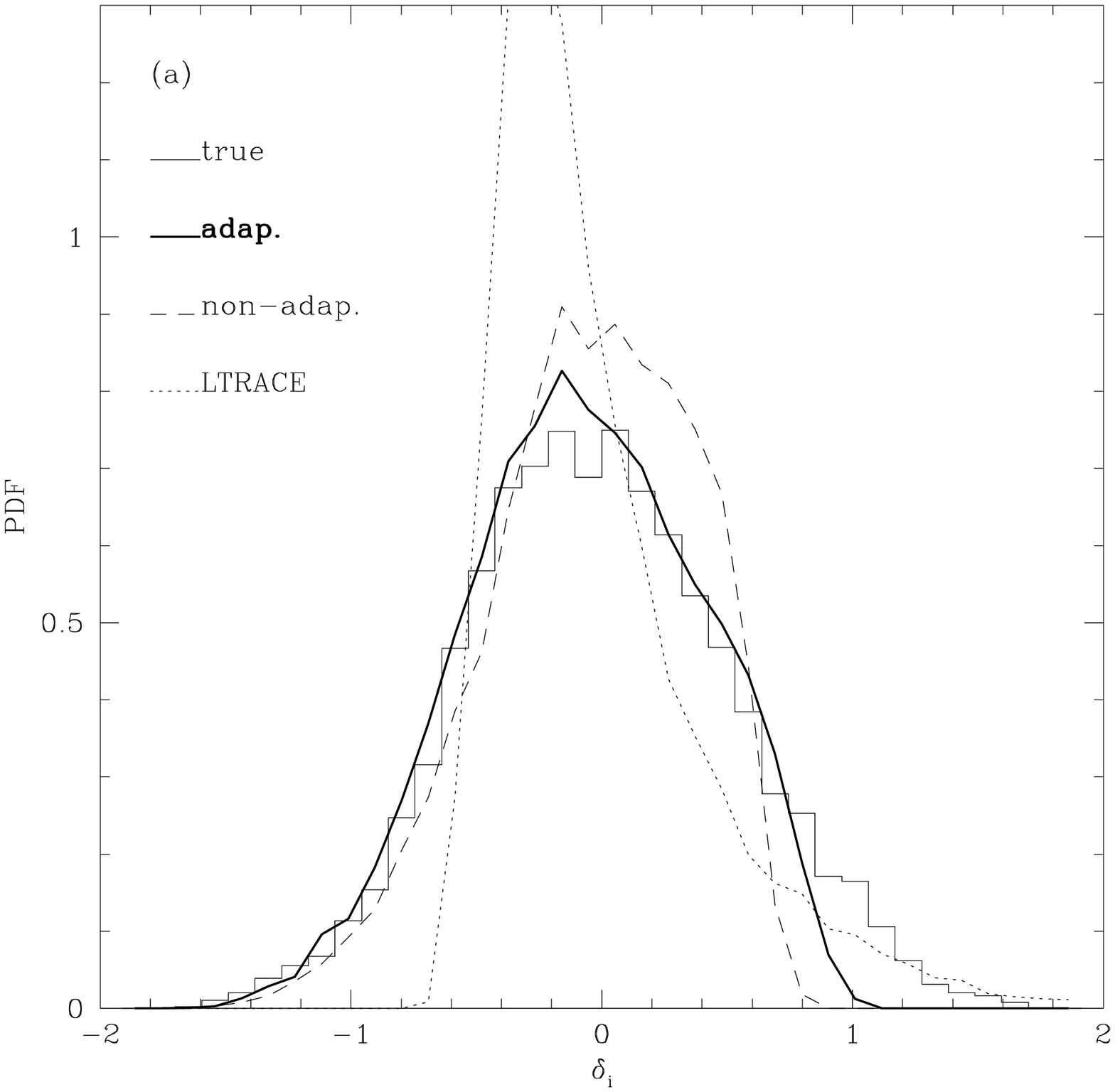,width=9cm}
\psfig{figure=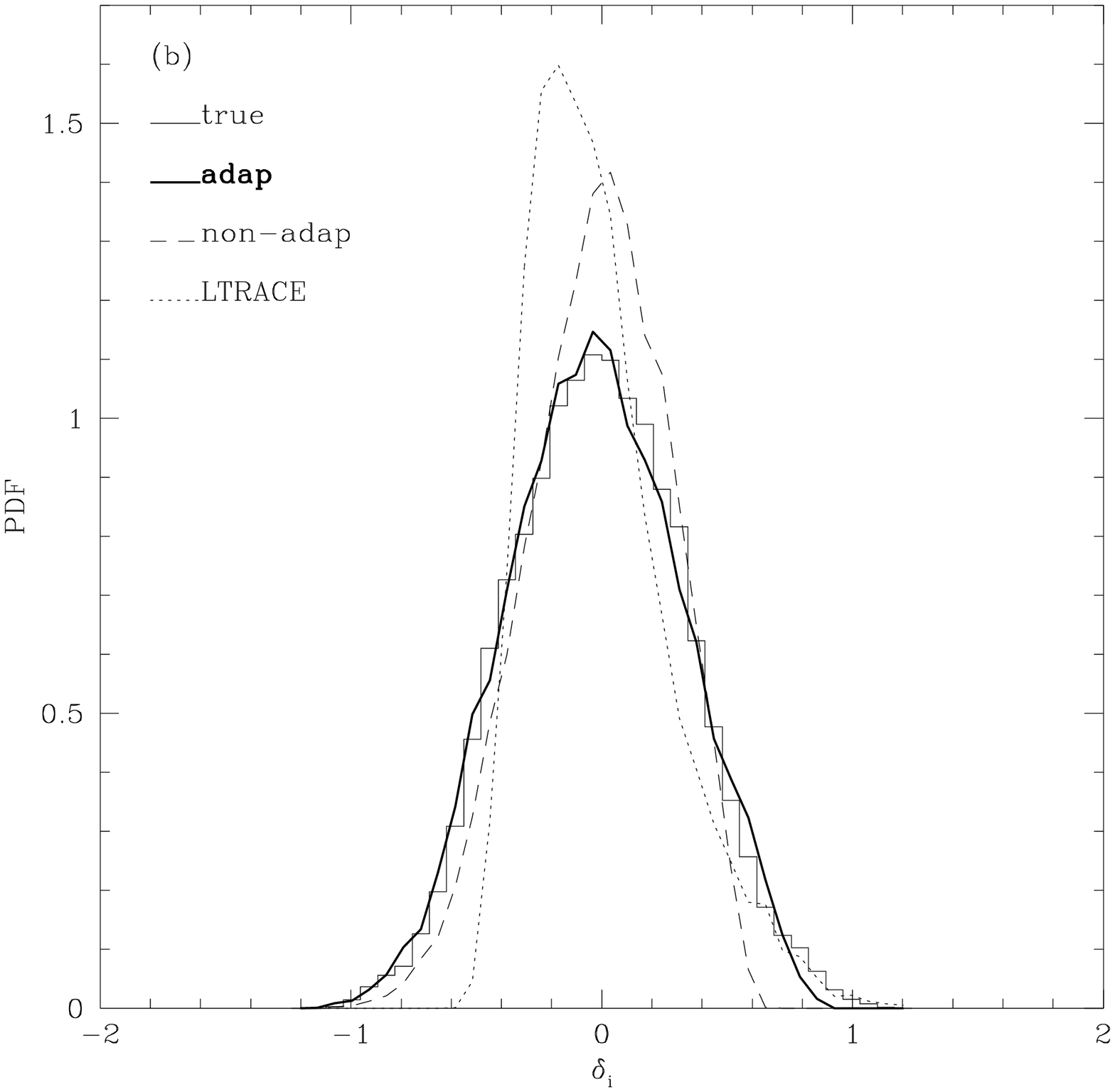,width=9cm}}
\caption{PDF of the linear density recovered by ZTRACE with adaptive
and non-adaptive smoothing compared to the true Gaussian PDF of the
N-body simulation. The dotted lines show the PDF's recovered by
LTRACE.  Panel (a) shows the PDF's within 50 \mpc and a smoothing of
5.15 \mpc; panel (b) shows the PDF's within 80 \mpc and a smoothing
of 7.65 \mpc.}
\end{figure*}

\subsection{Dependence on Cosmology}

The reconstruction in the case of an open universe is carried out in
exactly the same way as in the EdS universe.  The dependence on
cosmology enters the formalism only through the $f(\Omega)$ function
(Eq.~\ref{eq:redmap}), which determines the strength of the peculiar
velocities, and through the growing mode $D(t)$ (Eqs.~\ref{eq:zel},
\ref{eq:2nd} and \ref{eq:poi1}), which determines the rescaling of the
initial density to the present time (the linear density $\delta_l$).
As expected, the results obtained in the open case are
indistinguishable from those presented for the EdS model
in the previous sections.

It is interesting to apply the ZTRACE reconstruction to the evolved
density field in an open universe but assuming an EdS
cosmology.  In this way we can test the error made by ZTRACE as a
result of assuming the wrong cosmology.  In this test the initial
density field is rescaled to the present
time with the correct growing mode.  Fig. 12 shows the recovery of the
real-space density, LOS peculiar velocity and linear density when
ZTRACE is applied to a simulated PSCz catalogue within 80 \mpc\ (with
adaptive smoothing).  As expected, the LOS peculiar velocity is
recovered up to a factor $\Omega^{0.6}\simeq 0.48$ (Fig. 11b), which
means that the inferred LOS displacements are overestimated by nearly
a factor 2; in practice this number is degenerate with the linear bias
parameter $b$ (which is unity in the simulations). This overestimate
of the peculiar velocities should lead to an underestimate of
overdensities and an overestimate of underdensities
in both the real-space and the linear
density fields.  These effects are indeed present in Figs. 12a and c,
but are almost imperceptible.  This is not surprising because the bias
cannot be larger than the differences between the real-space and
redshift-space densities, which are small compared to the scatter in
the reconstruction.

This is another illustration of the intrinsic limits of reconstruction
algorithms applied to realistic galaxy surveys. As a consequence,
while the correct, unbiased, reconstruction of peculiar velocities
requires correct knowledge of the underlying cosmological model, both
the inferred real-space and the linear densities are robust with
respect to cosmology.  This has two consequences: (i) the recovery of
initial conditions and of their PDF is insensitive to the assumed
cosmology; (ii) the information on the cosmological model (actually
the parameter combination $\Omega^{0.6}/b$) is contained almost
exclusively in the peculiar velocities.

\begin{figure*}
\centerline{
\psfig{figure=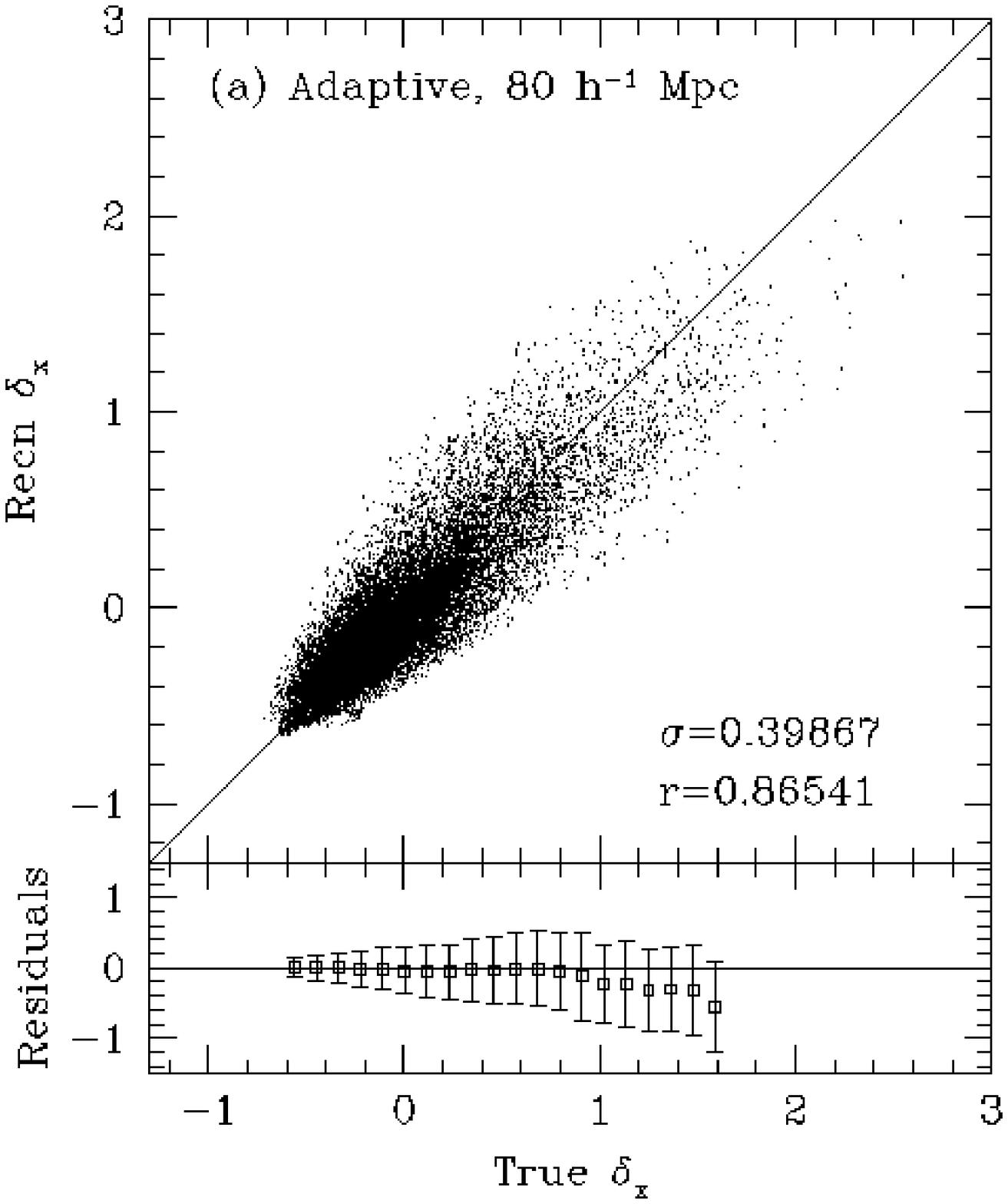,width=6cm}
\psfig{figure=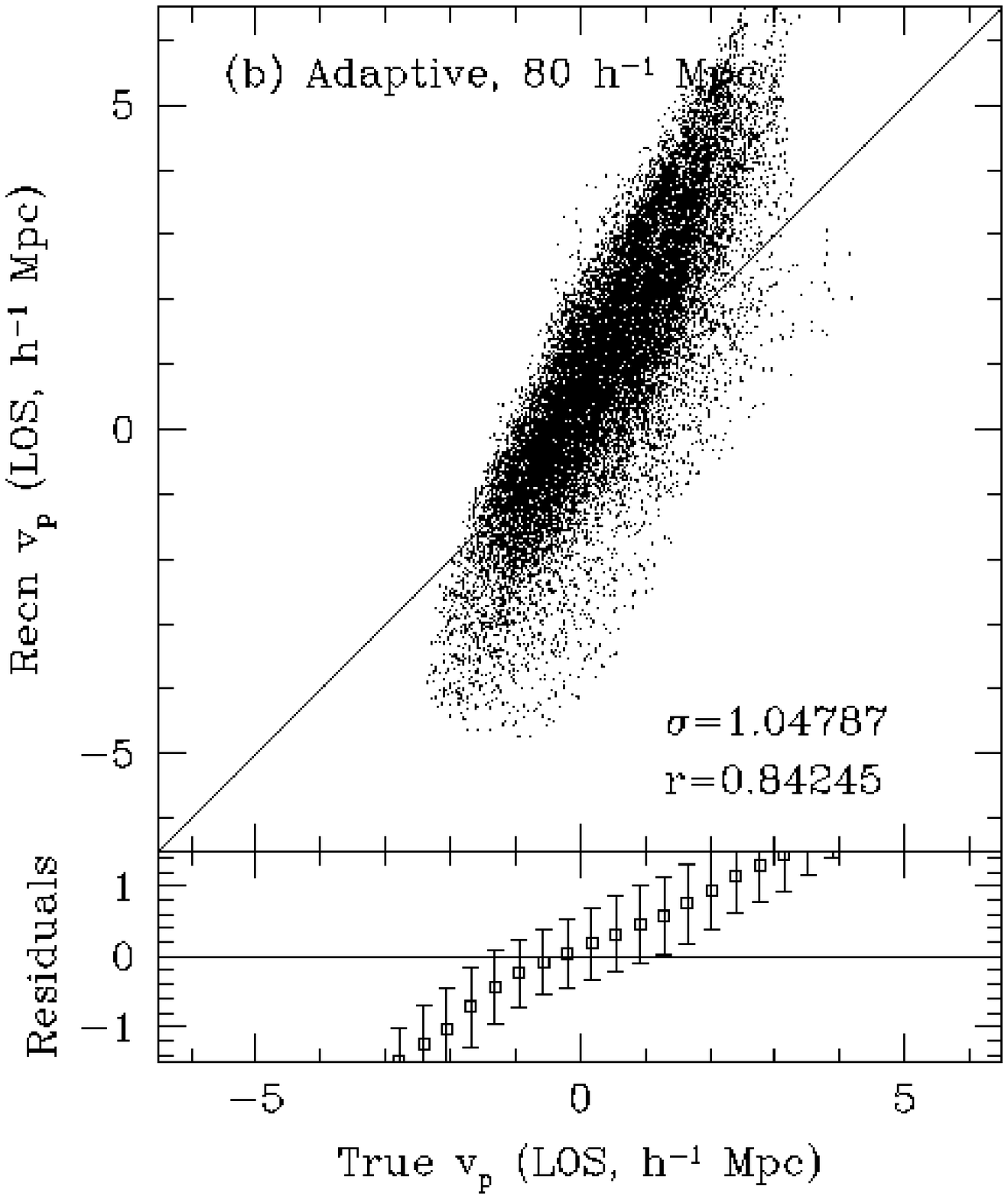,width=6cm}
\psfig{figure=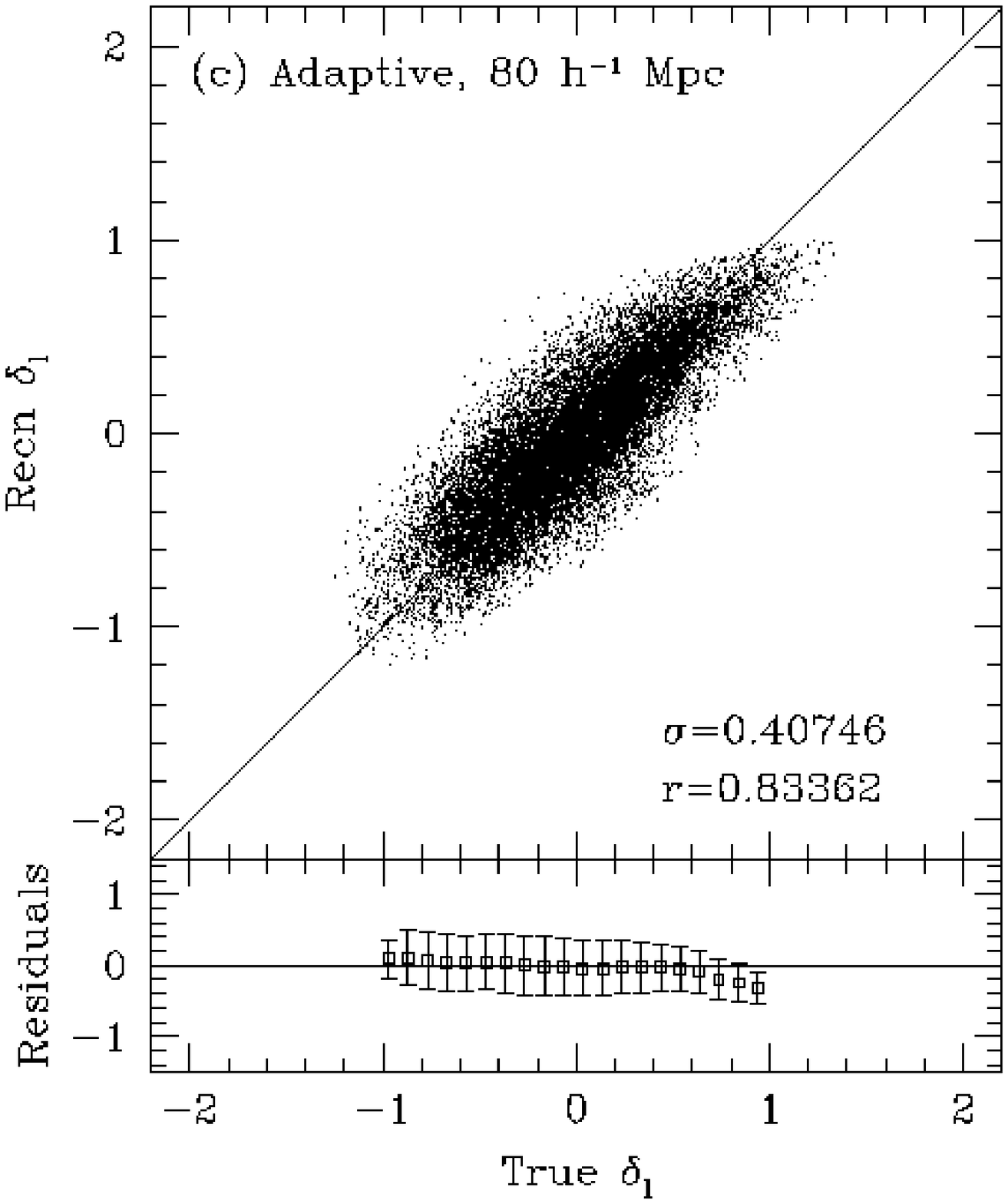,width=6cm}}
\caption{ZTRACE reconstruction using a PSCz catalogue generated from
the open N-body simulation, but where we have assumed an EdS
cosmology.  The figures show results within 80 \mpc\ using an adaptive
smoothing reference radius of 7.65 \mpc.}
\end{figure*}

\section{Conclusions}

We have described a reconstruction algorithm, ZTRACE, based on a
self-consistent solution of the Zel'dovich dynamics of large-scale
structure.  Given an input (observable) density field in redshift
space, this algorithm reconstructs the real-space density field, the
LOS peculiar velocities and the initial conditions which evolve into
this redshift-space density field according to the Zel'dovich
approximation.  The method can be easily extended to second order in
Lagrangian perturbation theory, though we have demonstrated using
N-body simulations that this extension does not give any significant
improvement in the reconstruction.

The ZTRACE algorithm has been tested by recovering the initial
conditions of N-body simulations of  CDM dominated EdS and open
universes with scale-invariant initial power spectra.  The
redshift-space density was estimated using all of the particles in 
the simulation and from simulated galaxy catalogues with the selection
function of the IRAS PSCz survey under the assumption that
galaxies trace mass.  We have also constructed a test version of the
algorithm (XTRACE), which requires the evolved real-space density
field as input, and a linear-theory version (LTRACE).

The major source of noise and bias in the reconstructions comes from
the density field estimates. For simulated PSCz catalogues with
smoothings of several \mpc, the dispersion in the density field
estimates is comparable to the dispersion in the density field itself.
The most significant problem revealed by our tests arose from the fact
that smoothing in Eulerian space does not commute with the dynamics;
it mixes the larger Lagrangian scales involved in overdensities with
the smaller ones involved in underdensities.  The problem can be
solved, at the expense of increased noise in the density estimates, by
applying a mass-preserving adaptive smoothing to the density field.

As a consequence of noise in the density field estimates, the ability
of ZTRACE to recover the real-space density and the LOS peculiar
velocities is not dramatically better than applying a simple algorithm
based on linear theory.  However, the use of adaptive smoothing allows
ZTRACE to reconstruct the initial density field in an unbiased
although noisy way, provided that the linearly extrapolated density does not
exceed a value of unity. Higher extrapolated densities cannot be
recovered accurately because of the presence of multi-stream and
triple-valued regions, which violate the single-stream assumption of
the Zel'dovich approximation.  The PDF of the initial conditions is
correctly reproduced by the reconstruction with ZTRACE to high
accuracy, except for the tail of high overdensities at $\delta_l>1$.

We have shown that the Gaussian PDF of the initial conditions used in
the N-body simulations is recovered from the input catalogues in which
galaxies are assumed to trace the mass.  If the relationship between
the galaxy and the mass density is more complicated, as proposed, for
example, by Mo \& White (1996), Catelan et al. (1998) or Dekel \&
Lahav (1998), then the PDF of the initial conditions recovered by
assuming that galaxies trace the mass may be non-Gaussian.  In a
future paper, the ZTRACE method will be applied to the reconstruction
of the initial conditions of our local Universe from the IRAS PSCz
catalogue and we will investigate how the reconstructed PDF can be
used to constrain models of non-linear bias.

\section*{Acknowledgments}

The Hydra team (H. Couchman, P. Thomas, F. Pearce) has provided the
code for N-body simulations.  Volker Springel has kindly provided a C
version of the code for adaptive smoothing.  The authors thank Tom
Theuns for many useful discussions.  P.M. has been supported by the EC
Marie Curie contract ERB FMBI CT961709. G.E. thanks PPARC for the award
of a Senior Fellowship.

\bsp

\label{lastpage}

\end{document}